\documentclass[11pt]{article}
\usepackage{mathtools}
\usepackage{amsmath}
\usepackage{amsfonts}
\usepackage{dsfont}
\usepackage{amssymb}
\usepackage{graphicx}
\usepackage[all]{xy}
\usepackage[authoryear]{natbib}
\usepackage{xcolor}
\usepackage{multirow}
\usepackage{hyperref}
\usepackage{eurosym}

\usepackage{newtxtext,newtxmath,amsmath}
\usepackage{cleveref}
\usepackage{tikz}

\usepackage[labelsep=period, labelfont=bf,justification=raggedright, singlelinecheck=false, tableposition=top, figureposition=top]{caption}
\usepackage{subcaption}
\usepackage{float}
\floatstyle{plaintop}
\restylefloat{table}
\restylefloat{figure}
\usepackage{hyperref}
\usepackage{booktabs}

\def\actuarial#1{%
  \vbox{
    \offinterlineskip
    \tabskip=0pt
    \mathsurround=0pt
    \halign{##&\vrule##\cr
      \noalign{\hrule}%
      &height 1pt\cr
      $\scriptstyle#1$&\cr
    }%
  }%
}

\DeclareMathAlphabet{\mathcalligra}{T1}{calligra}{m}{k}
\DeclareMathAlphabet{\mathpzc}{OT1}{pzc}{m}{it}

\definecolor{deepblue}{RGB}{0,129,188}
\hypersetup{
unicode=false,
pdftoolbar=true,
pdfmenubar=true,
pdffitwindow=false,
pdfstartview={FitH},
pdfnewwindow=true,
colorlinks=true,
linkcolor=deepblue,
citecolor=deepblue,
filecolor=black,
urlcolor=black,
pdfkeywords={annuities, collaborative data, dependence, family history, genealogy, grandparents-grandchilden, information, joint life insurance, parents-children, whole life insurance}}

\begin{document}

\begin{titlepage}

\begin{center}
\bf\LARGE 

Modeling Joint Lives within Families
\par\end{center}

\bigskip

\begin{center}
\Large by 
\par\end{center}

\renewcommand*{\thefootnote}{\fnsymbol{footnote}}
\begin{center}
\Large
\bigskip
\textbf{Olivier Cabrignac}\\[1ex]
\large
SCOR\\
5 Avenue Kléber, 75795 Paris, France\\
ocabrignac@scor.com
\bigskip

\Large
\bigskip
\textbf{Arthur Charpentier}\\[1ex]
\large
Universit\' e du Qu\'ebec \`a Montr\'eal (UQAM)\\
201, avenue du Pr\'esident-Kennedy, \\
Montr\'eal (Qu\'ebec), H2X 3Y7, Canada
\\ charpentier.arthur{@}uqam.ca
\bigskip

\Large
\bigskip
\textbf{Ewen Gallic}\\[1ex]
\large
Aix-Marseille Univ., CNRS, EHESS, Centrale Marseille, AMSE \\
5-7 boulevard Maurice Bourdet CS 50498\\
13205 Marseille Cedex 01, France
\\ ewen.gallic@univ-amu.fr

\par\end{center}

\setcounter{footnote}{0}

\vfill
\begin{center}
\large 
June 2020
\par\end{center}

\vfill
\indent A. Charpentier  acknowledges the support of the Natural Sciences and Engineering Research Council of Canada Grant NSERC-2019-07077. E. Gallic acknowledges the support of the French National Research Agency Grant ANR-17-EURE-0020. Initial work on the dataset was funded by the ACTINFO chair, of the Institut Louis Bachelier. O. Cabrignac points out that this article is not meant to represent the position or opinion of SCOR.

\end{titlepage}

\begin{abstract}

Family history is usually seen as a significant factor insurance companies look at when applying for a life insurance policy. Where it is used, family history of cardiovascular diseases, death by cancer, or family history of high blood pressure and diabetes could result in higher premiums or no coverage at all. In this article, we use massive (historical) data to study dependencies between life length within families. If joint life contracts (between a husband and a wife) have been long studied in actuarial literature, little is known about child and parents dependencies. We illustrate those dependencies using 19th century family trees in France, and quantify implications in annuities computations. For parents and children, we observe a modest but significant positive association between life lengths. It yields different estimates for remaining life expectancy, present values of annuities, or whole life insurance guarantee, given information about the parents (such as the number of parents alive). A similar but weaker pattern is observed when using information on grandparents.




\vfill
\end{abstract}


\vfill
\indent The authors thank the participants of the Online International Conference in Actuarial Science, Data Science and Finance (OICA) for stimulating questions. More particularly thanks to Tim J. Boonen and Montserrat Guillen for feedbacks. The authors also wish to thank Jer\^ome Galichon and Geneanet for kindly providing the dataset.

\newpage

{
\small%
\tableofcontents
\normalsize%
}
\newpage

\section{Introduction}

\begin{figure}[htb!]
    \centering
    \includegraphics[width=.872\textwidth]{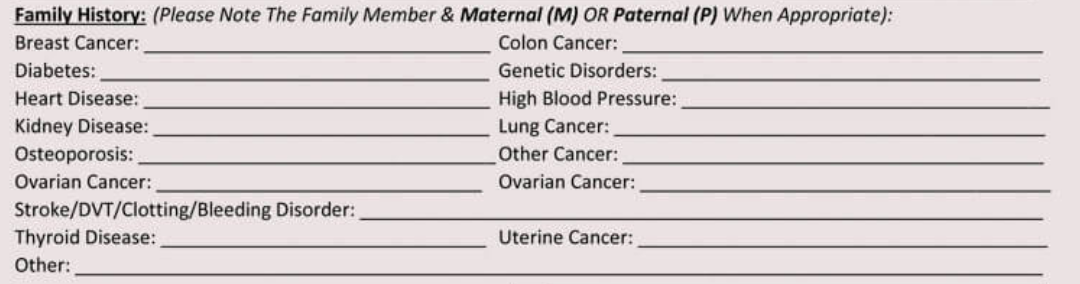}
    
    \vspace{.3cm}
    \includegraphics[width=.872\textwidth]{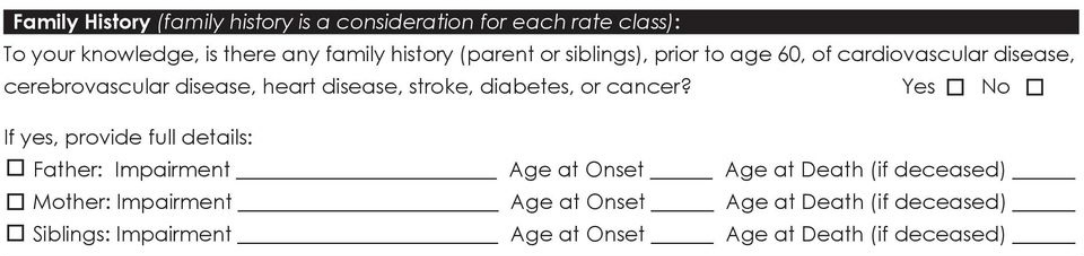}    

    \vspace{.3cm}
    \includegraphics[width=.872\textwidth]{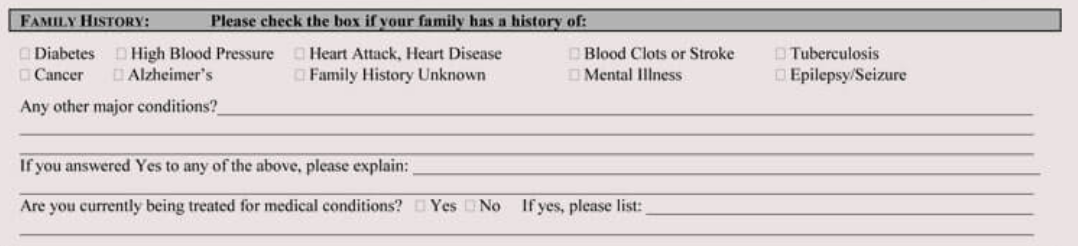}    

    \vspace{.3cm}
    \includegraphics[width=.872\textwidth]{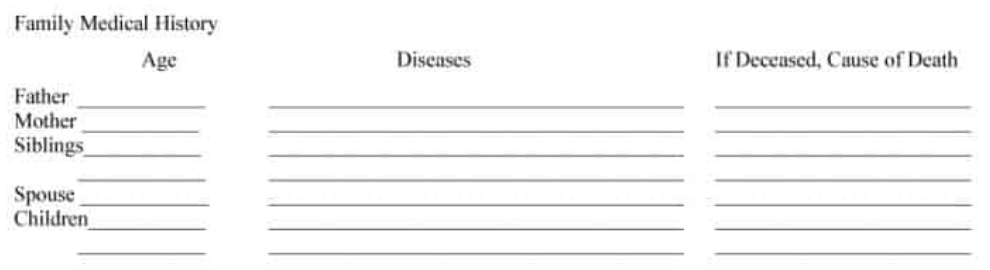}  
     \caption{Excerpts of medical history classical forms, with the {\em family medical history} section (source: \href{https://printabletemplates.com/medical/medical-history-form/}{\sffamily https://printabletemplates.com/})}
    \label{fig:ins:forms}
\end{figure}

Family history is usually seen as a significant factor insurance companies look at when applying for a life insurance policy.\footnote{Where it is allowed, \textit{e.g.}, in North America or in the United Kingdom, for instance. Some data privacy regulation may, however, restrict access to those data. It should also be noted that such practice is not allowed in most European countries, as discussed in \cite{Schmitz297}.}  As shown in Figure \ref{fig:ins:forms}, family history of cardiovascular diseases, cancers (ovarian, breast, colon, lung, etc), or family history of high blood pressure and diabetes are usually asked in medical forms. Family can include parents (father and mother), siblings, spouse, children, and if deceased, the age at death can be asked. As discussed in \citet{CutlerZeckhauser} or \citet{PardoSchott}, the information provided can yield higher insurance premiums, or no coverage at all.

Assuming that family history should have an impact on premiums means that the risk of the policyholder is correlated with information related to siblings, parents, etc. 
\cite{FreesCarriereValdez1996} considered the case of insurance products with dependent mortality, on spouses, showing that life lenghts with a married couple are (positively) correlated, and that this correlation should reflect in standard joint life annuities. The literature has not limited itself to looking at life course relationships within couples. The idea of the existence of a ``\textit{longevity inheritance}'', as \cite{Pearl} named it, has given rise to a considerable body of work to study longevity ties within families. However, most studies have relied on modest sample sizes.

In this article, we use a massive dataset of family trees to study correlations between life lengths between relatives. This dataset concerns individuals born in metropolitan France at the beginning of the 19th century and their descendants. Due to the historical nature of the data, and in contrast with the individuals on which \cite{FreesCarriereValdez1996} focused, the grandchildren of individuals born at the beginning of the 19th century have all died by now. Their age at the time of death is \textit{de facto} observable. We therefore avoid dealing with incomplete and partial data when studying mortality.\footnote{For example, since we face complete data regarding age at death, a convenient way to approximate life expectancy is to simply use the average age at death, no modeling assumption is required here.} In a first step, we use this rich data to explore joint mortality within couples. The analysis is based on $135,128$ pairs of individuals born in the 19th century. It focuses solely on demographic characteristics. In a second and third steps, knowledge of the descendants makes it possible not only to investigate the links between children and parents, but also between children and their grandparents. To this end, we use $174,318$ observations linking individuals to their parents where information about both parents is available, and $59,463$ individuals for whom the birth and death dates of the 4 grandparents are known. In each of the three cases, we begin by studying the correlation of life spans. Then we move from demographics to life insurance actuarial present values.

In line with what is reported in the literature, a positive association in mortality within couples is observed with our data.  {This dependence has a significant impact on joint life insurance product, consistent with results already obtained in previous literature (but usually with smaller datasets). Similarly, we also find a small but significant relationship between children and parents life lengths. With both parents still alive when twenty years old, life expectancy is relatively higher than when one or neither parent is still alive. Similar results are observed at older ages (30 and 40 years). Simulations based on historical data show that knowing whether a person's parents are still alive at any age has significant effects from an insurer's perspective. For example, for a young male individual (about 30 years old), knowing that both parents are still alive translates into an increase of about 2 years in his life expectancy relative to a person whose both parents are deceased. And at the same age, annuities should be 4\% higher than the average population when both parents are alive, and 4\% lower when both parents deceased. For a whole life insurance, the order of magnitude is almost the same (but opposite as insurance premiums are decreasing with life expectancy). Finally, the links are much less important between a person's mortality and that of their grandparents. As a result, the implications in terms of insurance are weaker and more questionable, with $\pm 1\%$, at best.}

The remainder of the article is structured as follows. Section~\ref{sec:description:data} concisely describes the datasets. Section~\ref{sec:mortality_models_to_insurance} first recalls classical notations and concepts used when modeling (univariate and then joint) mortality and then presents various insurance products. In Section~\ref{sec:couple}, based on 19th century data, joint life dependencies between husbands and wives is investigated. Section~\ref{sec:parents:grandparents} discusses inter-generational dependencies for life lengths. More specifically, the links between an individual's remaining lifetime and information about his or her parents are first studied (these include, for example, determining the remaining life span of a 40 years old person, given that both his or her parents are either still alive, or dead). Then, a similar analysis is conducted to study the relationship between an individual's remaining lifetime and the characteristics of his or her grandparents. Section~\ref{sec:conclusion} concludes.

\section{Description of the Data}\label{sec:description:data}

The data used in this paper come from a genealogy website, Geneanet.\footnote{\href{https://www.geneanet.org/}{\sffamily https://www.geneanet.org/}.} On this website, each user is invited to build their own family tree and can decide to share it with the rest of the community. We rely on these shared trees. Each user's tree contains, with varying degrees of completeness, information about the family members who make up the tree. This information concerns the events that can be found on civil and religious registers, \textit{i.e.}, birth, marriage, if any, and death. More specifically, we are interested in the dates of birth and death of individuals, as well as their family relationships. The sample we have contains individuals born in metropolitan France between 1800 and 1804, their descendants (and their parents) -- keeping only those born before 1900.\footnote{The scope of Geneanet's data mainly concerns European countries, and more particularly France.} It is well established that information regarding parents is quite well represented, as long as the information exists, but information related to other first-degree relatives (siblings, children) -- not to say second-degree relatives (cousins, grand-children) is more sparse. Thus, the extraction we have can have possible bias, as discussed in \citet{charpentier_gallic}. 

As the data focus on individuals born before 1900, all of them are dead by now. We do not face any censored-observation problem. However, some challenges arise with these data. Since each user creates their own family tree, an ancestor common to the trees of multiple users may be present several times in the raw data. To avoid redundancy, we refer to \citet{charpentier_gallic, charpentier_2020_genealogy} for more explanations about the methodology. In a nutshell, an important challenge was to deal with natural overlap of various family trees, and to identity an ancestor appearing in several trees as a unique individual.\footnote{That process of merging trees had to take into account typos in names, and (partially) missing information about dates.} In \citet{charpentier_gallic}, we proved that while there were bias in collaborative genealogical data (compared with official demographic data), more specifically on infant and young age mortality, these data provided very accurate mortality information (on force of mortality or remaining life expectancy, for ages higher than 20). In \citet{charpentier_2020_genealogy}, the goal was to study family migration, from a starting ancestor, and the challenge was to get a forward genealogical analysis (from ancestor to descendants) from backward genealogical data (from descendant to ancestors).

Once the trees have been matched and the data cleaned, a reference dataset is obtained containing, for each individual in the trees, information about: their birth and death dates, and a link to their parents' identifier. Two databases can then be created to study: ($i$) the dependencies within couples (in Section~\ref{sec:couple}) , and ($ii$) the inter-generational dependencies for life lengths (in Section~\ref{sec:parents:grandparents}). The remainder of this section focuses on providing more detail on these two datasets.

\subsection{Husband and Wife Dependencies}

There is an intensive literature about the positive association of life spans between couples, and more specifically on bereavement. \cite{Riley} suggests several sociological explanation of the positive association among life lengths, reinforced by mortality excess following the death of the first one. 
Some articles also focus on medical (mainly psychological) aspects. For example, 4,486 widowers of 55 years old (and older) have been followed up for nine years since the death of their wives in 1957 in \cite{Parkes}. In the first six months about 25\% of the deaths were from the same diagnostic group as the wife's death, and a higher mortality was observed. The authors concluded that there is no evidence suggesting that the proportion is any different among widows and widowers who have been bereaved for more than six months. \cite{Parkes} coined {\em broken heart syndrom} to describe that short term (positive) correlation. This was confirmed in \cite{Jagger}, but on a smaller dataset (with 344 elderly persons who were living with a spouse and who were part of a survey of a population of people aged 75 years and over). Nevertheless, \cite{Kastenbaum} claims that a large part of the excess of mortality following the death of the first one can be related to common health problems, rather than psychological trauma following the loss (see also more recently \cite{Espinosa} that reached a similar conclusion). 

We create a first set of data from family trees to see if the same type of associations that have been described in the literature between the lifespan of the members of a couple can be observed with our data. It is important to note that the term {\em couple} used in this study refers to a very specific definition. We do not rely on the definition of {\em married couple} (from census types of datasets), as in \cite{Glick} for example. Like studies based on genealogical data, such as \cite{Pearson}, we study here mothers and fathers. This comes from the structure of the data. Couples are observed as {\em parents}, and it is rather uncommon with genealogy data to have couples without children. We thus make two assumptions: ($i$) two people are defined here as a couple when they had a child together, and ($ii$), we further assume that parents were living together. Consequently, two persons who lived in a union but did not have a child (or without information on possible children available in the data) are not present in the dataset. This could lead to potential biases, but the main purpose of the section devoted to the study of joint mortality within couples is to see if the dataset in our hands provides similar results than previous studies.

Before going any further, it seems important to provide more details on the structure of the data. For individuals observed as {\em parents}, a table in which each row corresponds to a couple can be created. Each row contains the dates of birth and death of a father and a mother, as shown in the example provided in Table~\ref{tab:data:1} (with only 6 random observations). It should be noted that information for all four dates may not be fully available. Two situations arise. First, the date of birth or death of one or both members of the couple may simply be missing. When such a case occurs, we remove the couple from the data. Secondly, some dates may be incomplete: the month and day may not be stated. In such a case, the date is converted into July 1st of that year, which corresponds to the average date assuming uniform birth over the calendar year\footnote{In that dataset, 14.3\% of dates for men and 16.1\% of dates for women were incomplete, usually more for death than birth (6.6\% and 7.5\% for birth dates for men and women respectively, and 10.2\% and 11.3\% of death dates). We assume here that the impact on various quantities (including correlation) would be rather small}.
Finally, we end up with a dataset of $n=135,128$ couples, where husbands and wives were born between 1800 and 1870.

\begin{table}[htp!]
    \centering\small
    \begin{tabular}{cccc|ccc}\hline\hline\\[-.75em]
    & \multicolumn{3}{c}{Father/Husband} & \multicolumn{3}{c}{Mother/Wife}\\[.25em]
    \cmidrule(lr){2-4}\cmidrule(lr){5-7}\\[-.75em]
      &  birth & death & age & birth & death  & age  \\[.25em]
  \hline\\[-.75em]
1 & 1804-03-18 & 1880-09-21 & 76.5 & 1804-02-29 & 1854-08-26 & 50.5\\
2 & 1836-12-27 & 1902-04-06 & 65.3 & 1832-02-06 & 1901-09-11 & 69.6\\
3 & 1804-03-30 & 1870-01-17 & 65.8 & 1800-07-10 & 1868-07-12 & 68.0\\
4 & 1804-02-22 & 1876-01-06 & 71.9 & 1803-05-04 & 1881-05-16 & 78.0\\
5 & 1800-09-11 & 1837-08-15 & 36.9 & 1800-00-00 & 1836-08-11 & 36.1\\
6 & 1800-00-00 & 1843-03-08 & 42.7 & 1800-00-00 & 1865-03-15 & 64.7\\[.25em]
 \hline\hline\\[-.75em]
    \end{tabular}
    \begin{minipage}{\textwidth}
		\vspace{1ex}
		\scriptsize\underline{Note:} Random extraction of 6 rows of our entire dataset. In the case where only the year is mentioned, the date is eventually converted into July 1st of that year (mid-year, average date assuming uniform birth over the calendar year).
\end{minipage}
    \caption{Dataset for the joint life model, father/husband and mother/wife.}
    \label{tab:data:1}
\end{table}

When this dataset is created, it is straightforward to calculate the ages at death of each member of the couple, that are denoted $t_{\text{f}}$ and $t_{\text{m}}$ for the father and the mother, respectively. As shown in Table~\ref{tab:desc_stat_couples}, husbands and wives have a very similar life expectancy. (about 62 years old). As confirmed in \cite{Beltran8993} the differences between men and women's life expectancies (admitted as a fact, nowadays) began to emerge only in the late 1800s, in most modern countries. It may be noted that the average ages at death we observe are relatively high for the period; it should be remembered that the individuals studied do not concern the entire population, but only those who have had children. Hence, the average is not driven down by the high infant mortality observed in 19th century France. The age difference in the couples studied shows that men are, on average, slightly older than their wives by two years, albeit with a significant standard deviation. In fact, in a majority of couples ($58\%$), the man is older than his wife.

\begin{table}[htb!]
    \centering
    \begin{tabular}{lrrrrrrr}
\hline\hline\\[-.75em]
Variable & Mean & SD & Min & Max & $Q_1$ & $Q_2$ & $Q_3$\\[.25em]
\hline\\[-.75em]
Husband age at death $t_{\textrm{f}}$ & 63.2 & 15.5 & 15 & 105 & 52.5 & 65.4 & 75.0\\
Wife age at death $t_{\textrm{m}}$ & 62.4 & 17.3 & 15 & 105 & 50.1 & 65.2 & 75.9\\
Age difference & -2.2 & 12.1 & -80 & 80 & -5.0 & -1.0 & 1.0\\
\hline\hline
\end{tabular}
\begin{minipage}{\textwidth}
		\vspace{1ex}
		\scriptsize\underline{Note:} Husbands and wives were born between 1800 and 1870 ($n=135,128$). $SD$ stands for standard deviation, $Q_1$, $Q_2$, and $Q_3$ stand for the first, second, and third empirical quartiles. The age difference indicates the number of years separating the man and the woman. Negative values indicate that the man is older than the woman.
\end{minipage}
    \caption{Descriptive statistics for husbands and wives.}
    \label{tab:desc_stat_couples}
\end{table}

\subsection{Ancestors and Children Dependencies}

We explained in the introduction that family history can be important in insurance. Almost a century ago, \cite{Pearl} mentioned that ``{\em even a business so precise in some particulars as life insurance, which has, from its beginnings, acted on the assumption that the duration of life of an individual's near kinsfolk is of importance in estimating the nature of its risk accepted in insuring him, has made singularly little effort to determine exactly the weight of this factor.}''

We explore family history in two parts, by analysing the relationship between an individual's lifespan and that of: ($i$) his or her parents, and ($ii$) his or her grandparents.

\subsubsection{Parents and Children Dependencies}

The historical reference on dependencies between life lengths of parents and children is probably \cite{Pearson} (with also a lot of concern about brothers and sisters - we refer to \cite{BandeenRoche} for a modern perspective on that issue). In that study, they also use 19th century genealogical data (which would give us an order of comparison, even if most computations were performed on about 1,000 pairs).

The idea of ``{\em inheritance of longevity}” (as \citealp{Pearl} named it) received a lot of attention, and has been intensively discussed in the literature over the years, with various types of data. For example, 
\cite{Gudmundsson} based on Icelandic population, concluded that longevity was inherited within families, in their view probably because of shared genes.
\cite{Hjelmborg} looking at twin data, concluded that genetic influences on the lifespan were minimal before age 60 and only increase after that age. \cite{Kowald}, on the other hand, rejected any idea that mortality in old age is genetically programmed. Consistent with that view, a Swedish study of men born in 1913, found that a number of social and behavioural factors measured at age 50 were better predictors of longevity than their parents' survivorship, predicted longevity \cite{WilhelmsenEtAl2011}.
Caution must be exercised when advancing the idea of genetic transmission, especially when relying on genealogical data that provide no evidence of the veracity of the relationship between individuals. Indeed, as pointed out in \cite{Bellis_2005_JECH}, the median rates of paternal divergence as measured in studies conducted mainly during the second half of the 20th century amount to $3.7\%$, with some studies even advancing values as high as $30\%$ (considering tests performed on selected populations for reasons other than disputed paternity).

\cite{Mayer} studied up to six family pedigrees, from 1650 to 1874. An important issue in the literature is to understand if that correlation (or transmission) is due to genetics, or associated life styles, or social class. For example, \cite{Ruby1109} explains that ``{\em the majority of that correlation was also captured by correlations among non genetic (in-law) relatives, suggestive of highly assortative mating around life span-influencing factors (genetic and/or environmental)}''.  \cite{Philippe} mentions that spouse life spans ``{\em correlate as much or more than those of genetic relatives}''. See also \cite{Garibotti}, \cite{PIRAINO2014105} or \cite{temby_smith_2014}, who try to distinguish genetic effects to socioeconomic status in family history. \cite{Philippe2} suggests that there could be an overestimation of the possible positive correlation, presents a feature of some {\em elite sub-class}, but not of the general population of a community. Almost all studies are based on rather small sample. For example, \cite{Abbott} studied 7,103 progeny, sons and daughters of 1,766 men or women, who were alive in 1922–1930. \cite{Matroos} had 2,370 middle-aged children, while \cite{Vaillant} used a cohort of (only) 184 men.

Nevertheless, most studies confirm a significant but weak association within families. \cite{Bocquet-Appel} analyzed familial correlations of longevity at Arthez d'Asson, for individuals born between 1686 and 1899. At birth, the correlation is rather small (0.103), but it increases with the age of the son. For example, at 20 years old, the correlation is larger (0.167). As \cite{Vaupel1988} wrote it 
``{\em the life spans of parents and children appear only weakly related, even though parents affect their children’s longevity through both genetic and environmental influences}'' (see also \cite{Vaupel1979} or \cite{Vaupel855}, with similar conclusions).

To address the issues developed in this literature, a second set of data was created from the genealogy data. By using the parents' identifiers, and then those of the parents' parents, it is possible to create a table in which each observation provides the dates of birth and death of individuals, their parents, and their grandparents. As in the case of couples, dates may be missing or incomplete. If the treatment is the exact same for the case of incomplete dates, the treatment of missing dates differs depending on whether the focus is on parents or grandparents. For parents, as for couples, if one of the dates of birth or death is not available, then the individual is removed from the observations. For grandparents, if both dates of birth and death of all grandparents are missing, then the individual is removed from the observations. It should therefore be noted that if the information needed to calculate the age at death of at least one of the grandparents is available, the individual is kept.

The sample thus allows us to focus on the relationship between the age at death of an individual, $t_{\text{c}}$, and that of his or her parents, $t_{\text{f}}$ and $t_{\text{m}}$, for the father and the mother, respectively. As reported in Table~\ref{tab:desc_stat_parents} the sample is composed of $174,318$ observations for which full information on ages at death is available. It consists of two parts of $90,828$ men and $83,490$ women. The average age at death of individuals is 44.5 years (43.4 for men and 45.7 for women). We wish to compare this measure with that of the parents. This can be done by looking separately at the age at death of each parent, \textit{i.e.}, $t_{\text{f}}$ and $t_{\text{m}}$. It is also possible to construct indicators to represent the age at death of parents in the household. Three are under consideration. First, average age at death of the parents, $\text{mean}\{t_{\text{f}},t_{\text{m}}\}$, which is equal to $64.5$. It can be noted that it is relatively higher than that of the children. This can be explained by the fact that individuals who died at a young age are not taken into account: if parents are present in the data, they have necessarily had children (survivor bias). The sample of parents and children is constructed starting from children born between 1800 and 1900. Second, we look at the age at death of the first to die, $\min\{t_{\text{f}},t_{\text{m}}\}$, which quite a bit lower, $56.4$ years on average. And third, we consider the age at death of the last survivor, $\max\{t_{\text{f}},t_{\text{m}}\}$, which is equal to $72.2$ on average.

\begin{table}[htb!]
    \centering\scriptsize
    \begin{tabular}{lrrrrrr}
\hline\hline\\[-.75em]
& Mean & SD & Min & Max & $Q_1$ & $Q_3$\\
\cmidrule(lr){2-7}\\[-.75em]
& \multicolumn{6}{c}{Men ($n=90,828$)}\\[.25em]
\cmidrule(lr){2-7}\\[-.75em]
Individual ($t_{\text{c}}$) & 43.4 & 30.1 & 0 & 104.4 & 10.2 & 70.0\\
Father ($t_{\text{f}}$) & 64.7 & 14.6 & 15 & 104.0 & 54.8 & 75.7\\
Mother ($t_{\text{m}}$) & 64.2 & 16.4 & 15 & 104.8 & 53.1 & 76.8\\
First to die ($\min\{t_{\text{f}},t_{\text{m}}\}$) & 56.6 & 14.7 & 15 & 100.0 & 45.6 & 68.2\\
Last survivor ($\max\{t_{\text{f}},t_{\text{m}}\}$) & 72.3 & 12.0 & 15 & 104.8 & 65.7 & 80.8\\
Average parents ($\text{mean}\{t_{\text{f}},t_{\text{m}}\}$) & 64.4 & 11.8 & 15 & 100.9 & 56.5 & 73.3\\
\cmidrule(lr){2-7}\\[-.75em]
& \multicolumn{6}{c}{Women ($n=83,490$)}\\[.25em]
\cmidrule(lr){2-7}\\[-.75em]
Individual ($t_{\text{c}}$) & 45.7 & 32.0 & 0 & 104.7 & 9.9 & 74.7\\
Father ($t_{\text{f}}$) & 64.3 & 14.8 & 15 & 104.0 & 54.2 & 75.6\\
Mother ($t_{\text{m}}$) & 64.0 & 16.6 & 15 & 104.4 & 52.5 & 76.7\\
First to die ($\min\{t_{\text{f}},t_{\text{m}}\}$) & 56.2 & 14.8 & 15 & 100.8 & 45.1 & 67.9\\
Last survivor ($\max\{t_{\text{f}},t_{\text{m}}\}$) & 72.1 & 12.2 & 15 & 104.4 & 65.3 & 80.8\\
Average parents ($\text{mean}\{t_{\text{f}},t_{\text{m}}\}$) & 64.2 & 11.9 & 15 & 101.4 & 56.1 & 73.0\\
\hline\\[-.75em]
& \multicolumn{6}{c}{Men \& Women ($n=174,318$)}\\[.25em]
\cmidrule(lr){2-7}\\[-.75em]
Individual ($t_{\text{c}}$) & 44.5 & 31.1 & 0 & 104.7 & 10.0 & 72.2\\
Father ($t_{\text{f}}$) & 64.5 & 14.7 & 15 & 104.0 & 54.5 & 75.6\\
Mother ($t_{\text{m}}$) & 64.1 & 16.5 & 15 & 104.8 & 52.8 & 76.7\\
First to die ($\min\{t_{\text{f}},t_{\text{m}}\}$) & 56.4 & 14.7 & 15 & 100.8 & 45.3 & 68.0\\
Last survivor ($\max\{t_{\text{f}},t_{\text{m}}\}$) & 72.2 & 12.1 & 15 & 104.8 & 65.5 & 80.8\\
Average parents ($\text{mean}\{t_{\text{f}},t_{\text{m}}\}$) & 64.3 & 11.8 & 15 & 101.4 & 56.3 & 73.2\\[.25em]
\hline\hline
\end{tabular}
\begin{minipage}{\textwidth}
		\vspace{1ex}
		\scriptsize\underline{Note:} this table provides descriptive statistics of the ages contained in the dataset of children and parents. $n$ stands for the number of observations, $SD$ is the standard deviation, $Q_1$ and $Q_3$ are the first and third empirical quartiles, respectively.
\end{minipage}
    \caption{Age at death of the individuals and that of their parents, according to the gender of the children.}
    \label{tab:desc_stat_parents}
\end{table}

Table~\ref{tab:desc_stat_parents_decades_408} shows the distribution of the number of individuals by decade. Because of the way in which the sample was constructed, starting with individuals born in France between 1800 and 1804, a relatively high proportion of individuals in the raw data belong to the cohorts $(1790-1800]$ and $(1800-1810]$.\footnote{The 1790-1800 cohort is composed of individuals born in 1800 only.} However, incomplete information on the dates of birth and death of both parents for these people is only rarely available. As a result, the data used in the analysis contain few individuals from these cohorts. Also due to the constitution of the sample, the number of individuals in the $(1810,1820)$ cohort is very small, since we must wait until individuals born between 1800 and 1804 have had children before they can take on the role of parents in the database. For subsequent cohorts the observed average values of the ages at death of the children or their parents remain within the same orders of magnitude as for the rest of the sample.

\begin{table}[htb]
    \centering\scriptsize
    \begin{tabular}{lrrrrr}
\hline\hline\\[-.75em]
Cohort & $n$ & Age individual $t_{\text{c}}$ & Age father $t_{\text{f}}$ & Age mother $t_{\text{m}}$ & Prop. Women ($\%$)\\
\hline\\[-.75em]
(1790,1800] & $476$ & 42.5 & 62.0 & 59.4 & 47.7\\
(1800,1810] & $3,332$ & 39.9 & 62.8 & 59.9 & 47.8\\
(1810,1820] & $250$ & 47.4 & 58.7 & 58.4 & 42.8\\
(1820,1830] & $30,445$ & 42.3 & 63.4 & 62.3 & 46.5\\
(1830,1840] & $41,345$ & 38.8 & 65.2 & 64.4 & 46.9\\
(1840,1850] & $12,492$ & 36.9 & 66.4 & 66.1 & 47.4\\
(1850,1860] & $11,409$ & 40.6 & 62.7 & 61.4 & 48.7\\
(1860,1870] & $19,311$ & 42.8 & 63.1 & 62.1 & 49.2\\
(1870,1880] & $14,813$ & 48.8 & 64.2 & 63.2 & 50.0\\
(1880,1890] & $15,862$ & 53.2 & 64.8 & 65.4 & 48.8\\
(1890,1900] & $24,583$ & 56.2 & 66.0 & 68.3 & 48.4\\[.25em]
\hline\hline
\end{tabular}
\begin{minipage}{\textwidth}
		\vspace{1ex}
		\scriptsize\underline{Note:} This table reports some key descriptive statistics for the individuals contained in the dataset used to study the dependence between children and parents. $n$ stands for the number of observations, Prop. Women gives the proportion of women among the individuals, for each cohort.
		
\end{minipage}
    \caption{Information about individuals per cohort.}
    \label{tab:desc_stat_parents_decades_408}
\end{table}

\subsubsection{Grandparents and Children Dependencies}

As mentioned recently in \cite{Choi2020}, ``{\em little is known about whether and how intergenerational relationships influence older adult mortality}'', especially between children and their grandparents. Our sample makes it possible to investigate this question. We adopt the following notations: $t_{\text{gff}}$ and $t_{\text{gmf}}$ for the grandfather and grandmother on the father's side, and $t_{\text{gfm}}$ and $t_{\text{gmm}}$ for the grandfather and grandmother on the mother's side.

As with the relationship between parents and children, Table~\ref{tab:desc_stat_gparents} reports some descriptive statistics on the age at death of the grand parents, keeping only the individuals for whom birth and death dates are known for at least one grandparent (which does not necessarily imply that the birth and death dates of both \textit{parents} are complete).\footnote{It should be noted that the number of individuals for whom information on all four grandparents is available is small relative to the sample size: $31,096$ men and $28,367$ women. See Table~\ref{tab:desc_stat_gparents_4gp} in the Appendix. On average, the raw data provide information only on $1.6$ grandparents (1st quartile is 1 and 3rd quartile is 2).} As can be seen, the distribution of males is higher than that of females ($831,479$ versus $740,461$). The average age at death of grandchildren is $43.7$ ($41.9$ for men and $45.8$ for women). As is the case for what is observed with parents, the age at death of grandparents is much higher in our data, ranging from $62.8$ to $64.6$ years on average. This difference is mainly explained by some survivor bias: grandparents have been parents, so at least, they reach the 20's, while their grandchildren include a lot of individuals who died very early. For instance, 25\% of the grand children did not live beyond 6 years old.   
The three indicators that make it possible to subsequently synthesize the relationship between an individual's lifespan and that of his or her grandparents (\textit{i.e.}, last survivor, first to die and average age at death) provide the same orders of magnitude as those for parents.

\begin{table}[H]
    \centering\scriptsize
    \begin{tabular}{lrrrrrrrr}
\hline\hline\\[-.75em]
& Mean & SD & Min & Max & $Q_1$ & $Q_2$ & $Q_3$ & No. Missing\\
\cmidrule(lr){2-9}\\[-.75em]
& \multicolumn{8}{c}{Men ($n=831,479$)}\\[.25em]
\cmidrule(lr){2-9}\\[-.75em]
Individual ($t_{\text{c}}$) & 41.9 & 31.0 & 0 & 104.7 & 5.4 & 45.3 & 70.4 & 0\\
Maternal grandfather & 64.1 & 14.6 & 15 & 104.7 & 54.1 & 66.1 & 75.2 & 526,191\\
Maternal grandmother & 62.8 & 16.0 & 15 & 104.4 & 51.8 & 65.3 & 75.1 & 511,685\\
Paternal grandfather & 64.6 & 14.3 & 15 & 104.7 & 55.1 & 66.7 & 75.4 & 492,622\\
Paternal grandmother & 63.3 & 15.7 & 15 & 104.4 & 52.8 & 65.8 & 75.2 & 489,805\\
Last survivor & 67.6 & 14.4 & 15 & 104.7 & 59.3 & 70.3 & 78.1 & 0\\
First to die & 59.6 & 15.5 & 15 & 104.7 & 48.2 & 61.0 & 71.4 & 0\\
Average grandparents & 63.6 & 13.6 & 15 & 104.7 & 55.1 & 65.1 & 73.5 & 0\\[.25em]
\cmidrule(lr){2-9}\\[-.75em]
& \multicolumn{8}{c}{Women ($n=740,461$)}\\[.25em]
\cmidrule(lr){2-9}\\[-.75em]
Individual ($t_{\text{c}}$) & 45.8 & 33.1 & 0 & 104.9 & 7.0 & 52.5 & 76.4 & 0\\
Maternal grandfather & 64.1 & 14.6 & 15 & 104.7 & 54.2 & 66.1 & 75.1 & 461,575\\
Maternal grandmother & 62.8 & 16.0 & 15 & 104.8 & 51.9 & 65.3 & 75.1 & 446,549\\
Paternal grandfather & 64.6 & 14.4 & 15 & 104.7 & 55.0 & 66.6 & 75.5 & 444,459\\
Paternal grandmother & 63.3 & 15.7 & 15 & 104.9 & 52.9 & 65.8 & 75.2 & 439,981\\
Last survivor & 67.7 & 14.4 & 15 & 104.9 & 59.4 & 70.3 & 78.2 & 0\\
First to die & 59.5 & 15.5 & 15 & 104.9 & 48.1 & 60.9 & 71.3 & 0\\
Average grandparents & 63.6 & 13.6 & 15 & 104.9 & 55.1 & 65.1 & 73.5 & 0\\[.25em]
\cmidrule(lr){2-9}\\[-.75em]
& \multicolumn{8}{c}{Men \& Women ($n=1,571,940$)}\\[.25em]
\cmidrule(lr){2-9}\\[-.75em]
Individual ($t_{\text{c}}$) & 43.7 & 32.0 & 0 & 104.9 & 6.1 & 48.4 & 73.3 & 0\\
Maternal grandfather & 64.1 & 14.6 & 15 & 104.7 & 54.2 & 66.1 & 75.2 & 987,766\\
Maternal grandmother & 62.8 & 16.0 & 15 & 104.8 & 51.8 & 65.3 & 75.1 & 958,234\\
Paternal grandfather & 64.6 & 14.4 & 15 & 104.7 & 55.0 & 66.7 & 75.4 & 937,081\\
Paternal grandmother & 63.3 & 15.7 & 15 & 104.9 & 52.8 & 65.8 & 75.2 & 929,786\\
Last survivor & 67.6 & 14.4 & 15 & 104.9 & 59.4 & 70.3 & 78.2 & 0\\
First to die & 59.5 & 15.5 & 15 & 104.9 & 48.1 & 61.0 & 71.4 & 0\\
Average grandparents & 63.6 & 13.6 & 15 & 104.9 & 55.1 & 65.1 & 73.5 & 0\\[.25em]
\hline\hline
\end{tabular}
\begin{minipage}{\textwidth}
		\vspace{1ex}
		\scriptsize\underline{Note:} this table provides descriptive statistics of the ages contained in the dataset of grandchildren and their grandparents. $n$ stands for the number of observations, $SD$ is the standard deviation, $Q_1$, $Q_2$ and $Q_3$ are the first, second, and third empirical quartiles, respectively, and No. Missing refers to the number of missing values.
\end{minipage}
    \caption{Age at death of the individuals and age at death of their grandparents, according to the gender of the grandchildren.}
    \label{tab:desc_stat_gparents}
\end{table}

Table~\ref{tab:desc_stat_gparents_decades_526} gives a better idea of the distribution of the data according to the birth cohort of individuals for whom relationships with their grandparents are studied. The table gives the number of observations for each cohort as well as the average age at death of grandchildren (regardless of gender) and each grandparent. The upper part gives, the information for grandchildren for whom the birth and death dates of at least one grandparent are known. The lower part is restricted to grandchildren whose birth and death dates of all four grandparents are known. Again, due to the initial sample design, very few people are present for the first three cohorts (however, there are still a substantial number of individuals in the sample, with $n=59,463$ individuals for whom full detailed information about the four grand-parents is available).

\begin{table}[htb]
    \centering\scriptsize
    \begin{tabular}{lrrrrrrr}
\hline\hline\\[-.75em]
Cohort & $n$ & $t_{\text{c}}$ & $t_{\text{mgf}}$ & $t_{\text{mgm}}$ & $t_{\text{pgf}}$ & $t_{\text{pgm}}$ & Prop. Women ($\%$)\\
\hline\\[-.75em]
& \multicolumn{7}{c}{All individuals with information on at least one grandparent}\\[.25em]
\cmidrule(lr){2-8}\\[-.75em]
(1790,1800] & $1,277$ & 45.7 & 66.7 & 65.6 & 66.8 & 65.0 & 46.9\\
(1800,1810] & $6,375$ & 42.8 & 66.9 & 66.3 & 67.4 & 66.5 & 48.2\\
(1810,1820] & $5,611$ & 42.8 & 61.6 & 60.3 & 62.9 & 61.4 & 47.7\\
(1820,1830] & $178,975$ & 40.5 & 63.8 & 62.2 & 63.8 & 62.3 & 46.8\\
(1830,1840] & $252,818$ & 37.1 & 64.6 & 62.9 & 64.8 & 63.0 & 46.6\\
(1840,1850] & $119,729$ & 35.5 & 64.1 & 62.3 & 65.2 & 63.2 & 46.9\\
(1850,1860] & $145,099$ & 36.9 & 63.3 & 62.5 & 63.8 & 62.4 & 46.8\\
(1860,1870] & $216,136$ & 40.3 & 64.8 & 63.9 & 64.8 & 63.8 & 46.8\\
(1870,1880] & $201,357$ & 47.6 & 65.1 & 64.0 & 65.6 & 64.7 & 47.6\\
(1880,1890] & $202,352$ & 51.5 & 63.4 & 62.4 & 64.5 & 63.6 & 47.8\\
(1890,1900] & $242,211$ & 54.7 & 63.3 & 62.6 & 63.8 & 63.4 & 47.4\\
\hline\\[-.75em]
& \multicolumn{7}{c}{All four grandparents known}\\[.25em]
\cmidrule(lr){2-8}\\[-.75em]
(1790,1800] & $1$ & 52.4 & 59.2 & 72.5 & 56.8 & 46.6 & 0.0\\
(1800,1810] & $9$ & 47.4 & 69.2 & 56.8 & 66.9 & 64.6 & 44.4\\
(1810,1820] & $128$ & 45.8 & 61.4 & 60.0 & 60.6 & 61.6 & 38.3\\
(1820,1830] & $19,468$ & 40.9 & 64.1 & 63.0 & 63.8 & 62.7 & 47.1\\
(1830,1840] & $26,651$ & 37.7 & 64.8 & 63.4 & 64.8 & 63.1 & 47.2\\
(1840,1850] & $7,093$ & 35.2 & 65.1 & 63.4 & 65.1 & 63.3 & 48.1\\
(1850,1860] & $1,229$ & 36.7 & 65.7 & 63.8 & 63.5 & 62.6 & 51.3\\
(1860,1870] & $2,090$ & 38.5 & 65.1 & 64.3 & 63.8 & 63.1 & 51.0\\
(1870,1880] & $1,294$ & 46.4 & 65.3 & 64.1 & 64.5 & 63.0 & 52.9\\
(1880,1890] & $610$ & 49.9 & 63.5 & 62.1 & 63.6 & 61.6 & 51.6\\
(1890,1900] & $890$ & 54.5 & 63.1 & 62.1 & 62.5 & 61.8 & 50.6\\
\hline\hline
\end{tabular}
\begin{minipage}{\textwidth}
		\vspace{1ex}
		\scriptsize\underline{Note:} This table reports some key descriptive statistics for the individuals contained in the dataset used to study the dependence between children and grandparents. $n$ stands for the number of observations, $t_{\text{c}}$, $t_{\text{gfm}}$, $t_{\text{gmm}}$, $t_{\text{gff}}$, and  $t_{\text{gmf}}$ stand for the age at death of individuals, maternal grandfather, maternal grandmother, paternal grandfather, paternal grandmother, respectively. Prop. Women gives the proportion of women among the individuals, for each cohort. 
\end{minipage}
    \caption{Information about individuals per cohort for the dependencies between individuals and their grandparents when partial information is known on grandparents (top) and when all the grand parents are known (bottom).}
    \label{tab:desc_stat_gparents_decades_526}
\end{table}

\clearpage

\section{From Mortality Models to Insurance Premiums}\label{sec:mortality_models_to_insurance}

In this section, we introduce general notation that we will use when discussing the impact of joint life dependencies, withing families. We start with general notations to model univariate mortality, and then introduce copulas which describe the joint distribution. Inference is also discussed in that first part. Then, we present various insurance products, that we will price in various contexts, such as a life insurance and pension annuities (as well as joint life related guarantees). We might stress here that we focus on contemporary insurance guarantees: the goal is not to discuss historical insurance prices in the early 19th century, but to understand the impact of possible dependencies between family relatives on insurance prices based on the massive (historical) data we have.\footnote{Due to recent regulation on personal data in most countries, running such a study on contemporary data would be much more difficult. If mortality \textit{per se} changed a lot, we study the impact of dependencies on insurance prices based on order of magnitudes of correlations obtained on historical data.}

\subsection{Univariate Mortality}

The lifetime of a newborn is modelled by a positive variable $T$, with cumulative distribution function $F$ (with $F(t)=\mathbb{P}[T\leq t]$ for any positive $t$) and survival function $S$ (with $S(t)=\mathbb{P}[T> t]$). Let $T_x$ denote the remaining lifetime of the person at age $x$, in the sense that $T_x = (T-x)\vert T>x$. Conditional cumulative distribution function is $F_x$ (with $F_x(t)=\mathbb{P}[T_x\leq t]=\mathbb{P}[T-x\leq t\vert T>x]$) which is also denoted ${}_tq_x$ in actuarial literature, while the survival distribution function is $S_x$ (with $S_x(t)=\mathbb{P}[T_x> t]=\mathbb{P}[T-x> t\vert T>x]$) which is also denoted ${}_tp_x$. One can write, for $t>0$
$$
{}_tp_x = \mathbb{P}[T-x> t\vert T>x]=\frac{\mathbb{P}[T> t+x]}{\mathbb{P}[ T>x]}=\frac{S(x+t)}{S(x)},
$$
with the convention that ${}_0p_x=1$. Assuming that $T$ is an absolutely continuous random variable allows us to consider the density of $T(x)$, denoted $f_x$ which satisfies
$$
f_x(t) = {}_tp_x \cdot \mu(x+t),
$$
where $\mu$ is the hazard rate, also called force of mortality in demographic and actuarial applications. In the life table terminology, $L_x$ denotes the size of a cohort-type group, at age $x$, with $L_0=100,000$. Note that $L_x = L_0\cdot{}_xp_0$.

And finally, curtate life expectancy for individual $(x)$ is defined as
$$
e_x = \mathbb{E}\big(\lfloor T_x\rfloor\big)=\mathbb{E}\big(\lfloor T-x\rfloor\vert T>x\big)=\sum_{t=0}^\infty t{}_tp_x \cdot q_{x+t}=\sum_{t=1}^\infty{}_tp_x,
$$
for some integer $x\in\mathbb{N}$, using notations from \cite{Bowers:2e:1997}, where we count here the expected number of future years completed by $(x)$ prior to death (to contrast, the complete expectation of life, $\mathbb{E}(T_x)$, is denoted $\mathring{e}_x$).

\begin{figure}[htb!]
    \centering
    \includegraphics[width=\textwidth]{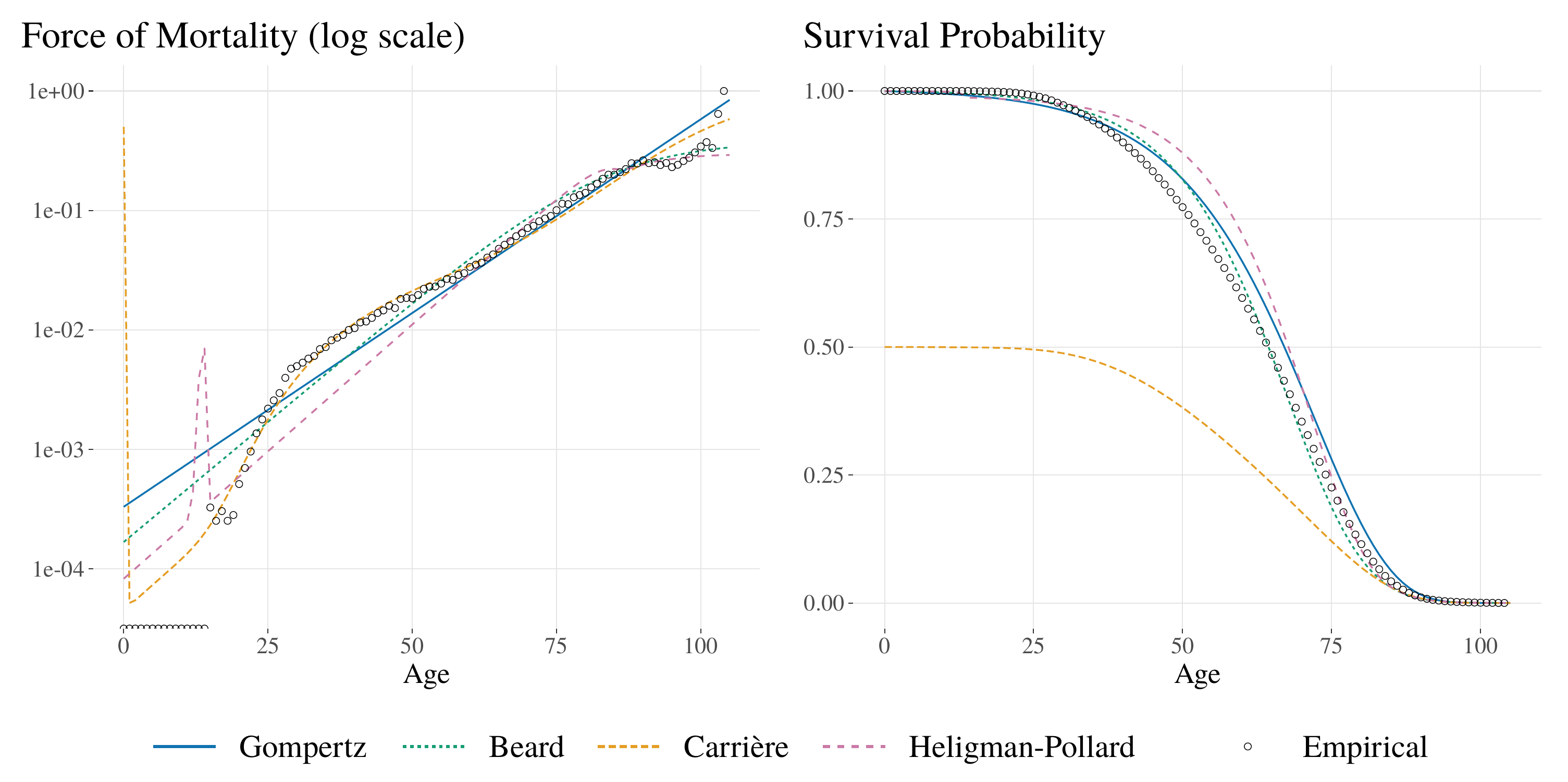}
    \caption{Marginal mortality, with force of mortality for males on the left, $\mu_x$, and the survival function on the right.}
    \label{fig:mortality:male}
\end{figure}

In the context of small amount of data, it is natural to use parametric models to compute complex quantities. This will be the case with joint life models, to derive more robust estimates (but probably more model sensitive).
Classical parametric models for mortality are Gompertz distribution, with $\mu(x) = Ae^{Bx}$ (from \citealp{Gompertz1825}) or Beard with $\mu(x) = Ae^{B^x} / (1 + KAe^{B^x}) $ (from \citealp{beard1971}). \cite{carriere1992} suggested to use mixtures of distribution, with $S(x) = {\psi}_1S_1(x)+\ {\psi}_2S_2\left(x\right)+\ {\psi }_3S_3\left(x\right)$ from standard survival families (that will be the Carriere model). Finally, the Heligman-Pollard with $q(x)/p(x) =  A^{{\left(x+B\right)}^C}+De^{-E{\left({\log x\ }-{\log F\ }\right)}^2}+GH^x$ (from \citealp{heligman1980}). The later was initially fitted on Australian mortality and performed fairly well at all age. That is also what we observe on our data. The first part $A^{{\left(x+B\right)}^C}$ is a rapidly declining exponential, that reflects the fall in mortality during the early childhood years (this component of mortality has three parameters, $A$, which is nearly equal to $1-q_0$, $B$ which is a location factor, and $C$ which measures the rate of mortality decline in childhood). The third term $GH^x$ is the Gompertz model, that reflects the near geometric rise in mortality at the adult ages. And finally, the second term $De^{-E{\left({\log x\ }-{\log F\ }\right)}^2}$ reminds of a lognormal model, that reflects   accident mortality (as discussed in \citealp{heligman1980}, $F$ indicates location, $E$ the spread, and $D$ the severity). The adjustment of these four distributions on male mortality based on our data can be visualized graphically. In Figure \ref{fig:mortality:male}, we can compare the parametric models to raw values (force of mortality $\mu(x)$ and survival probabilities ${}_xp_0$ or $L_x/L_0$). The mixed distribution suggested in \cite{carriere1992} provides a very good fit (and not the other standard models). Computations were performed using the \texttt{MortalityLaws} R package (see \citealp{MortalityLaws})

\subsection{Modeling the Dependence Structure}\label{subsec:dependance_structure}

A classical model to describe the joint distribution of a pair of lifetimes $(T_1,T_2)$, 
-- for remaining lifetimes of two individuals, which could be either $(T_f,T_m)$ for the father and the mother, or $(T_m,T_c)$ for the mother and a child of hers -- is to use a copula function, so that the joint survival function
$$
S(t_1,t_2)=\mathbb{P}[T_1> t_1,T_2> t_2] = C\big(S_{1}(t_1),S_{2}(t_2)\big),
$$
where $C:[0,1]^2\rightarrow[0,1]$ is a copula function (see \citealp{Joe:1997} or \citealp{Nelsen}, but in the context of survival lifetimes, using the survival copula makes more sense.).\footnote{Note that $C$ is usually called the survival copula of the pair $(T_1,T_2)$.} Some parametric family can be considered for $C$. The independent and the comonotonic copulas are defined as $C_{\perp}(u,v)=uv$ and $C_{+}(u,v)=\min\lbrace u,v\rbrace$. In Appendix \ref{appendix:copula}, popular parametric copulas are mentioned. In a nutshell, for inference, we use here pseudo observations based on ranks, $(\widehat{u}_{i},\widehat{v}_{i})$ where
$$
\widehat{u}_{i} = \widehat{S}_{1}(t_{1,i})\text{ where } \widehat{S}_{1}(t)=\frac{1}{n}\sum_{j=1}^n \boldsymbol{1}(t_{1,j}> t),
$$
for the first type of individuals (say {\em father} for joint life in a couple), and a similar expression for the second type\footnote{Here we assume that data consist in pairs $(t_{1,i},t_{2,i})$ of ages at dead, but actually, we have the {\em dates} of deaths, so a dynamic model could be considered, this point will be addressed in the next section.}.

We define the empirical copula $\widehat{C}_n$ as the cumulative distribution function of $(\widehat{u}_{i},\widehat{v}_{i})$'s
$$
\widehat{C}_n(u,v) = \frac{1}{n}\sum_{i=1}^n \boldsymbol{1}\big(\widehat{u}_{i}\leq u,\widehat{v}_{i}\leq v\big),
$$
or some smooth version $\widetilde{C}_n(u,v)$ using some probit transformation, as in \cite{geenens}. In the case where copulas are non-symmetric, remember that since we look at the survival copula, the lower corner $(0,0)$ corresponds to small probabilities (and therefore large ages).

\subsection{Impact on Annuities and Life Insurance Premiums}\label{sec:annuities:couple}

In the section, we present various insurance products and their notations. These products will allow us to compute various quantities at time $t$, when the insurance contract is signed. In the case of a single life guarantee, we consider an individual age $x$ (at time $t$), while for multiple lives (here father and mother), both have age $x_f$ and $x_m$ respectively. As discussed in the next section, most quantities are rather stable over time, so notation $t$ will not be used here. Furthermore, as mentioned at the beginning of that section, we consider here `contemporary' insurance guarantees, even if we use historical data.

\subsubsection{Notations for Single Life Guarantees}

For financial application, let $\nu$ denote the discount factor associated with the (constant) annual rate $i$, in the sense that $\nu=(1+i)^{-1}$. Formally, we compute the expected present value, of future cash flow, for a sequence $c_1,\dots,c_n$, at time $t_1,\dots,t_n$, where the expected value is computed related to probabilities of paying, denoted generally $p_1,\dots,p_n$, 
$$
\sum_{k=1}^n \nu^{t_i}c_ip_i.
$$

For example, whole life insurance are contractual guarantees that promise a fixed amount (a normalized cash flow of \$1) {\em at the time of death}. The expected present value of such a contract is
$$
\overline{A}_{x} =\sum_{k=0}^{\infty} \nu^k~{}_kp_x q_{x+k},
$$
or, when payment is made at the end of the year of the death
$$
A_{x} =\sum_{k=0}^{\infty} \nu^{k+1}~{}_kp_x q_{x+k}.
$$
The $n$-year term insurance provides a fixed amount only if the death occurs in the next $n$ years, and the expected present value 
$$
A^1_{x:\actuarial{n}} =\sum_{k=0}^{n-1} \nu^{k+1}~{}_kp_x q_{x+k},
$$
again, with a payment at the time of death. A $n$-year endowment guarantees a payment at the end of the $n$ years if the person survives. The expected present value is here
$$
{}_nE_x=A^{~~~1}_{x:\actuarial{n}} = \nu^n~{}_np_x.
$$

For example, annuities are contractual guarantees that promise a {\em periodic income} (usually annually, with a normalized cash flow of \$1) over the lifetimes of individuals. The actuarial present value of the annuity of an individual age $(x)$ is
$$
a_{{x}}=\sum_{k=0}^\infty \nu^{k+1} {}_kp_{{x}},
$$ where payment is made at the end of the year if individual is still alive (as in \citet{Bowers:2e:1997}, $\ddot{a}_{{x}}$ is used for payments done at the beginning of the year -- also called {\em annuity-due}). 
Such a contract pays \$1 at the end of the years 1, 2, 3, $\dots$ as long as the individuals is alive. As previously, an $n$-year temporary life annuity guarantees yearly unit cash flows, until year $n$, over the lifetimes of individual $(x)$
$$
a_{{x}:\actuarial{n}}=\sum_{k=0}^{n-1} \nu^{k+1} {}_kp_{{x}}.
$$
We refer to \citet{Bowers:2e:1997} for technical distinctions among the various subtleties. Here, we simply compare life insurance expected present values, and annuities, in various scenarios. To that end, use numerical tools developed for the \texttt{lifecontingencies} R package, presented in \cite{Spedicato2013}.

\subsubsection{Notations for Conditional Single Life Guarantees}

In Section \ref{sec:parents:grandparents}, we will consider single life guarantees, using family history as conditional information, at the time of signature. For instance, we can consider the actuarial present value of the annuity of an individual age $(x)$, given some information $\star$ about parents or grand-parents, is
$$
a_{{x}}^{\star}=\sum_{k=0}^\infty \nu^{k+1} {}_kp_{{x}}^{\star},
$$
where $\star$ could mean that both parents are still alive when the child has age $x$, for example, or that only one of them is alive. 

\subsubsection{Notations for Multiple Life Guarantees}

In section \ref{sec:couple}, we will also consider joint and survivor annuities options on two live contracts. More precisely, as discussed earlier, we consider the case of married couples, where the two lives are the one of the father (denoted f) and the one of the mother (denoted m).
Let $T(x_f,x_m)$ denote the joint life status and $T(\overline{x}_f,\overline{x}_m)$ the last survivor, defined as
$$
T(x_f,x_m)=\min\lbrace T_m(x_f),T_f(x_m)\rbrace\text{  and }T(\overline{x}_f,\overline{x}_m)=\max\lbrace T_f(x_f),T_m(x_m)\rbrace.
$$
Survival probabilities are
$$
{}_tp_{x_f,x_m} = \mathbb{P}[T(x_f,x_m)> t]\text{ and }
{}_tp_{\overline{x}_f,\overline{x}_m} = \mathbb{P}[T(\overline{x}_f,\overline{x}_m)> t],
$$
while curtate life expectancies are
$$
e_{x_f,x_m} = \mathbb{E}\big(\lfloor T(x_f,x_m)\rfloor\big)=\sum_{t=1}^\infty{}_tp_{x_f,x_m}\text{ and }
e_{\overline{x}_f,\overline{x}_m} = \mathbb{E}[T(\overline{x}_f,\overline{x}_m)]=\sum_{t=1}^\infty{}_tp_{\overline{x}_f,\overline{x}_m}.
$$

A {\em joint-life annuity} pays benefits until the death of the first of the two annuitants, $T(x_f,x_m)$, for a husband/father $(x_f)$ and a wife/mother $(x_m)$. The standard joint-life annuity pays \$1 at the end of the years 1, 2, 3, $\dots$ as long as both spouses survive. Its actuarial present value is
$$
a_{{x}_f,{x}_m}=\sum_{k=1}^\infty \nu^k {}_kp_{{x}_f,{x}_m}.
$$

A {\em last-survivor annuity} pays a certain amount until the second (and last) death, $T(\overline{x}_f,\overline{x}_m)$. The standard joint-life annuity pays \$1 at the end of the years 1, 2, 3, $\dots$ as long as long as either spouses survives. Its actuarial present value is
$$
a_{\overline{x}_f,\overline{x}_m}=\sum_{k=1}^\infty \nu^k {}_kp_{\overline{x}_f,\overline{x}_m}.
$$

A {\em reversion annuity} starts after the first death $T(x_m,x_f)$ until the last one $T(\overline{x}_f,\overline{x}_m)$. Some one-way reversion can be considered, for instance from the husband to the wife, also called {\em widow's pension}: payments start at the death of $(x_f)$ until the death of $(x_m)$, and no payment is made if $(x_m)$ dies before $(x_f)$. The actuarial present value of the {\em reversion annuity} is $a_{{x}_f,{x}_m}-a_{\overline{x}_f,\overline{x}_m}$, while the actuarial present value of the {\em widow's pension} is $a_{{x}_m|{x}_f}=a_{{x}_m}-a_{{x}_f,{x}_m}$.


\section{Husband and Wife Dependencies}\label{sec:couple}

This section highlights, as observed in the literature, the positive links between the life expectancy of the members of a couple, and then examines the impacts on annuities and life insurance premiums.

\subsection{Empirical Relationship Between Lifespans Within Couples}

\begin{figure}[htb!]
  \centering
  \begin{subfigure}[t]{.45\textwidth}
		\centering
		\caption{$(T_{\text{f}},T_{\text{m}})$ without restrictions on $x$}\label{subfig:copula_couples_nonparam}
		\includegraphics[width=\textwidth]{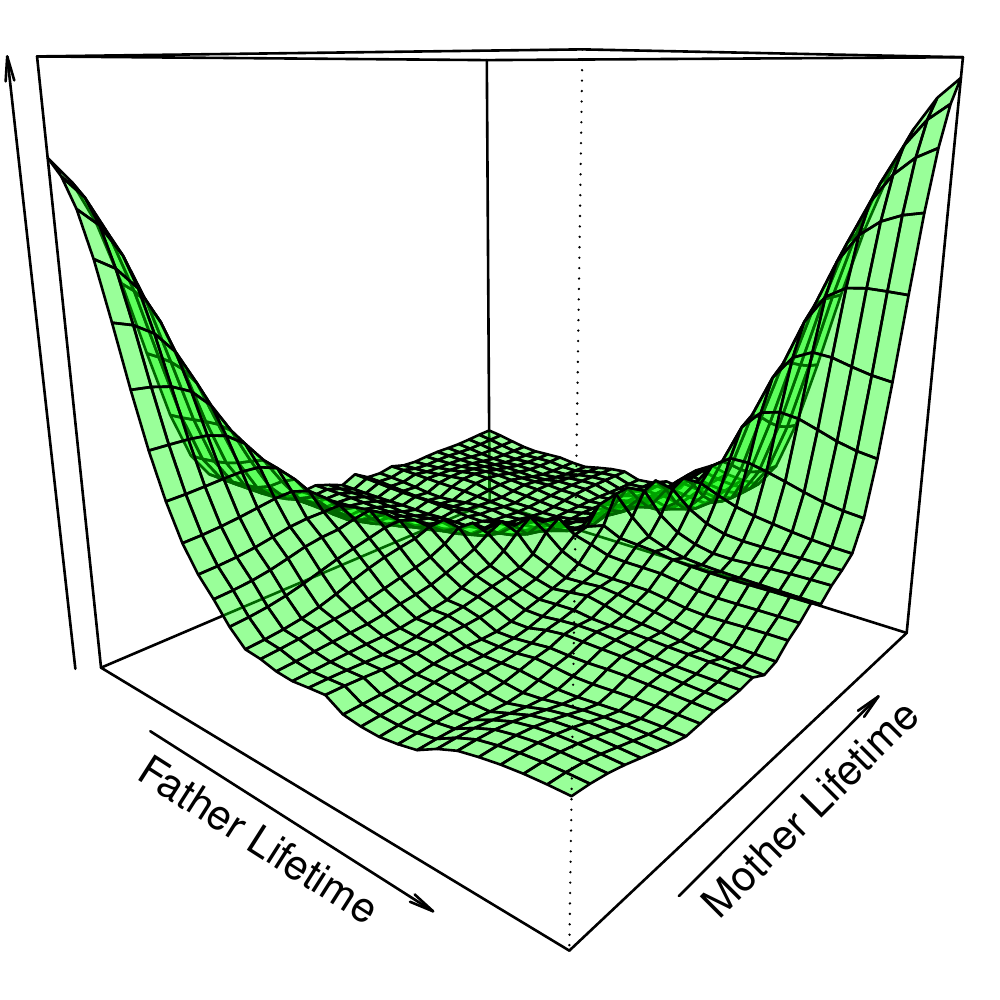}
\end{subfigure}
  \\
  \begin{subfigure}[t]{.31\textwidth}
		\centering
		\caption{$T_{\text{f}}\geq 25$}\label{subfig:copula_couples_resid_nonparam_f_25}
		\includegraphics[width=\textwidth]{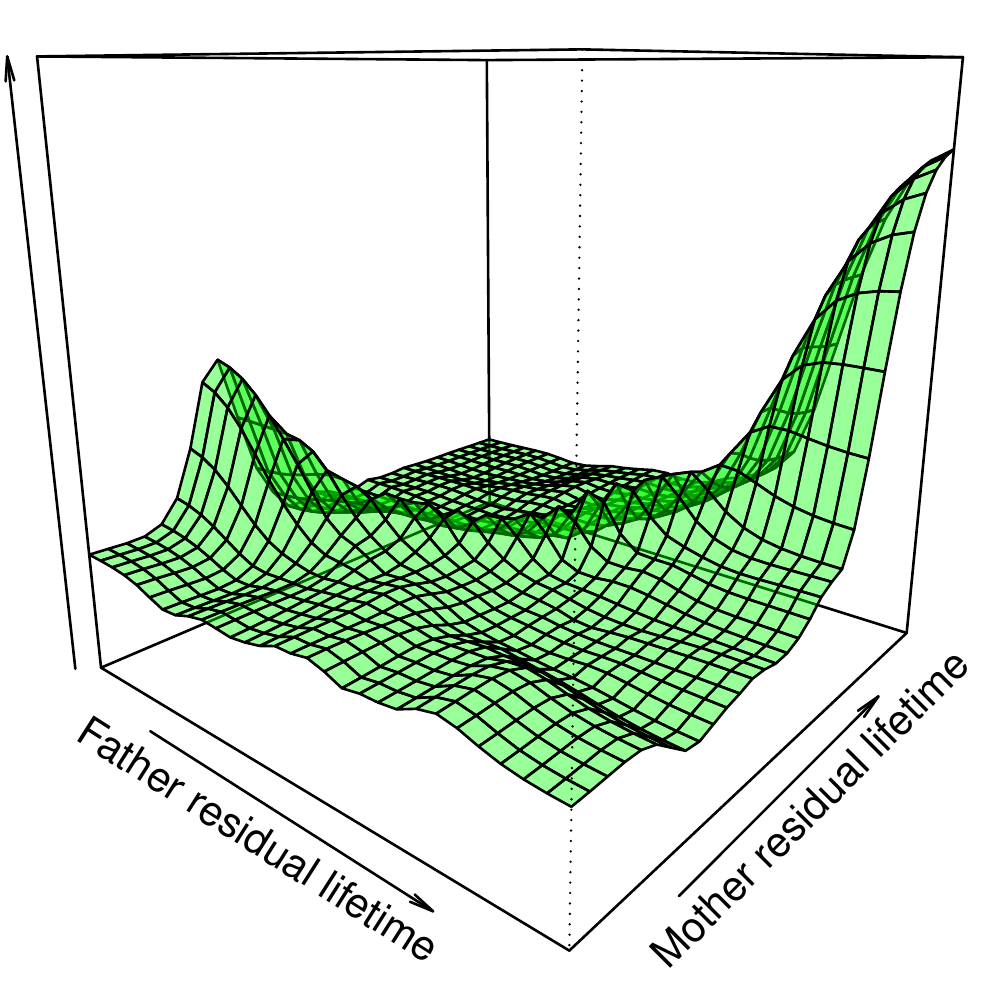}
	\end{subfigure}
	\hspace{.5em}
	\begin{subfigure}[t]{.31\textwidth}
		\centering
		\caption{$T_{\text{f}}\geq 35$}\label{subfig:copula_couples_resid_nonparam_f_35}
		\includegraphics[width=\textwidth]{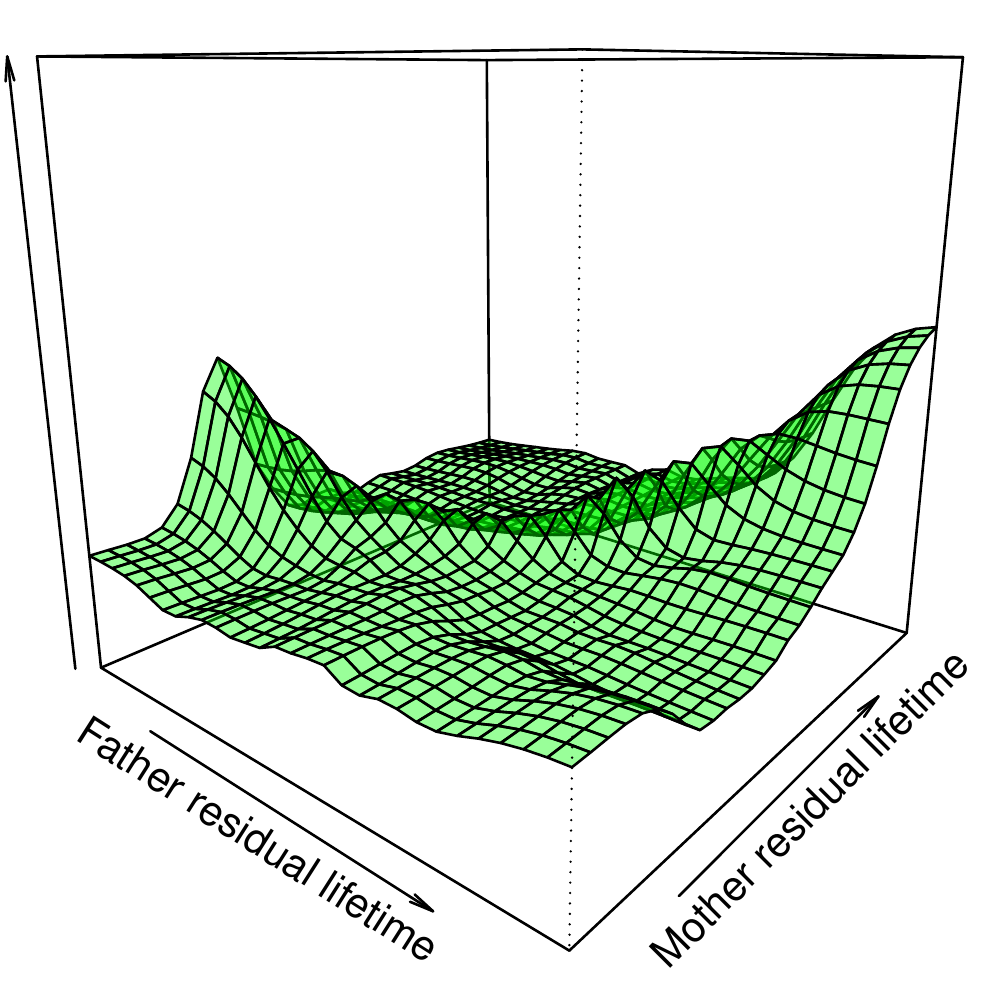}
	\end{subfigure}
	\hspace{.5em}
	\begin{subfigure}[t]{.31\textwidth}
		\centering
		\caption{$T_{\text{f}}\geq 45$}\label{subfig:copula_couples_resid_nonparam_f_45}
		\includegraphics[width=\textwidth]{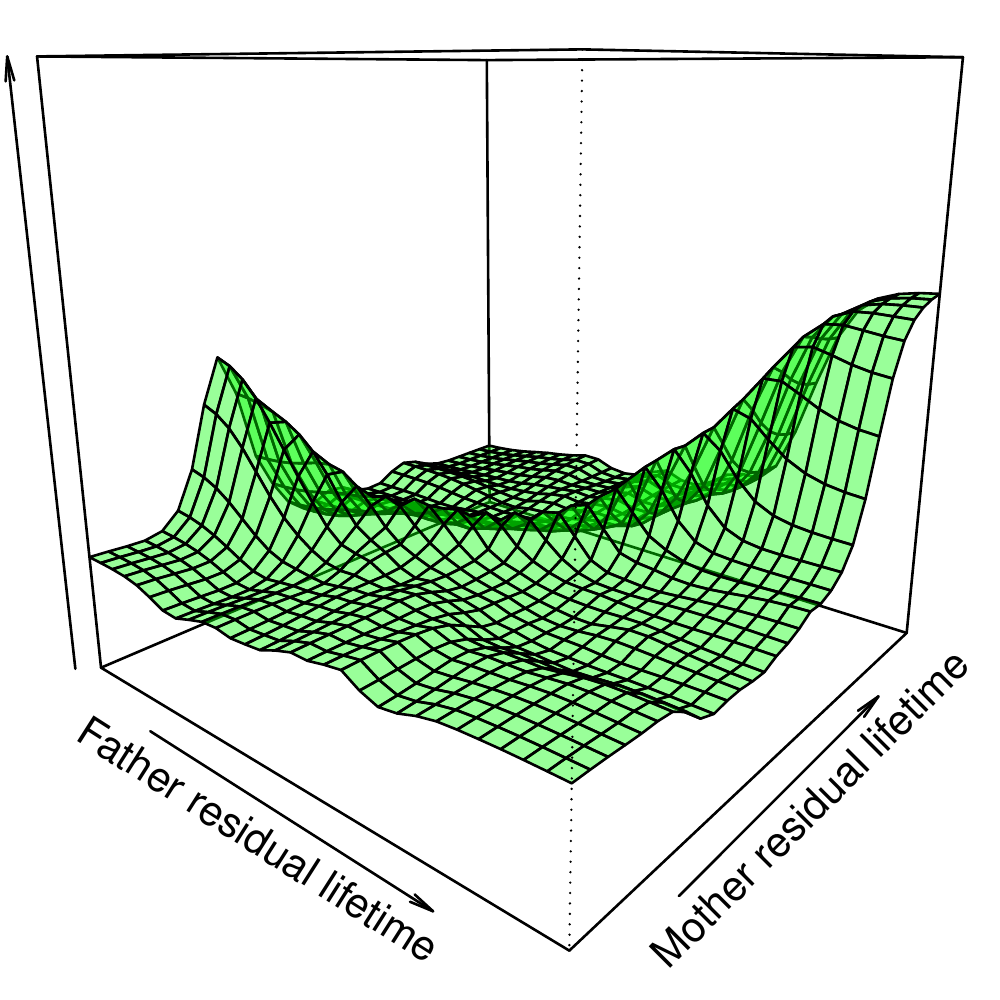}
	\end{subfigure}

	\begin{subfigure}[t]{.31\textwidth}
		\centering
		\caption{$T_{\text{m}}\geq 25$}\label{subfig:copula_couples_resid_nonparam_m_25}
		\includegraphics[width=\textwidth]{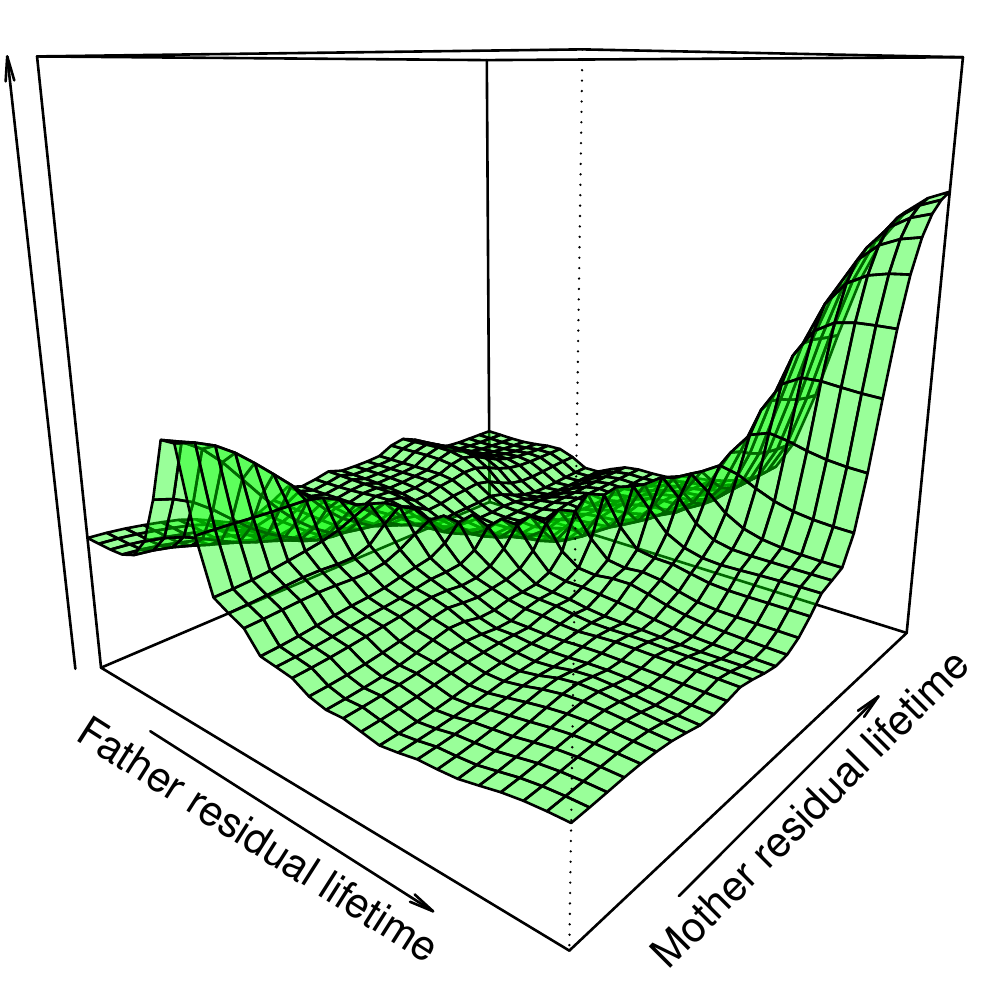}
	\end{subfigure}
	\hspace{.5em}
	\begin{subfigure}[t]{.31\textwidth}
		\centering
		\caption{$T_{\text{m}}\geq 35$}\label{subfig:copula_couples_resid_nonparam_m_35}
		\includegraphics[width=\textwidth]{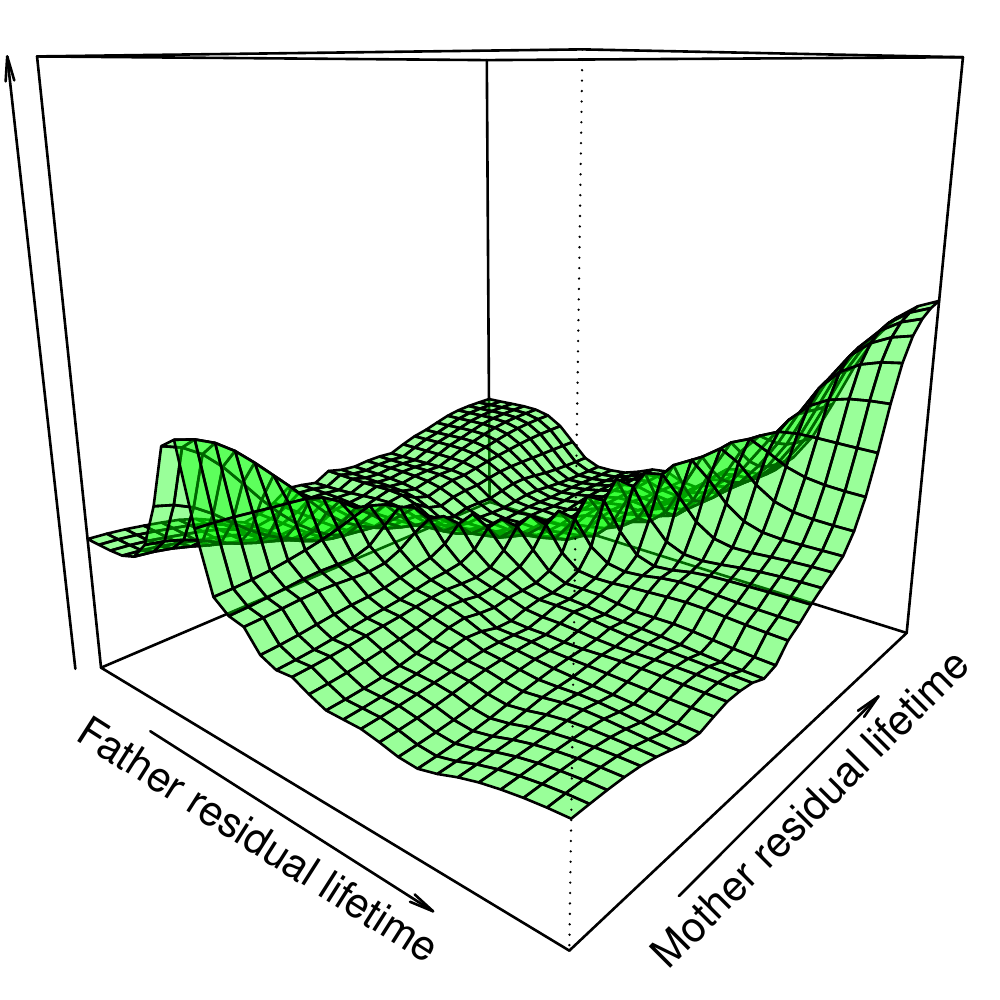}
	\end{subfigure}
	\hspace{.5em}
	\begin{subfigure}[t]{.31\textwidth}
		\centering
		\caption{$T_{\text{m}}\geq 45$}\label{subfig:copula_couples_resid_nonparam_m_45}
		\includegraphics[width=\textwidth]{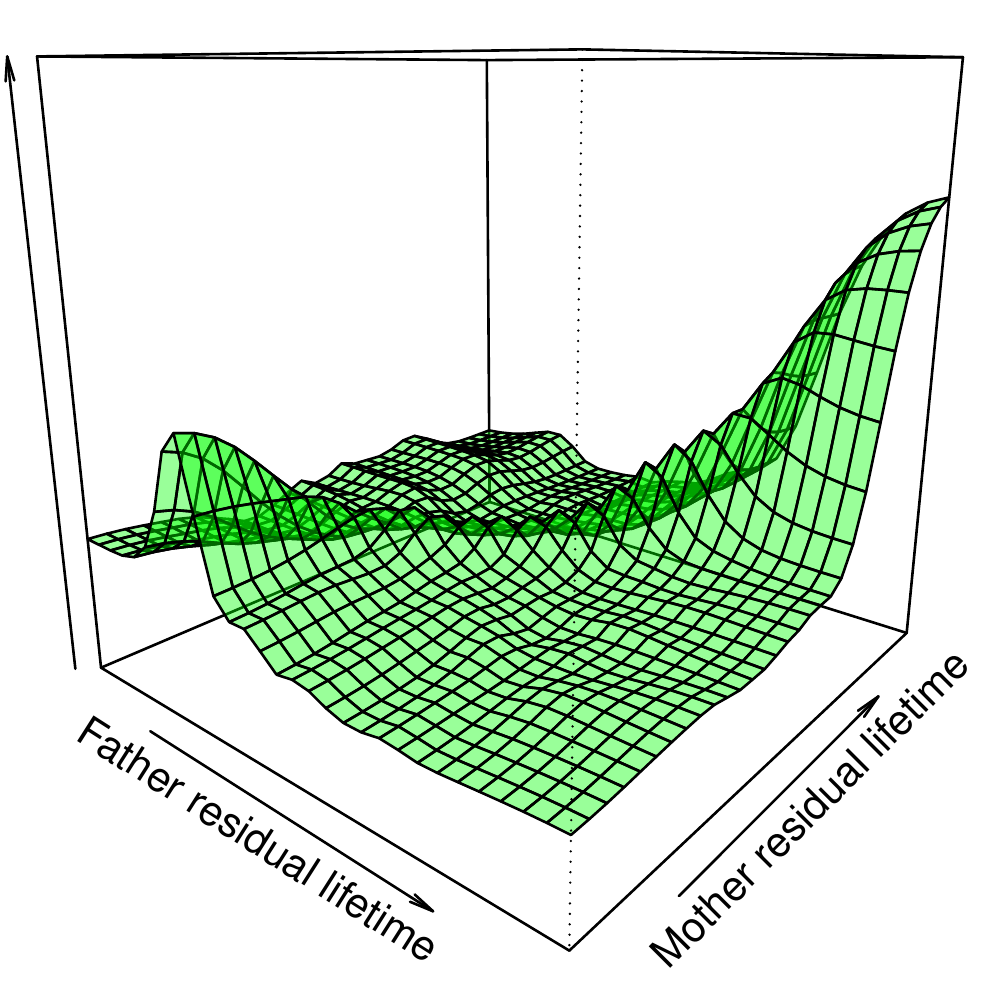}
	\end{subfigure}

  \begin{minipage}{\textwidth}
		\vspace{1ex}
		\scriptsize\underline{Note:} On top, nonparametric estimate of the (survical) copula density for $(T_{\text{f}},T_{\text{m}})$. Below, on top, copulas of $(T_{\text{f}},T_{\text{m}})$ given $T_{\text{f}}\geq x$ (and that both are still alive), for $x=25,35,45$. At the bottom, copulas of $(T_{\text{f}},T_{\text{m}})$ given $T_{\text{m}}\geq x$ (and that both are still alive), for $x=25,35,45$. The $z$-axis for copula density graphs is always $[0,5]$, which allows us to compare all distributions.
		
\end{minipage}
\caption{Nonparametric estimate of the copula density for $(T_{\text{f}},T_{\text{m}})$, given some restrictions on $x$.}\label{fig:cop:np}
\end{figure}

\begin{figure}[htb!]
  \centering
    \includegraphics[width=\textwidth]{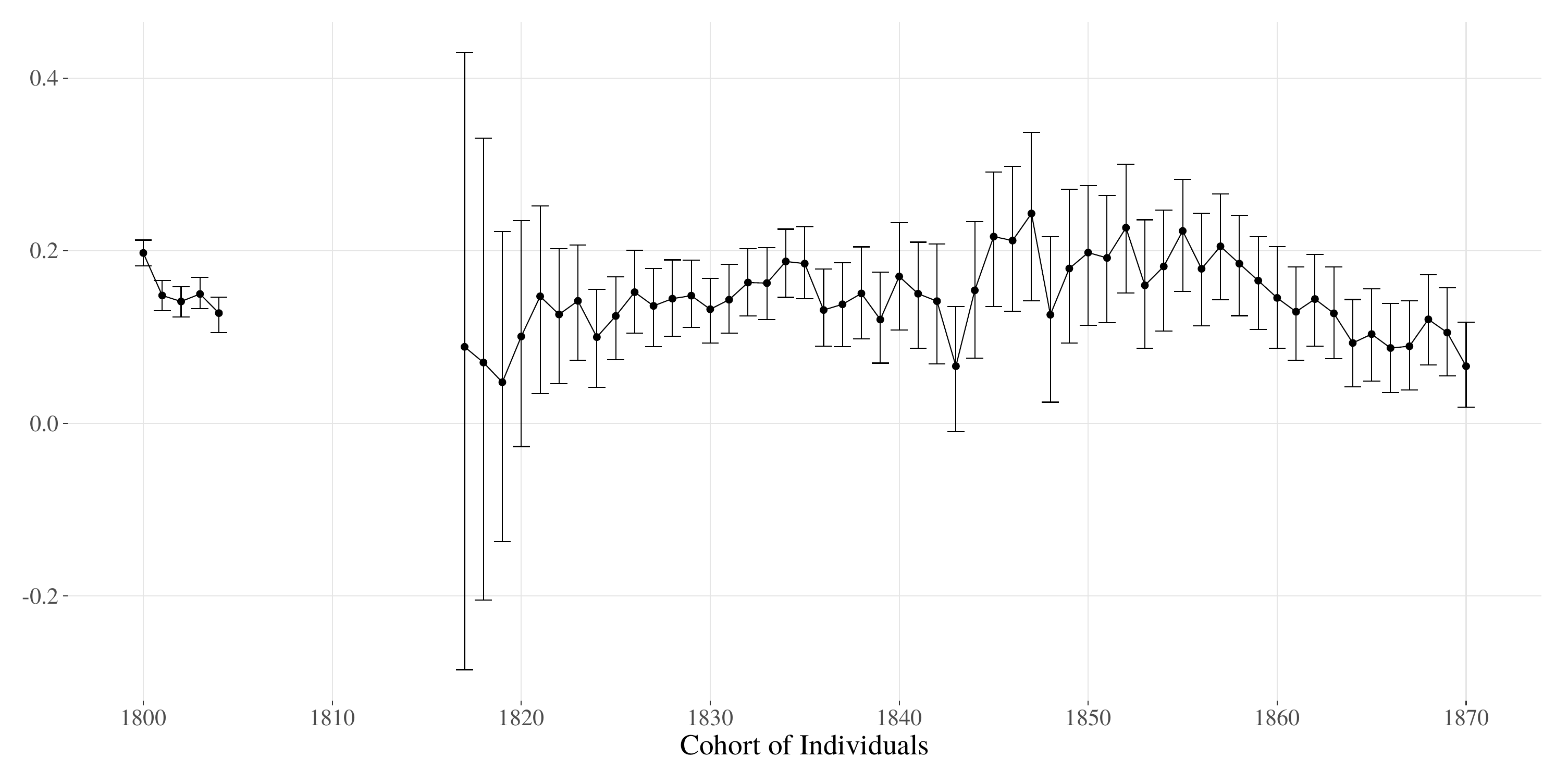}
  \begin{minipage}{\textwidth}
		\vspace{1ex}
		\scriptsize\underline{Note:} The dots represent the estimated Spearman correlation between the age at death of spouses for each cohort, using the birth year of the father to define the cohorts. The bars correspond to 95\% bootstrap confidence interval (based on $1,000$ resamples).
\end{minipage}
\caption{Spearman correlation between age at death of the spouses, by year of birth of the husband.}\label{fig:correl_couples}
\end{figure}

The nonparametric\footnote{Estimated parametric densities  -- Gaussian, Clayton and Gumbel -- are mentioned in the Appendices.} estimation of the (survival) copula of remaining lifetimes $(T_{x_f},T_{x_m})$, can be seen in Figure~\ref{fig:cop:np}. The nonparametric estimator suggest very similar behavior in the lower and in the upper tails, which would disqualify Gumbel and Clayton copula. Based on almost 15,000 individuals (but unfortunately most were censored data since a lot of people were still alive), \cite{FreesCarriereValdez1996} suggested to use Frank copula (exhibiting symmetric dependence between the lower and the upper tail), with $\widehat{\theta}=3.367$, corresponding to a $0.5$ Spearman correlation. \cite{Denuit2001} selected at random two cemeteries in Brussels (Koekelberg and Ixelles / Elsene) and they collected the ages at death of 533 couples buried there. Those data are very close to the ones we have, and they observed a $0.139$ Spearman correlation, and they used Gumbel copula, with parameter $\widehat{\theta}=0.104$. In our data -- see Figure \ref{fig:correl_couples} -- Spearman correlation was 0.156, with a 95\% confidence interval $(0.151 ; 0,161)$.

\begin{figure}[htb!]
  \centering
  \includegraphics[width=\textwidth]{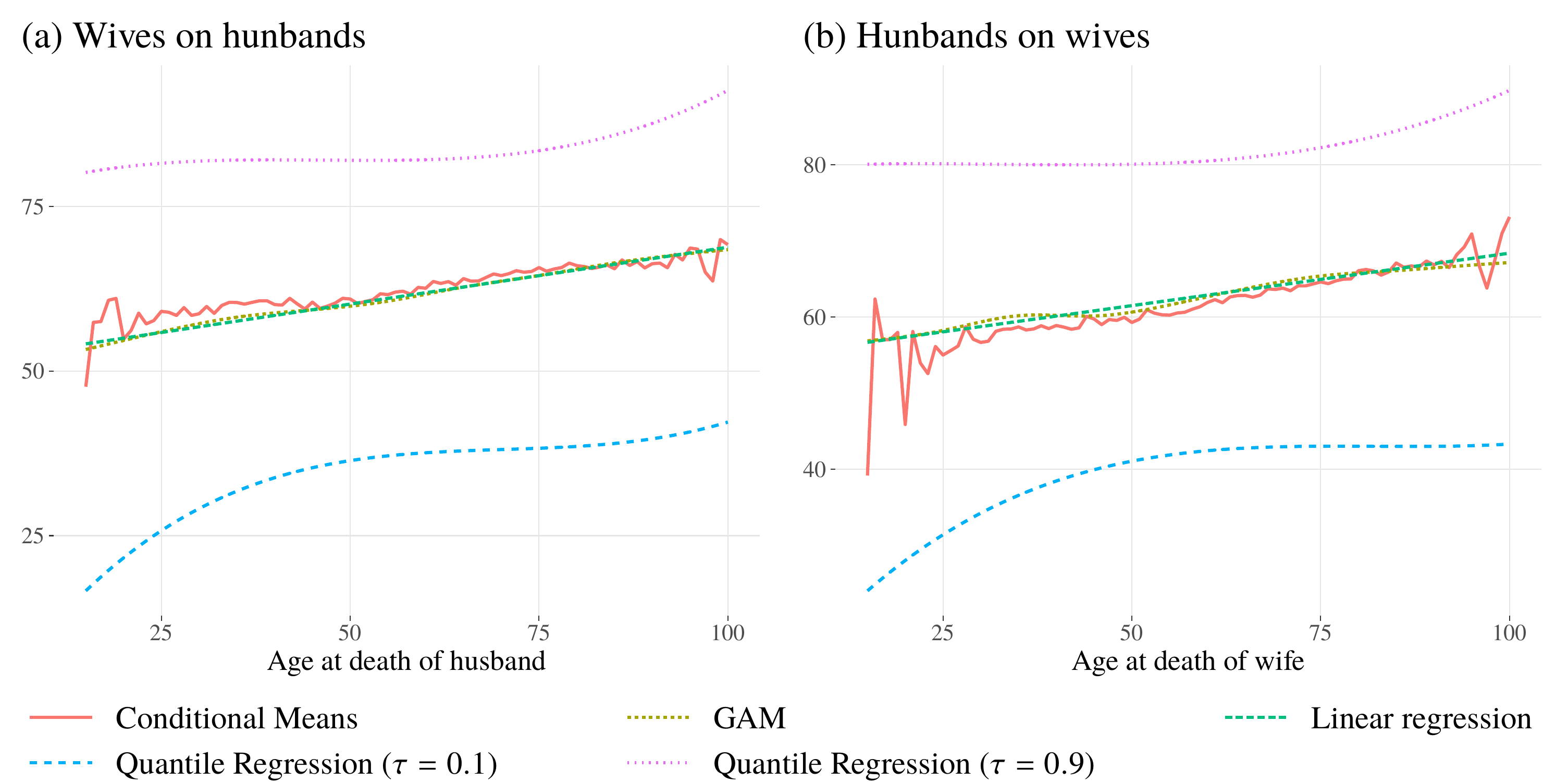}
  \begin{minipage}{\textwidth}
		\vspace{1ex}
		\scriptsize\underline{Note:} The blue and purple dashed lines correspond to the first and ninth deciles of quantile regression of age at death of an individual as a function of the age at death of their spouse.
\end{minipage}
\caption{Age at death of women as a function of age at death of their husband (left) and age at death of men as a function of the age at death of their wife (right).}\label{fig:qr_couples}
\end{figure}

The link between the age at death of a person and that of their spouse can be visualized in a simple way through several methods. Figure~\ref{fig:qr_couples} shows a woman's age at death as a function of her husband's age at death as well as the relationship in the other direction.  The positive correlation between these two quantities can thus be seen graphically, as in \cite{Pearson}.

\subsection{Annuities and Life Insurance Premiums Within Couples}

\begin{figure}[htb!]
  \centering
    \includegraphics[width=\textwidth]{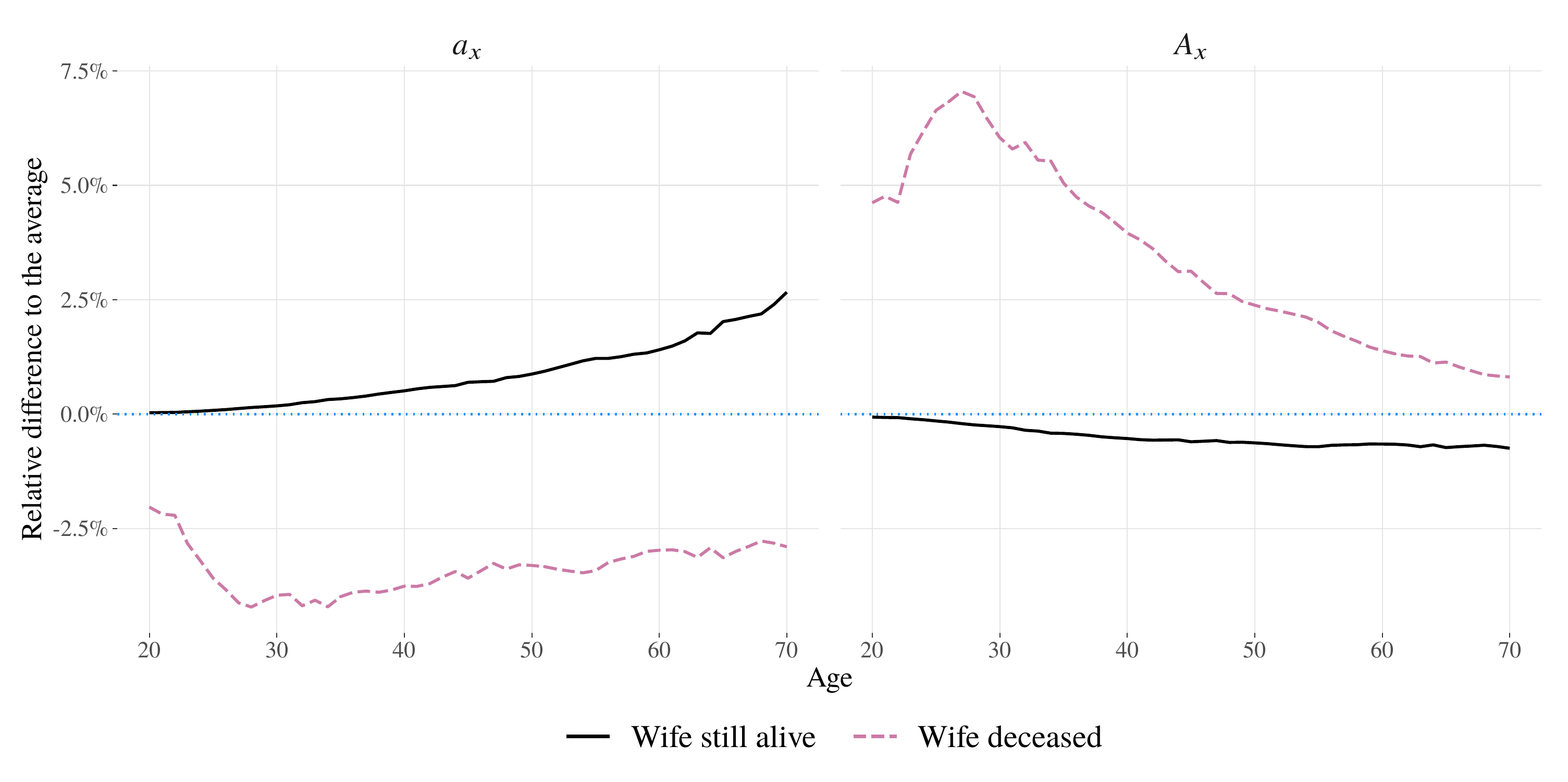}
  \begin{minipage}{\textwidth}
		\vspace{1ex}
		\scriptsize\underline{Note:} the horizontal blue dashed line corresponds to the average value calculated on all individuals, regardless of the death status of the spouse. The annuities are calculated for 100 terms and the expected present value for life insurance are calculated for 100-year coverage. The interest rate is assumed to be $3\%$.
\end{minipage}
\caption{Relative difference to the average (in \%) of present value of an annuity (left) and expected present value for a life insurance (right) depending on the age of the annuitant and on the death status of the wife at the time of the contract.}\label{fig:p_insur_couples_homme_pct}
\end{figure}

As mentioned previously, the positive relationship exhibited between the lifetime of the members of a couple can be used to derive bounds for most actuarial quantities. {To that end, we consider an annuity and a life insurance signed by a man. We compare their present value according to the age of the annuitant and distinguishing between cases where the policyholder's wife is alive or deceased at the time of signature. The results are shown in Figure~\ref{fig:p_insur_couples_homme_pct}, for an annuity (on the left) and for a life insurance (on the right). The values are expressed relative to the case where information on the wife's status is not taken into account, represented by the blue dashed line. This reference situation is compared to those where the current values of the two types of contracts are calculated by separating the men whose wives are still alive when the contract is signed (black solid line) from those whose wives are deceased (pink dashed line).}
In Figure~\ref{fig:p_insur_couples_homme_pct} (and all figures where relative differences are computed), the baseline is the entire population. For a male, age 60, the expected value of the pension should be 2\% larger if his wife is still alive, for example, but the premium of a life insurance is only $5$\textperthousand~ lower.

\begin{figure}[htb!]
  \centering
    \includegraphics[width=\textwidth]{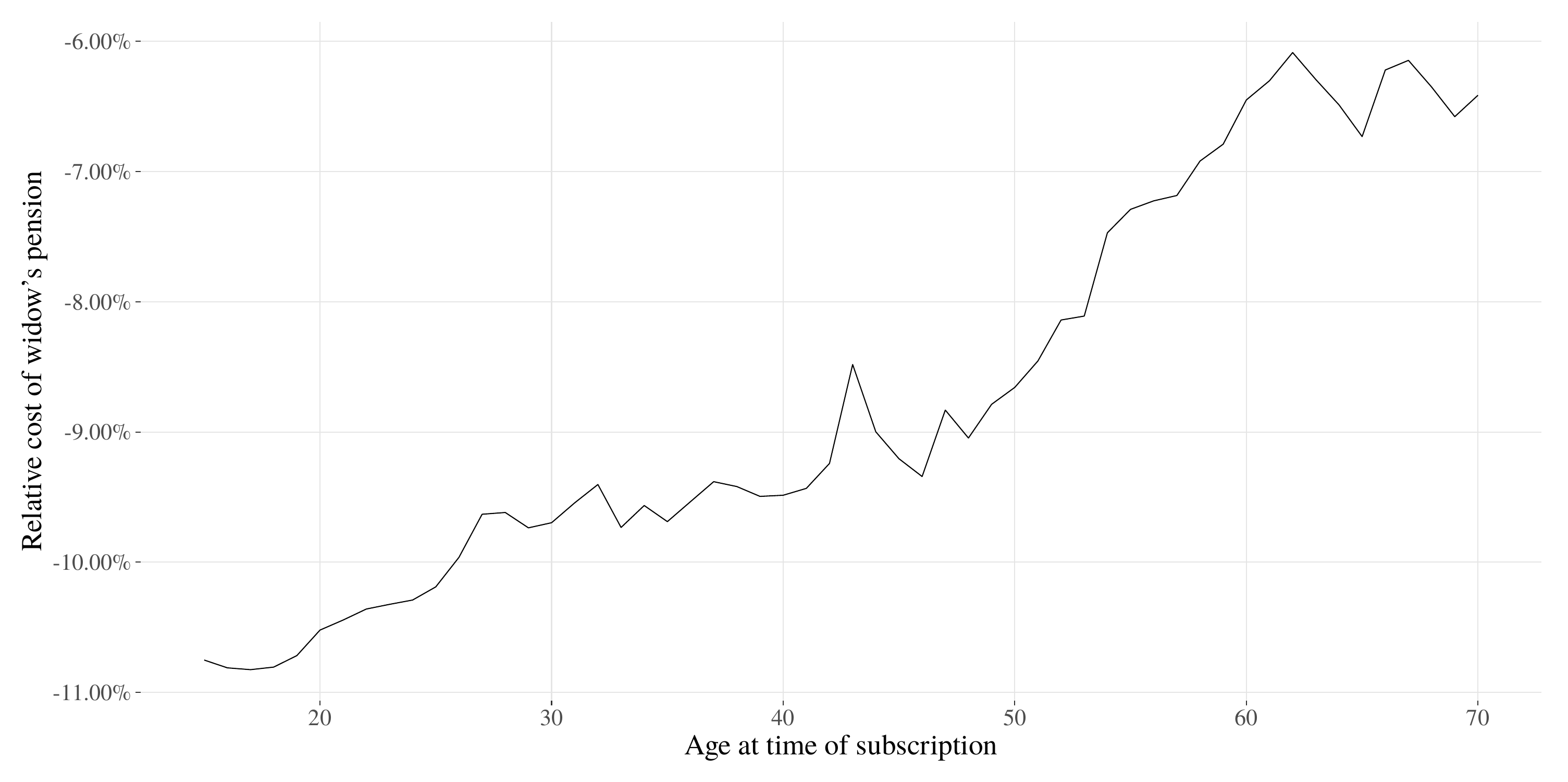}
\caption{Widow's pension, $a_{\text{m}|\text{f}}$ (relative to independent case $a^{\perp}_{\text{m}|\text{f}}$), as a function of $x_{\text{m}}$.}\label{fig:widows_pension}
\end{figure}

Figure~\ref{fig:widows_pension} compares the cost of a standard widow's annuity $a_{\text{m}|\text{f}}$ relative to the cost of the annuity under the assumption of (statistically) independence. As explained in the appendix, since we have a positive relationship between joint lives, widow's pension should be lower than under the assumption that joint lives are independent. For a mother in her 30's, the value of the widow's pension should be about 10\% lower (than the independent case), while it should be 7\%  lower if she is in her 60's.

\section{Parents and Children Dependencies}\label{sec:parents:grandparents}

This section first presents the results of the analysis of the links between the lifespan of an individual and that of his or her parents, and then considers the links with the lifespan of grandparents.

\subsection{Children Conditional on Parents}

{As noted in Section~\ref{sec:description:data}, several parametric models were fitted to estimate individual mortality. The adjustments of the Gompertz, Beard, Carrière and Hellingman-Pollard distributions on the force of mortality and survival probabilty, respectively, for men, women, fathers and mothers are displayed in the Appendix, in Figures~\ref{fig:force_mortality_parents} and~\ref{fig:survival_parents}. The graphs also report the observed values. The Carri\`ere model is the one that best fits the data.}

\subsubsection{Empirical Evidence of the Relationship Between an Individual's Lifespan and that of his or her Parents}

\begin{figure}[htb!]
  \centering
    \includegraphics[width=\textwidth]{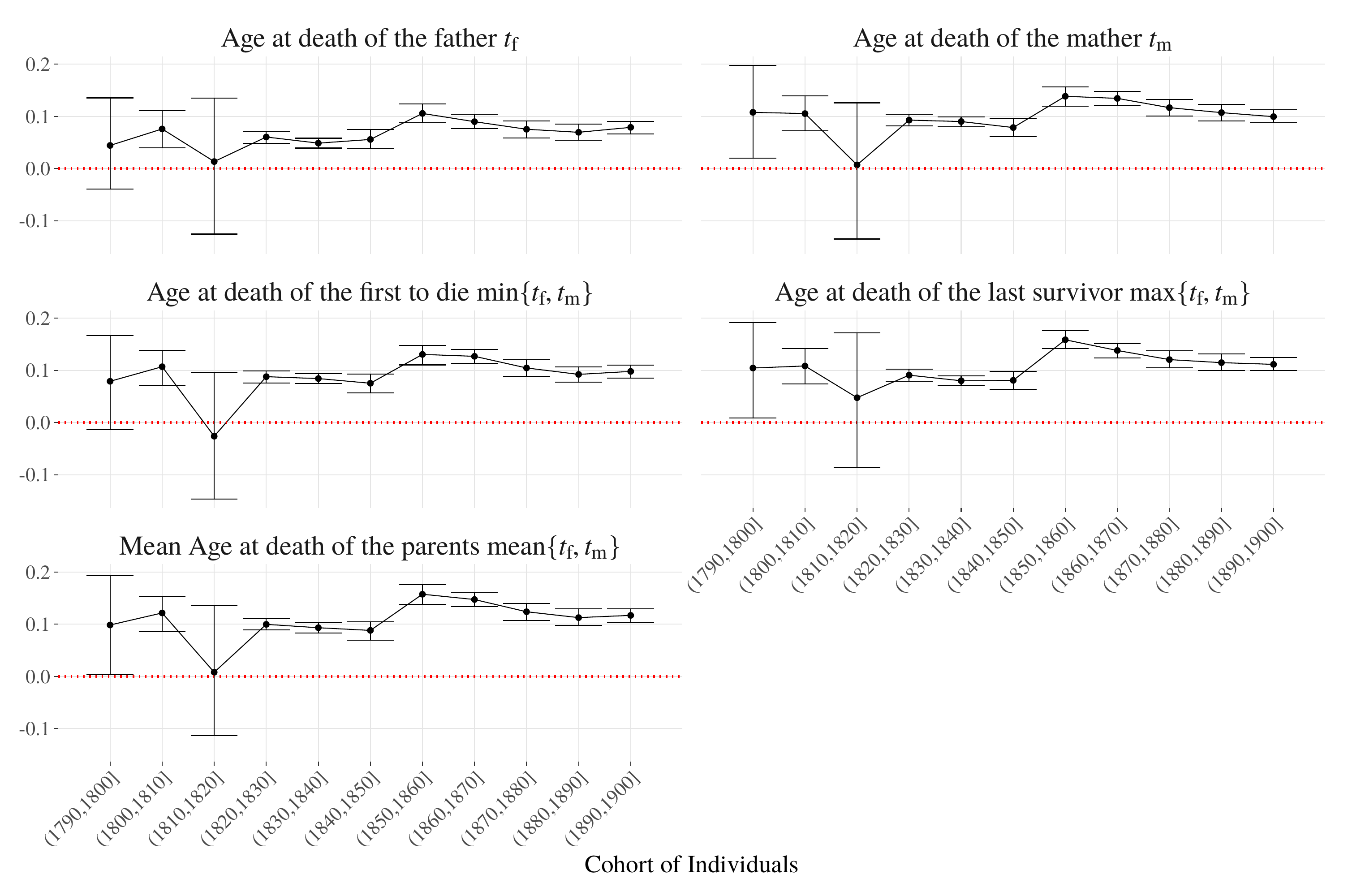}
  \begin{minipage}{\textwidth}
		\vspace{1ex}
		\scriptsize\underline{Note:} The dots represent the estimated Spearman correlation between the age at death of a child $t_\textrm{c}$ and that of their parents. The bars correspond to 95\% bootstrap confidence interval. The red horizontal line corresponds to a value of 0. The correlations for the $(1790,1800]$ and $(1810,1820]$ cohorts are calculated on 476 and 250 couples of parents only. For the other cohorts, the correlations are calculated from a much larger number of observations, ranging from $3,332$ (for the $(1800,1810]$ cohort) to $41,345$ (for the $(1830,1840]$ cohort).
	
\end{minipage}
\caption{Spearman correlation between age at death of individuals $t_{\text{c}}$ and age at death of their parents.}\label{fig:correl_parents}
\end{figure}

As was done in the case of couples, a first way to visualize the relationship between the age at death of children and their parent's age at death is to look at the correlation between $t_{\text{c}}$ and various variables: the age at death of the father $t_{\text{f}}$, the age at death of the mother $t_{\text{m}}$, the age at death of the first to die $\min\{t_{\text{f}},t_{\text{m}}\}$, the age at death of the last survivor $\max\{t_{\text{f}},t_{\text{m}}\}$, and the average age at death of the parents $\text{mean}\{t_{\text{f}},t_{\text{m}}\}$. As can be seen in Figure~\ref{fig:correl_parents}, this correlation is positive, albeit relatively weak and appears to be constant over cohorts. Overall, regardless of the cohorts, the Spearman correlation between $t_\textrm{c}$ and $\text{mean}\{t_{\text{f}},t_{\text{m}}\}$ is 0.125, with a 95\% bootstrap confidence interval of $[0.121;0.130]$.

\begin{figure}[htb!]
  \centering
  \begin{subfigure}[t]{.31\textwidth}
		\centering
		\caption{Age at death of the father $t_{\text{f}}$}\label{subfig:nonparam_copula-struc_both_age_father}
		\includegraphics[width=\textwidth]{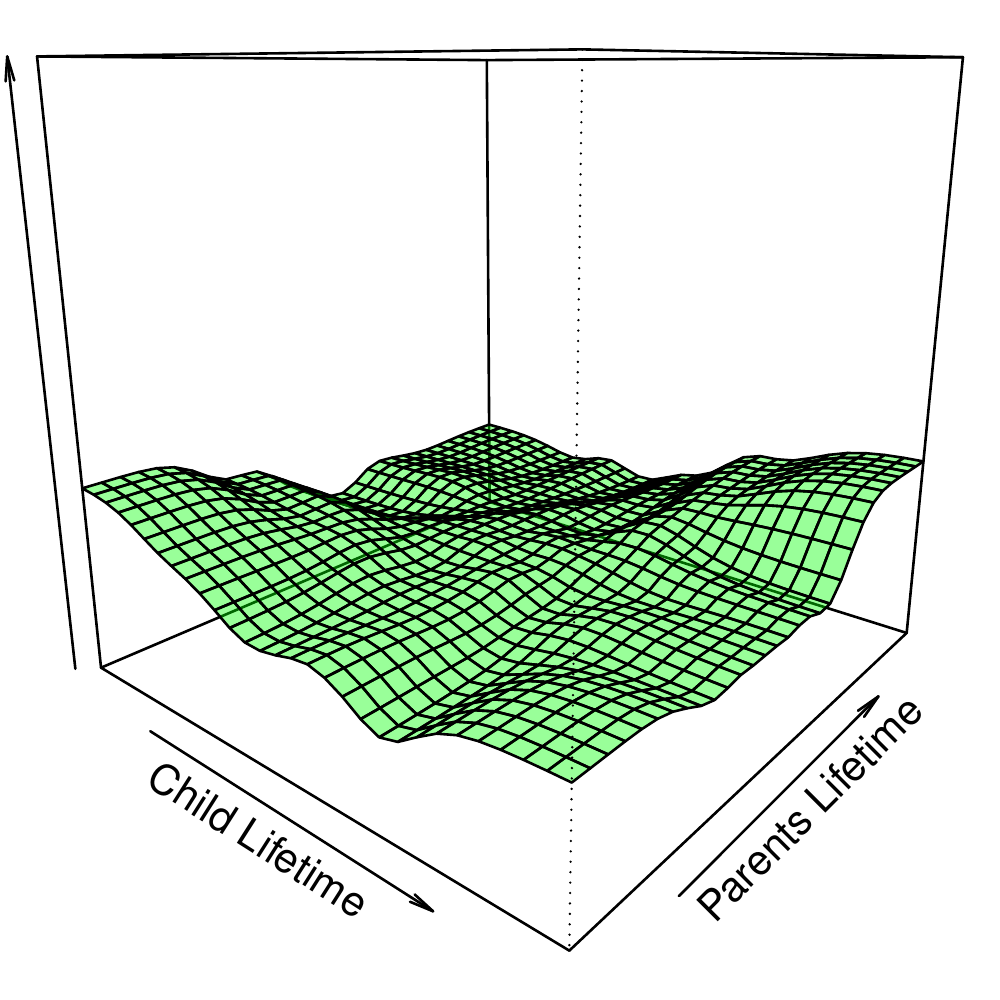}
\end{subfigure}
  \hspace{.5em}
  \begin{subfigure}[t]{.31\textwidth}
		\centering
		\caption{Age at death of the mother $t_{\text{m}}$}\label{subfig:nonparam_copula-struc_both_age_mother}
		\includegraphics[width=\textwidth]{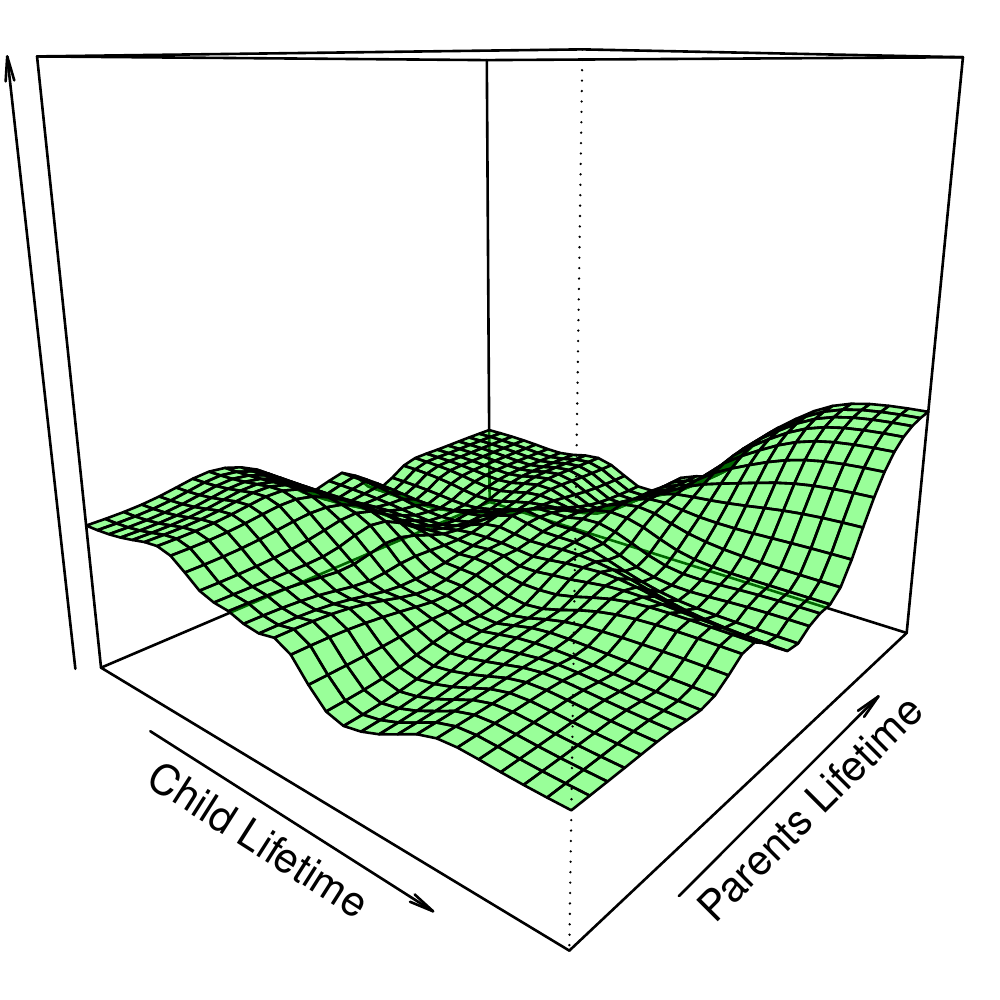}
	\end{subfigure}
	\hspace{.5em}
	\begin{subfigure}[t]{.31\textwidth}
		\centering
		\caption{Age at death of the first to die $\min\{t_{\text{f}},t_{\text{m}}\}$}\label{subfig:nonparam_copula-struc_both_age_parents_min}
		\includegraphics[width=\textwidth]{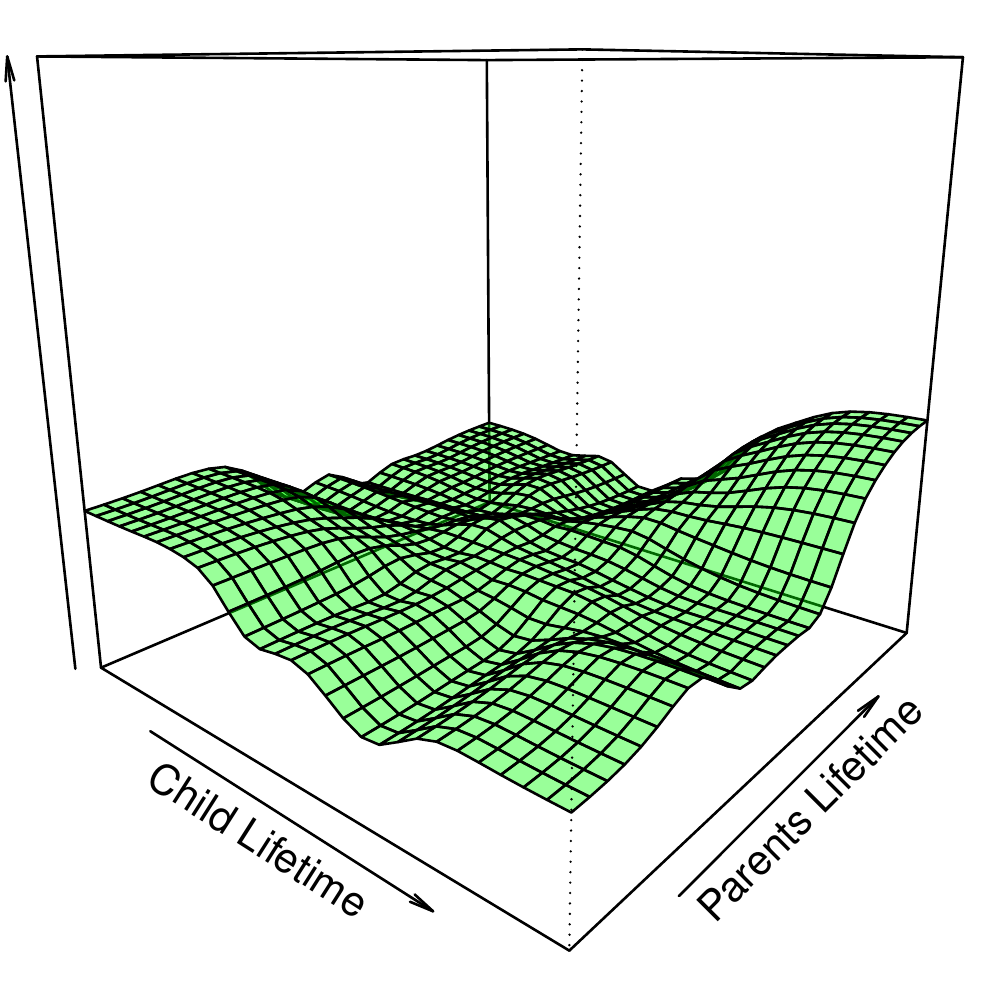}
	\end{subfigure}
	\begin{subfigure}[t]{.31\textwidth}
		\centering
		\caption{Age at death of the last survivor $\max\{t_{\text{f}},t_{\text{m}}\}$}\label{subfig:nonparam_copula-struc_both_age_parents_max}
		\includegraphics[width=\textwidth]{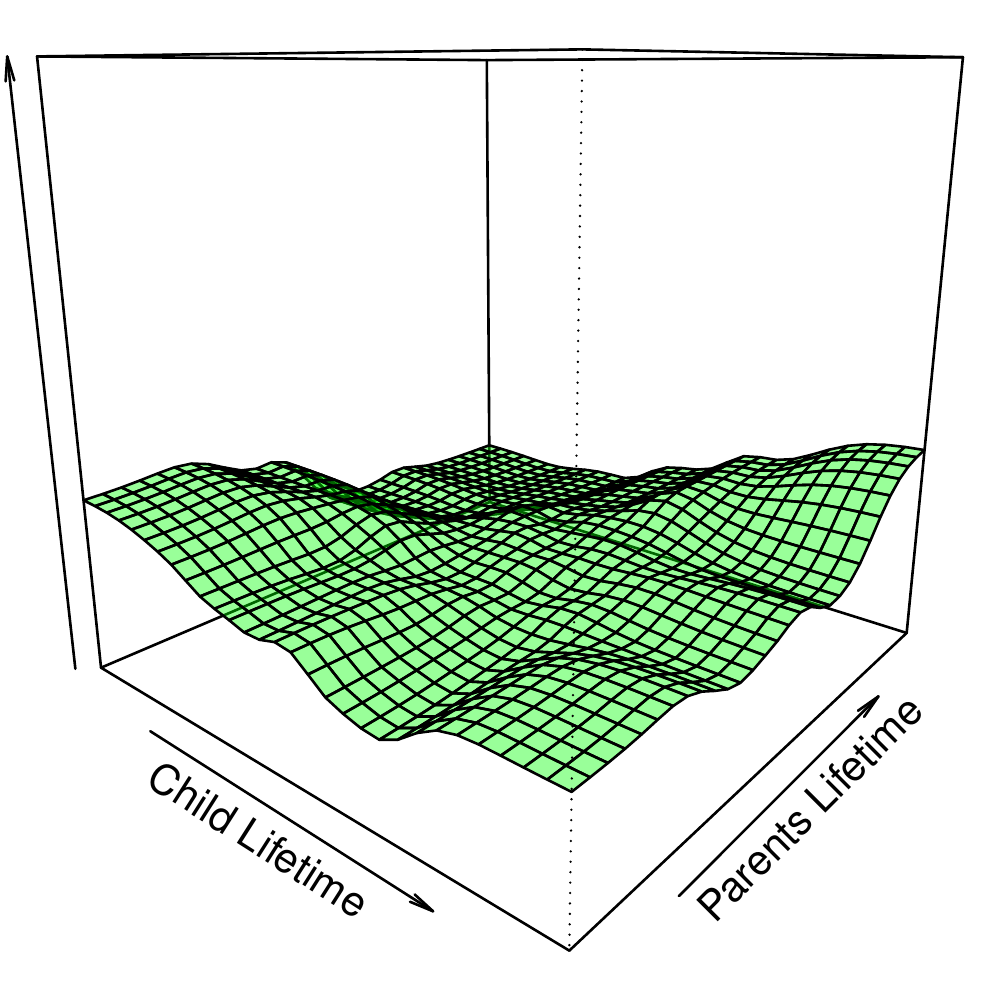}
	\end{subfigure}
	\hspace{1em}
	\begin{subfigure}[t]{.31\textwidth}
		\centering
		\caption{Mean Age at death of the parents $\text{mean}\{t_{\text{f}},t_{\text{m}}\}$}\label{subfig:nonparam_copula-struc_both_age_parents_mean}
		\includegraphics[width=\textwidth]{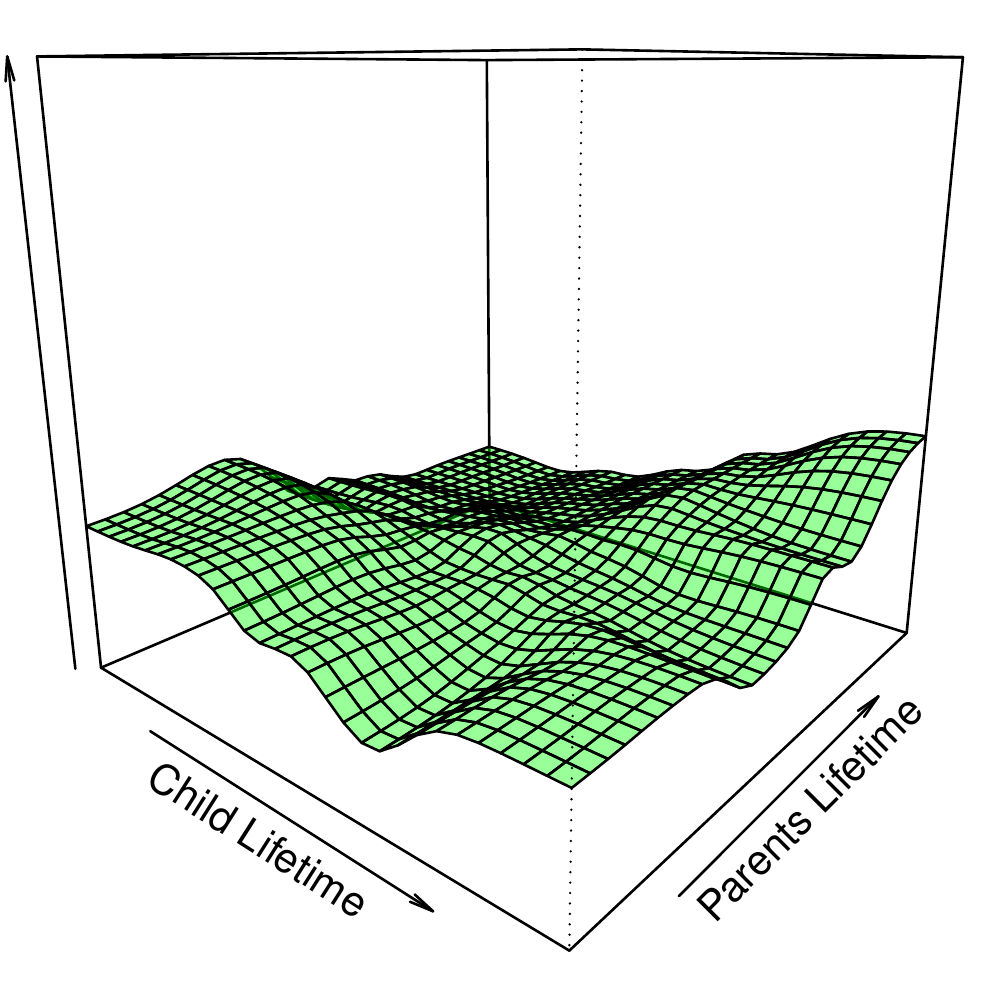}
	\end{subfigure}

  \begin{minipage}{\textwidth}
		\vspace{1ex}
		\scriptsize\underline{Note:} these graphs show the dependence between the age at death of a children ($t_{\text{c}}$) and the age at death of a parent, considering different possible measures for the parent: the age at death of the father $t_{\text{f}}$, the age at death of the mother $t_{\text{m}}$, the age at death of the first to die $\min\{t_{\text{f}},t_{\text{m}}\}$ and the age at death of the last survivor  $\max\{t_{\text{f}},t_{\text{m}}\}$.

\end{minipage}
\caption{Nonparametric estimation of the copula densities.}\label{fig:children:1}
\end{figure}

These small links between parents and children are also seen in Figure~\ref{fig:children:1}, which shows the relationships between $t_{\text{c}}$ and the same age at death variables as previously, this time using the copula estimation results. Regardless of the variable selected for the parents, the surfaces look the same.

\begin{figure}[htb!]
  \centering
    \includegraphics[width=\textwidth]{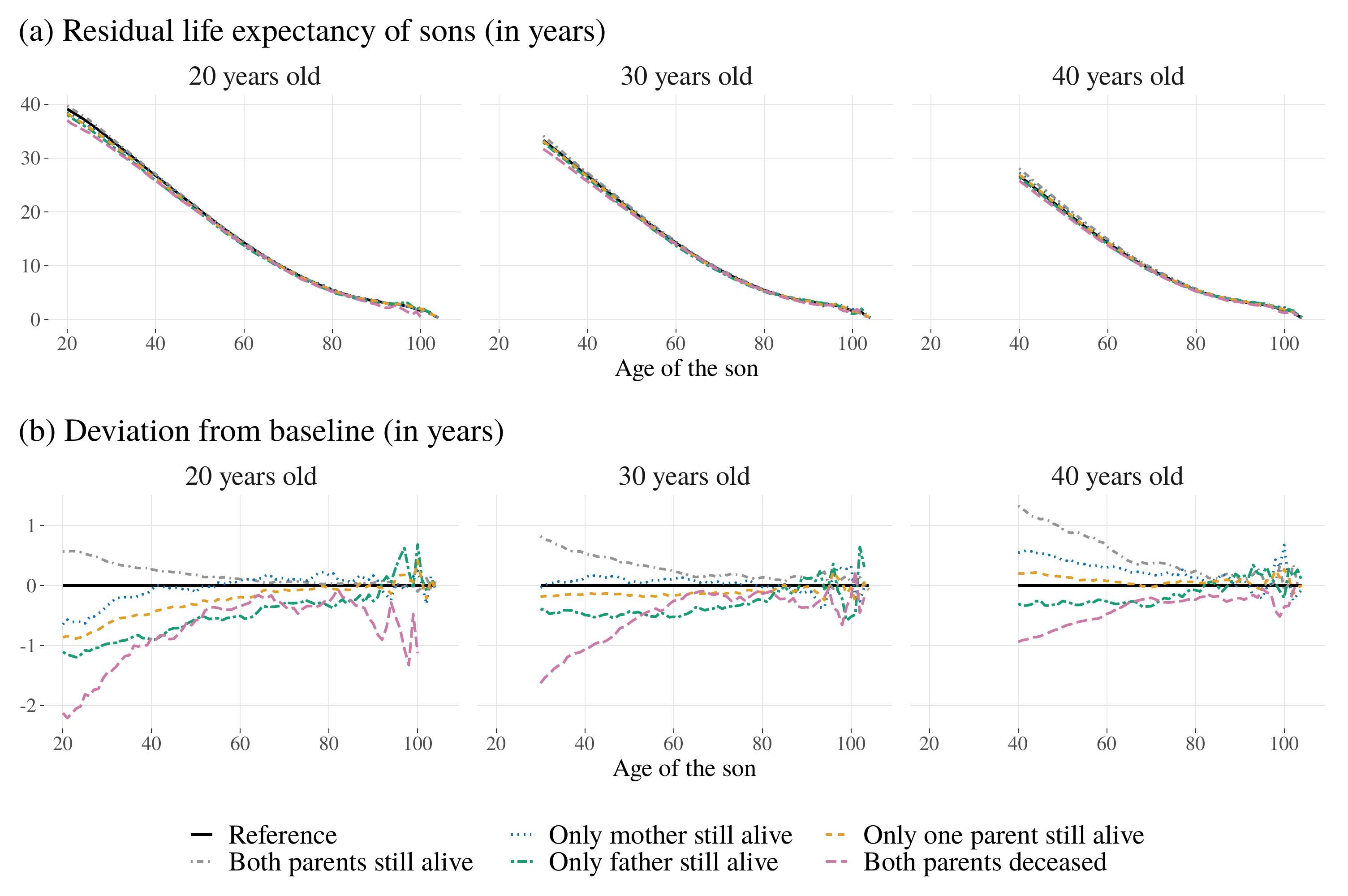}
  \begin{minipage}{\textwidth}
		\vspace{1ex}
		\scriptsize\underline{Note:} each panel reports the residual life expectancy of individuals (only men) according to their parents' death status (information not accounted for -- reference --, both still alive, only the father still alive, only the mother still alive, both deceased), at different times in the life of the individuals (at 20, 30, and 40 years old). Left panels thus indicates the residual life expectancy of men when both parents are still alive when they are 20 years old, when only one of them is alive, and so on. Top panels show the residuals life expectancy expressed in years. Bottom panels show the relative difference to the reference, expressed in years.
\end{minipage}
\caption{Residuals life expectancy depending on the death status of the parents, at different times in the lives of men.}\label{fig:life_ex_parents}
\end{figure}

The evolution of the residual life expectancy $e_x$ of men given some information about their parents can be visualized in Figure~\ref{fig:life_ex_parents}. Various situations are compared concerning the death's status of the parents at a given age of their son (at 20 years of age for the graphs on the left, 30 years of age for those in the middle and 40 years of age for those on the right): information not accounted for (baseline situation), both parents still alive, the father deceased and the mother still alive, the mother deceased and the father still alive, one of the two parents still alive regardless of gender, and both parents deceased. The upper graphs show the residual life expectancy expressed in years, while the lower graphs allow an easier comparison with the reference situation by showing the difference in years on the residual life expectancy of sons compared to the reference situation in which the information on the death of the parents is not taken into account. 

Irrespective of the age at which the information regarding parent's death is looked upon, \textit{i.e.}, 20, 30 or 40 years old, it can be noted that the residual life expectancy curve for males whose parents are both still alive is systematically above the other curves, while the curve for children with both parents dead is systematically below the other curves. The difference in life expectancy is, however, relatively small and lessens over the years. As reported in Table~\ref{tab:ex_parents}, a male child whose both parents were still alive when he was 20 years old had a life expectancy of 39.7 years, compared to only 37.0 if both parents were deceased. In comparison, the residual life expectancy of a 20 years old man, without taking into account information about his parents, is 39.1 years. At age 30, the difference was much smaller: compared to the baseline value of 33.3, a male individual whose parents were both alive at that time was expected to live an additional 0.8 year while a man whose parents were both deceased was expected to live 1.6 year less. Lastly, at age 40, an man was expected to live another 26.7 years, a bit more (1.3 year) if his parents were still both alive at that age and slightly less if both parents were deceased (about one year). Similar patterns are observed for female children.

\subsubsection{Annuities and Life Insurance Premiums Accounting for the Status of Parents}

\begin{figure}[htb!]
  \centering
    \includegraphics[width=\textwidth]{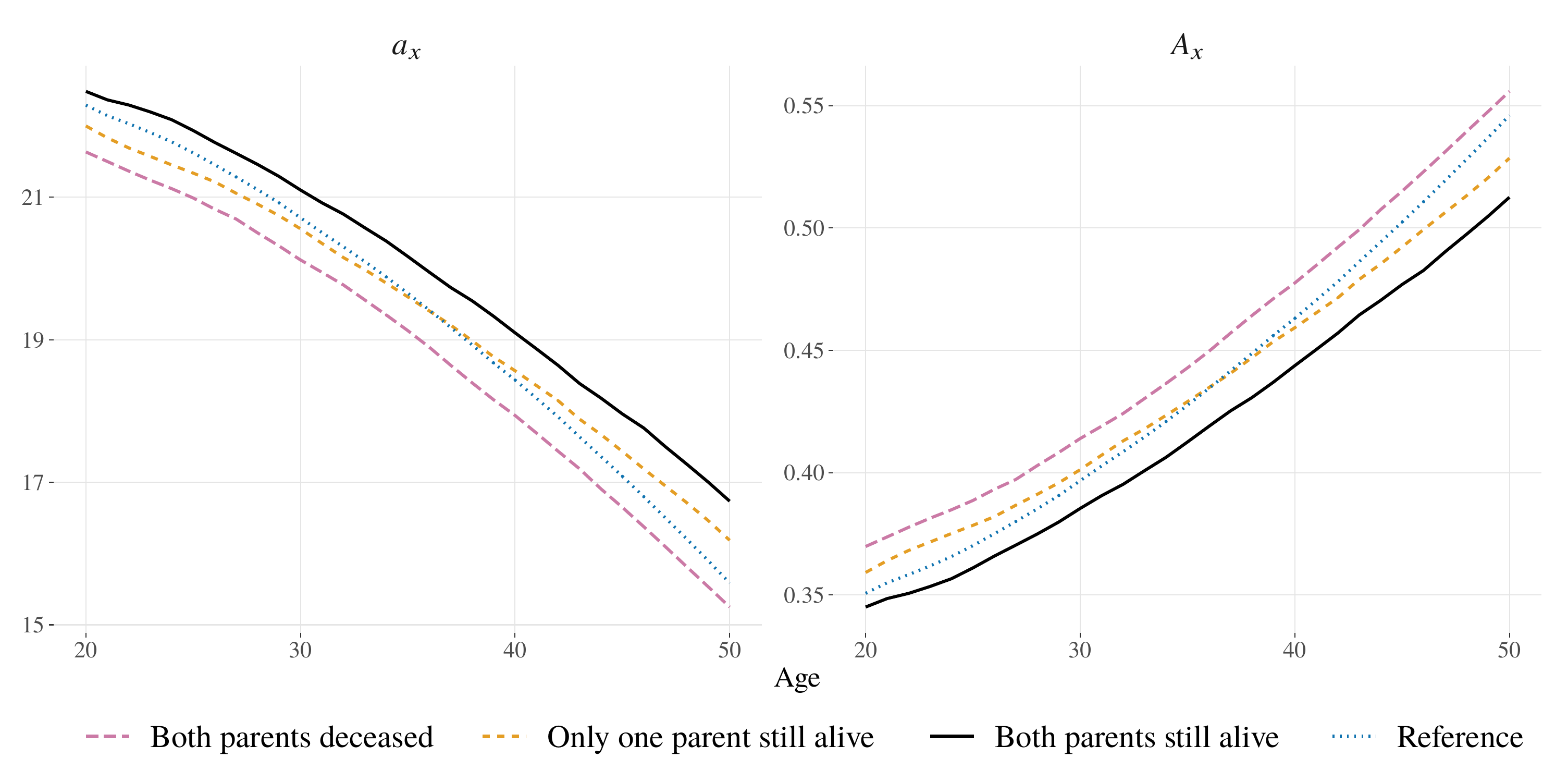}
  \begin{minipage}{\textwidth}
		\vspace{1ex}
		\scriptsize\underline{Note:} the annuities are calculated for 100 terms, the expected present value for life insurance are calculated for 100-year coverage.

\end{minipage}
\caption{Present value of an annuity (left) and expected present value for a life insurance (right) depending on the age of the annuitant and on how many parents are still alive at the time of the contract. The interest rate is assumed to be $3\%$.}\label{fig:p_insur_parents}
\end{figure}

\begin{figure}[htb!]
  \centering
    \includegraphics[width=\textwidth]{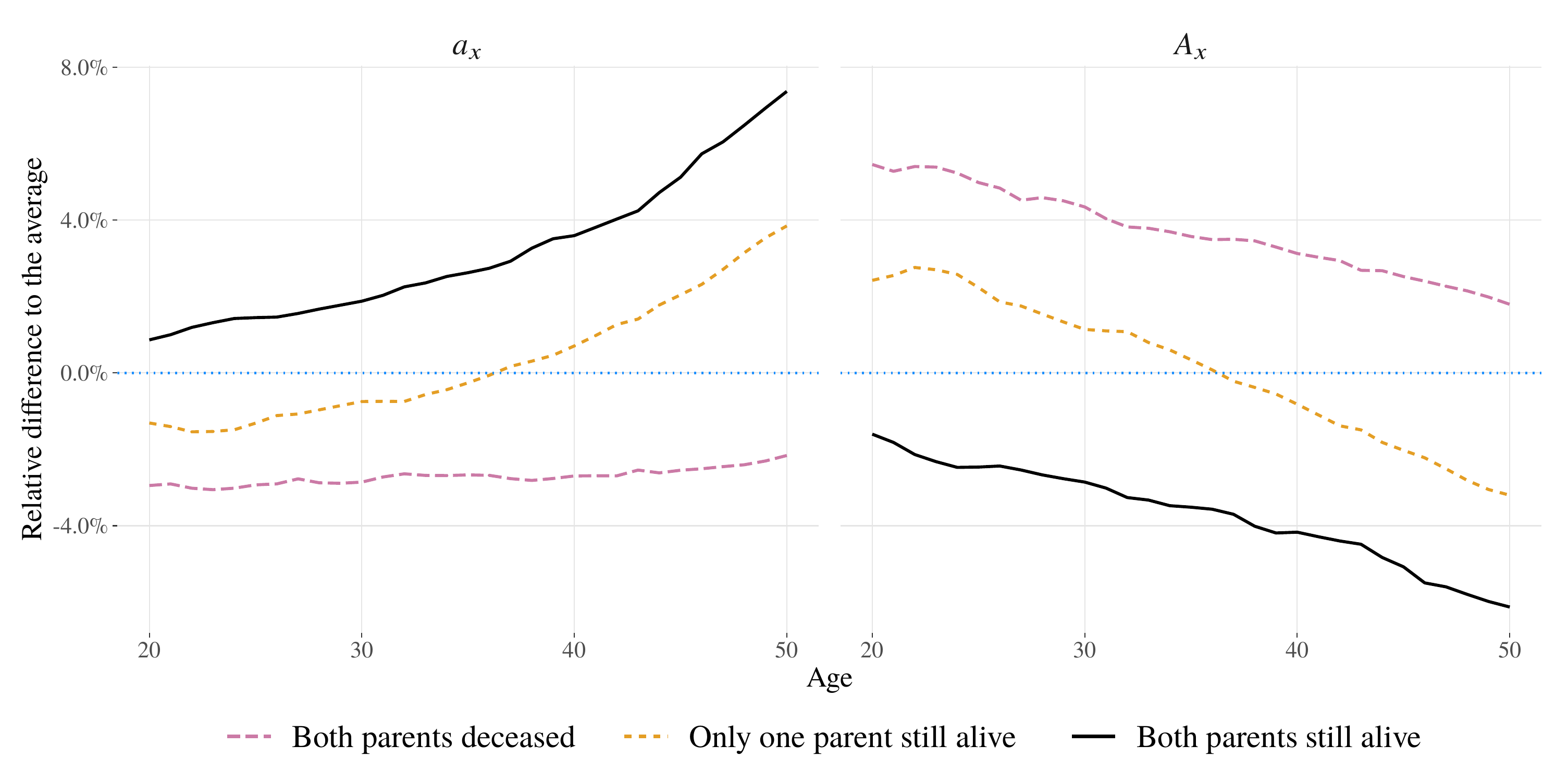}
  \begin{minipage}{\textwidth}
		\vspace{1ex}
		\scriptsize\underline{Note:} the horizontal blue dashed line corresponds to the average value calculated on all individuals, regardless of the death status of the parents. The annuities are calculated for 100 terms and the expected present value for life insurance are calculated for 100-year coverage. The interest rate is assumed to be $3\%$.
\end{minipage}
\caption{Relative difference to the average (in \%) of present value of an annuity (left) and expected present value for a life insurance (right) depending on the age of the annuitant and on how many parents are still alive at the time of the contract.}\label{fig:p_insur_parent_pct}
\end{figure}

In Figure~\ref{fig:p_insur_parents}, we observe the evolution of the present value of an annuity and the life insurance, as a function of the age of the insured, $x$, given information about his or her parents when buying the insurance contract. It can be noted that (empirical) monotonicity is consistent with theoretical results (decreasing with $x$ for the pension and increasing for the life insurance), and the ordering of the three cases (both parents deceased, one parent still alive and both parents still alive) is consistent with the positive dependence between all lifespans.
Figure~\ref{fig:p_insur_parent_pct} displays the relative difference (to the average baseline, as previously). It shows that present value of the annuity is consistently 3\% lower when both parents are deceased, whatever the age. If both parents are alive, the difference of the present value of that annuity is increasing with $x$, and is 3.7\% higher for an insured age $40$.

\subsection{Children Conditional on Grand-Parents}

We now extend our previous work to grandparents, where insured provide information about his or her four grandparents.

\subsubsection{Empirical Evidence of the Relationship Between an Individual's Lifespan and that of his or her Grandparents}

\begin{figure}[htb!]
  \centering
    \includegraphics[width=\textwidth]{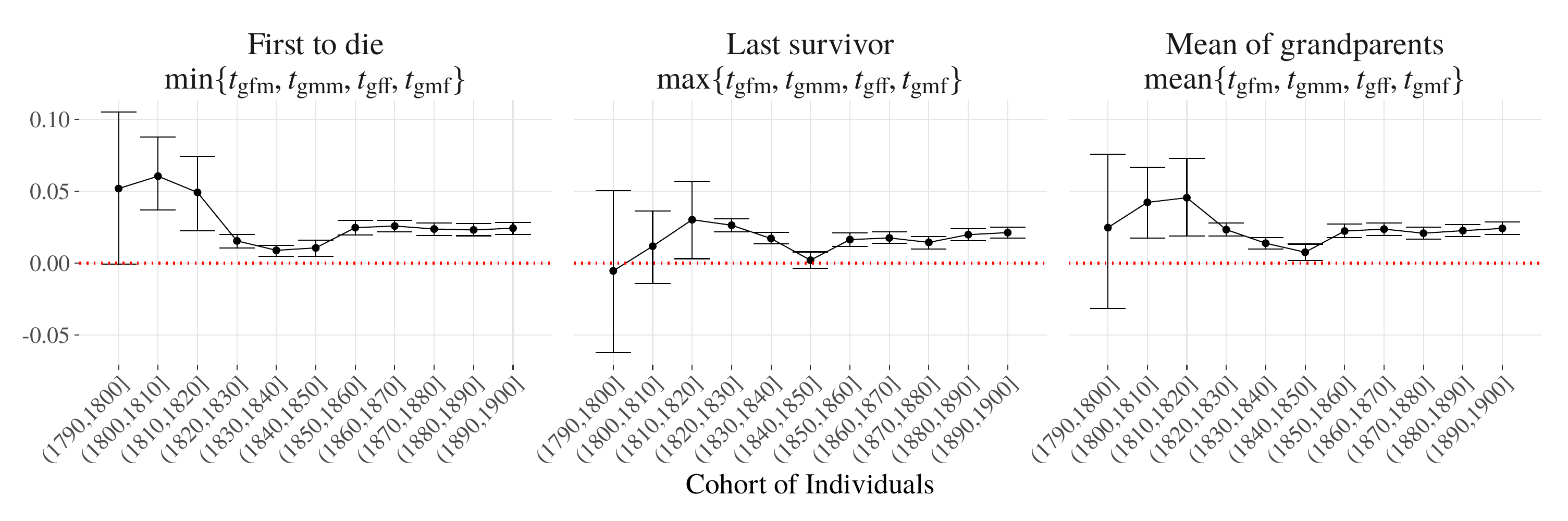}
  \begin{minipage}{\textwidth}
		\vspace{1ex}
		\scriptsize\underline{Note:} The dots represent the estimated Spearman correlation between the age at death of a child $t_\textrm{c}$ and that of their grandparents. The bars correspond to 95\% bootstrap confidence interval.
\end{minipage}
\caption{Spearman correlation between age at death of individuals $t_{\text{c}}$ and age at death of their grandparents.}\label{fig:correl_gparents}
\end{figure}

Spearman's correlation between the age at death of individuals $t_c$ and the average age at death of their grandparents $\text{mean}\{t_{\textrm{gfm}},t_{\textrm{gmm}},t_{\textrm{gff}}, t_{\textrm{gmf}}\}$ is even weaker than with the average age at death of their parents $\text{mean}\{t_{\textrm{f}},t_{\textrm{m}}\}$: $0.0251$, with a $95\%$ bootstrap confidence interval equal to $[0.0235, 0.0266]$. However, this confidence interval stresses that this correlation, although tenuous, is sinigicatively different from zero. In addition, as shown in Figure~\ref{fig:correl_gparents}, this correlation appears to be relatively stable over time. The same Figure also shows that the correlation between the age at death of an individual and that of grandparents is stable over time when the latter is measured by using only the age at death of the first or last to die.

\begin{figure}[htb!]
  \centering
  \begin{subfigure}[t]{.31\textwidth}
		\centering
		\caption{Age at death of the first to die $\min\{t_{\textrm{gfm}},t_{\textrm{gmm}},t_{\textrm{gff}}, t_{\textrm{gmf}}\}$}\label{subfig:nonparam_copula-struc_both_age_gparents_min}
		\includegraphics[width=\textwidth]{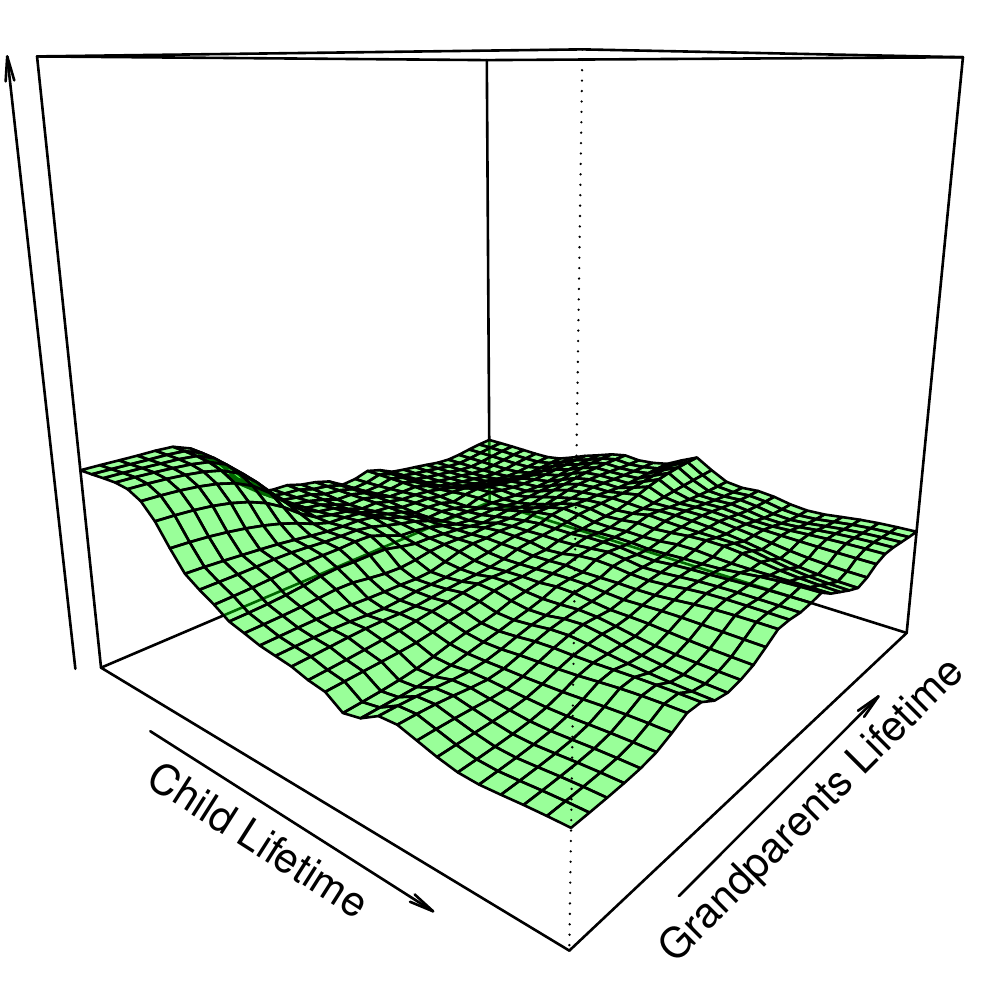}
\end{subfigure}
  \hspace{1em}
  \begin{subfigure}[t]{.31\textwidth}
		\centering
		\caption{Age at death of the last survivor $\max\{t_{\textrm{gfm}},t_{\textrm{gmm}},t_{\textrm{gff}}, t_{\textrm{gmf}}\}$}\label{subfig:nonparam_copula-struc_both_age_gparents_max}
		\includegraphics[width=\textwidth]{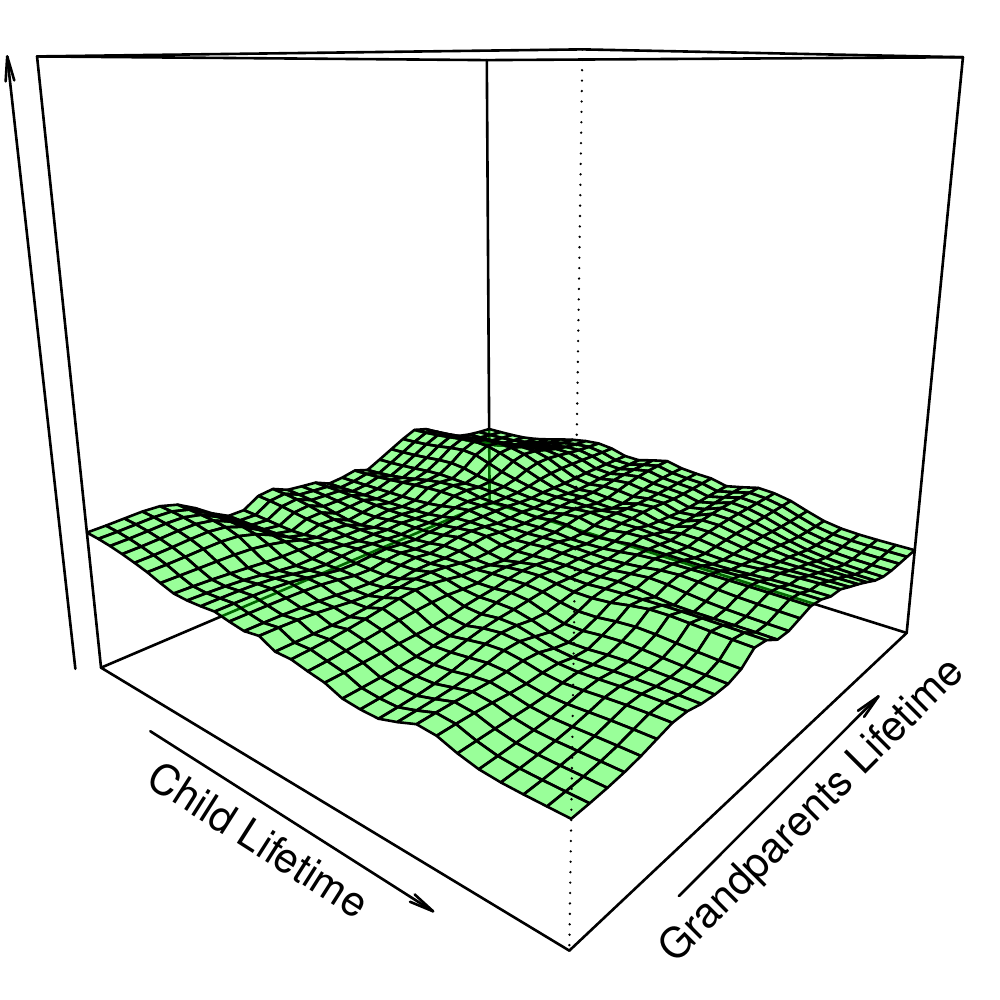}
	\end{subfigure}
	\begin{subfigure}[t]{.32\textwidth}
		\centering
		\caption{Mean Age at death of the grandparents $\text{mean}\{t_{\textrm{gfm}},t_{\textrm{gmm}},t_{\textrm{gff}}, t_{\textrm{gmf}}\}$}\label{subfig:nonparam_copula-struc_both_age_gparents_mean}
		\includegraphics[width=\textwidth]{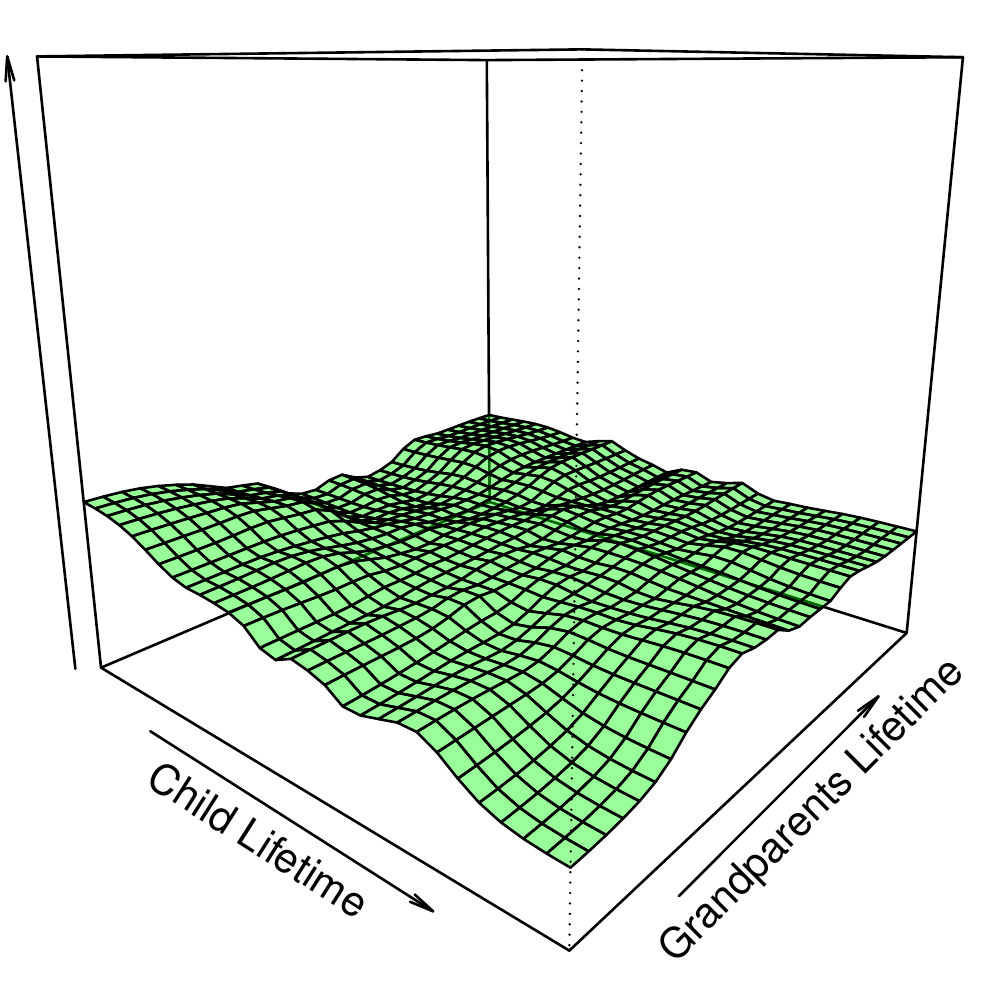}
	\end{subfigure}
  \begin{minipage}{\textwidth}
		\vspace{1ex}
		\scriptsize\underline{Note:} these graphs show the dependence between the age at death of a children ($t_{\text{c}}$) and the age at death of their grandparents, considering different possible measures for the grandparents: the age at death of the first to die $\min\{t_{\textrm{gfm}},t_{\textrm{gmm}},t_{\textrm{gff}}, t_{\textrm{gmf}}\}$, the age at death of the last survivor  $\max\{t_{\textrm{gfm}},t_{\textrm{gmm}},t_{\textrm{gff}}, t_{\textrm{gmf}}\}$ or the average age at death of the grandparents $\text{mean}\{t_{\textrm{gfm}},t_{\textrm{gmm}},t_{\textrm{gff}}, t_{\textrm{gmf}}\}$.
		
\end{minipage}
\caption{Nonparametric estimation of the copula densities for the grandparents.}\label{fig:cop:grandparents}
\end{figure}

In a similar way to what was presented for the parents, the relationship between $t_{\text{c}}$ and $\text{mean}\{t_{\textrm{gfm}},t_{\textrm{gmm}},t_{\textrm{gff}}, t_{\textrm{gmf}}\}$ can be studied using the results of copula estimates. These are graphed in Figure~\ref{fig:cop:grandparents}.

\begin{figure}[htb!]
  \centering
    \includegraphics[width=\textwidth]{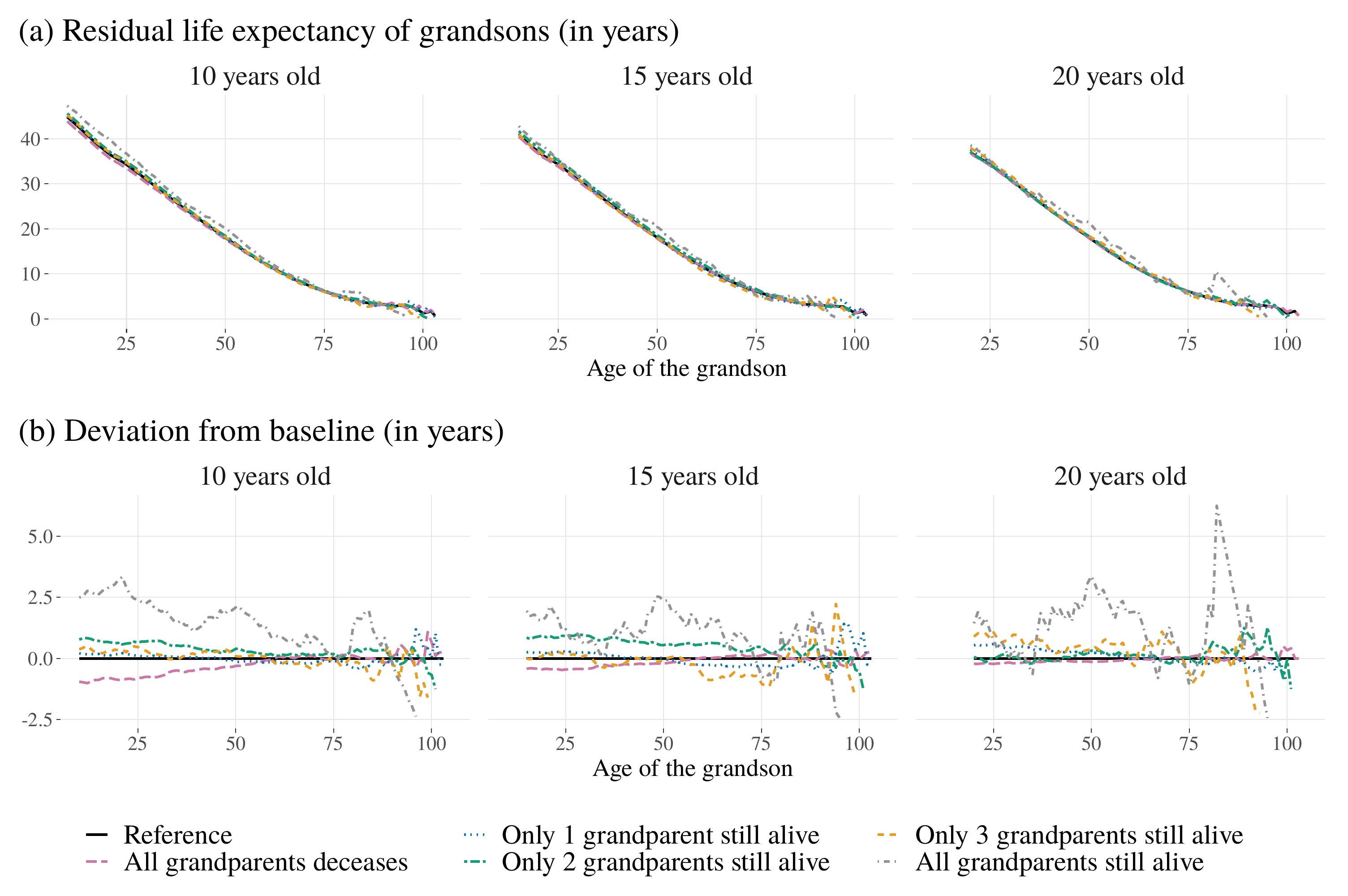}
  \begin{minipage}{\textwidth}
		\vspace{1ex}
		\scriptsize\underline{Note:} each panel reports the residual life expectancy of individuals (only men) according to their grandparents' death status (information not accounted for -- reference --, all four still alive, only one still alive, two still alive, three still alive, all four deceased), at different times in the life of the grandsons (at 10, 15, and 20 years old). Left panels thus indicates the residual life expectancy of grandsons when all four grandparents are still alive when they are 10 years old, when only one of the grandparents is still alive, and so on. Top panels show the residuals life expectancy expressed in years. Bottom panels show the relative difference to the reference, expressed in years.
\end{minipage}
\caption{Residual life expectancy depending on the death status of the grandparents, at different times in the lives of men.}\label{fig:life_ex_gparents}
\end{figure}

A comparison of the remaining life expectancy of an individual (male) according to the number of his grandparents still alive when he is 10, 15 or 20 years old is shown in Figure~\ref{fig:life_ex_gparents}, taking as a reference the situation in which knowledge of this information is not taken into account.\footnote{Because of the relatively low life expectancy in the 19th century, it is unfortunately not possible to form groups of individuals with 3 or 4 grandparents when the first ones are 30 or 40 years old. This is why the ages of interest here (10, 15 and 20 years) are lower than in the previous comparison with information on parents (20, 30 and 40 years).} When all four grandparents of a person (male) are still alive when he is 10 years old, the figure shows that his remaining life expectancy (grey dot-dashed line) is relatively higher than average (black solid line). On the contrary, when all four grandparents are dead, then the remaining life of the grandson (pink dashed line) is lower than the average, although the absolute difference is not as large. For both cases, the absolute deviation from the average decreases over the years. This difference between the two groups is much less marked if the number of grandparents still alive when the children are 15 years old is considered, but it should be noted that those who still have their four grandparents at that age still have a relatively higher remaining life expectancy than the average.

\subsubsection{Annuities and Life Insurance Premiums Accounting for the Status of Grandparents}

In this section, computations are based on a (much) smaller dataset, where we kept individuals for whom information about the four grand-parents was available.

\begin{figure}[htb!]
  \centering
    \includegraphics[width=\textwidth]{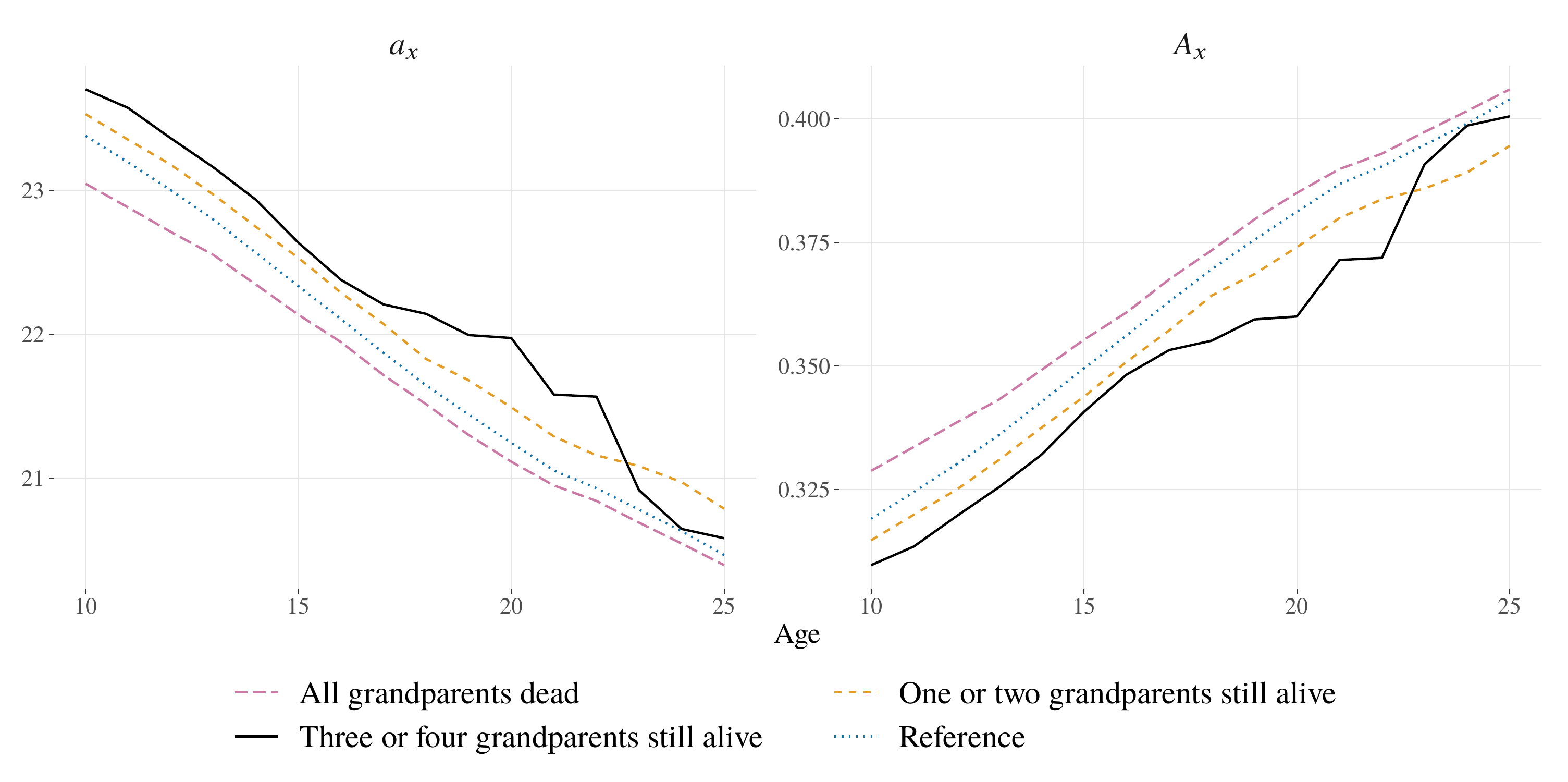}
  \begin{minipage}{\textwidth}
		\vspace{1ex}
		\scriptsize\underline{Note:} the annuities are calculated for 100 terms, the expected present value for life insurance and for endowment are calculated for 100-year coverage. The interest rate is assumed to be $3\%$.
\end{minipage}
\caption{Present value of an annuity (left), expected present value for a life insurance (middle) and for an endowment (right) depending on the age of the annuitant and on how many grandparents are still alive at the time of the contract.}\label{fig:p_insur_gparents}
\end{figure}

\begin{figure}[htb!]
  \centering
    \includegraphics[width=\textwidth]{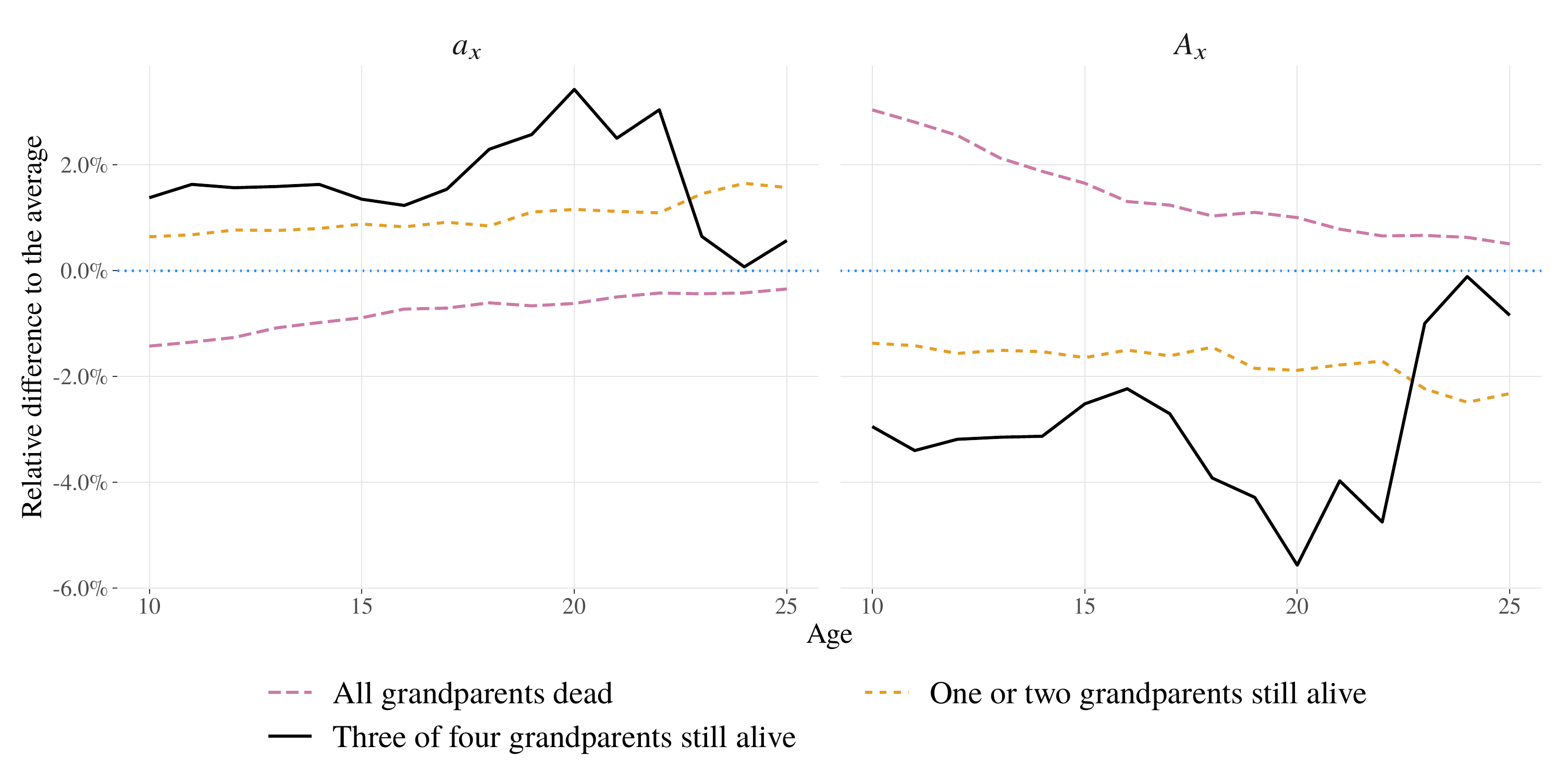}
  \begin{minipage}{\textwidth}
		\vspace{1ex}
		\scriptsize\underline{Note:} the horizontal blue dashed line corresponds to the average value calculated on all individuals, regardless of the death status of the grandparents. The annuities are calculated for 100 terms, the expected present value for life insurance and for endowment are calculated for 100-year coverage. The interest rate is assumed to be $3\%$.
\end{minipage}
\caption{Relative difference to the average (in \%) of present value of an annuity (left), expected present value for a life insurance (middle) and for an endowment (right) depending on the age of the annuitant and on how many grandparents are still alive at the time of the contract.}\label{fig:p_insur_gparent_pct}
\end{figure}

Figure~\ref{fig:p_insur_gparents} is the analogous of Figure~\ref{fig:p_insur_parents}, where the evolution of the present value of an annuity and the life insurance is represented as a function of the age of the insured, $x$, given information about his or her grandparents when buying the insurance contract. Again, the (empirical) monotonicity is consistent with theoretical results (decreasing with $x$ for the pension and increasing for the life insurance), and the ordering of the three cases (all grandparents deceased, one or two still alive and three or four still alive) is consistent with the overall positive dependence between all lifespans. But here, as seen in Figure~\ref{fig:p_insur_gparent_pct}, the relative difference is much smaller, (at most) 2\%, except perhaps when three or four grands parents are still alive and when $x$ is 
`large' (but in that case, the number of observations is much smaller, and the difference probably not significant).

\section{Conclusion}\label{sec:conclusion}

In this article, we use collaborative genealogy data to study joint mortality within families. Using $135,128$ observations of couples from people born during the 19th Century in France, we observe well-known results from the literature on dependency in mortality. Then we look at the potential links between individuals and their parents regarding mortality. To do this, contrary to what is usually observed in the literature, we rely on a substantial volume of data: $174,318$ observations. We find results similar to those obtained in studies with smaller samples, \textit{i.e.}, a very weak but significant association between an individual's lifespan and that of his or her parents. Our data also allow us to take the study a step further by looking at the relationship between an individual's longevity and that of his or her grandparents. As with the parents, a very weak but significant positive association is observed. We then look at the potential implications for annuities and life insurance. 

Here, we consider family history only through the prism of the age at death of ancestors. Furthermore, we do not try to explain the nature of the correlation, and it is rather difficult to understand what is due to genetics and heredity on the one side, and environmental factors on the other. This is all the more difficult since the reason for death is not stated in such a dataset. We can still wonder if family history (about the age at death of ancestors) is an important information: since it has a rather small impact (as we proved in this article), if the cause is mainly environmental, the correlation can be substantially captured by other common variables (such as the wealth).

\clearpage

\appendix

\setcounter{table}{0}
\renewcommand{\thetable}{\Alph{section}\arabic{table}}
\setcounter{figure}{0}
\renewcommand{\thefigure}{\Alph{section}\arabic{figure}}

\section{Models for Joint Life Dependencies}\label{appendix:copula}

\subsection{Parametric Copulas}

In order to provide a more accurate comparison with related literature, we consider some popular parametric copulas, in this section: Clayton copula is defined as $$C_{\text{c},\theta}(u,v)= {\displaystyle \left[u^{-\theta }+v^{-\theta }-1\right]^{-1/\theta }},$$
(with $\theta\geq 0$ from \citealp{Clay:1978}), Gumbel copula $$C_{\text{g},\theta}(u,v)={ \exp \!\left[-\left((-\log(u))^{\theta }+(-\log(v))^{\theta }\right)^{1/\theta }\right]},$$
(with $\theta\geq 1$, from \citealp{Clay:1978}) the normal (or Gaussian) copula $$C_{\text{n},\theta}(u,v)=\int_{-\infty}^{\Phi^{-1}(u)}\int_{-\infty}^{\Phi^{-1}(v)}{\displaystyle  {\frac {1}{2\pi {\sqrt {1-\rho ^{2}}}}}\exp \left(-{\frac {x^2-2\rho xy+y^2}{2(1-\rho ^{2})}}\right)}\mathrm{d}y\mathrm{d}x ,
$$
(with $\rho \in (-1,1)$), and finally Frank copula (from \cite{Frank})
$$C_{\text{c},\theta}(u,v)=
-{\frac {1}{\theta }}\log \!\left[1+{\frac {(\exp(-\theta u)-1)(\exp(-\theta v)-1)}{\exp(-\theta )-1}}\right]  ,
$$
(with $\theta\in \mathbb{R}$, with the independent copula with $\theta$ equals zero).

To estimate the parameters of the copula, instead of using the IFM method of \cite{Joe_Xu_1996}, we prefer the omnibus semiparametric procedure described in \cite{oakes}, where the copula is fitted on non-parametric pseudo observations $(\widehat{u}_{f,i},\widehat{u}_{m,i})$ where
$$
\widehat{u}_{f,i} = \widehat{S}_{f}(x_{f,i})\text{ where } \widehat{S}_{f}(x)=\frac{1}{n}\sum_{j=1}^n \boldsymbol{1}(x_{f,j}> x),
$$
for fathers, and a similar expression for mothers.

We define the empirical copula $\widehat{C}_n$ as the cumulative distribution function of $(\widehat{u}_{f,i},\widehat{u}_{m,i})$'s
$$
\widehat{C}_n(u,v) = \frac{1}{n}\sum_{i=1}^n \boldsymbol{1}\big(\widehat{u}_{f,i}\leq u,\widehat{u}_{m,i}\leq v\big),
$$
or some smooth version $\widetilde{C}_n(u,v)$ using some probit transformation, as in \cite{geenens}. 

If a positive dependence is observed, in the sense defined by \cite{Lehmann}, it can be used to derive bounds for most actuarial quantities.

\subsubsection{Positive Association between Lifetimes}\label{PQD}

As in \cite{Lehmann} -- see also \cite{scarsinishaked} for an exhaustive survey -- $X$ and $Y$ are said to be positively quadrant dependent (PQD) if and only if 
$$
F_X(x)\cdot F_y(y) \leq \mathbb{P}[X\leq x,Y\leq y] \text{ for all }x,y\in\mathbb{R}_+,
$$
or equivalently
$$
S_X(x)\cdot S_y(y) \leq \mathbb{P}[X>x,Y>y]\text{ for all }x,y\in\mathbb{R}_+.
$$
The later can be written equivalently
$$
C_\perp(u,v) \leq C(u,v)\text{ for all }u,v\in[0,1].
$$
An interesting interpretation of PQD association of lifetimes is given in \cite{Denuit_2004_JFE}: if lifetimes of a husband age $(x_m)$ and spouse age $(x_f)$ are positively quadrant dependent, then for all $t\in\mathbb{R}_+$,
$$
\mathbb{E}[T_{x_f}|T_{x_m}>t] \geq \mathbb{E}[T_{x_f}]\text{ and }
\mathbb{E}[T_{x_m}|T_{x_f}>t] \geq \mathbb{E}[T_{x_m}].
$$
The interpretation of those inequality is that knowing that one of the two spouses is still alive, at some time, increases the remaining lifetime of the other one.

\cite{gijbels} compares several test for positive quadrant dependence, and in this section, we use a Kolmogorov-Smirnov test, as in \cite{scaillet}. More specifically, we want to test $H_0:C\geq C_\perp$, and we use $$S_n=\displaystyle{\sqrt{n}\sup_{(u,v)}\lbrace C_\perp(u,v)-\widehat{C}_n(u,v) \rbrace},$$ where $\widehat{C}_n$ is the empirical copula.\footnote{and the $p$-value is approximated using standard bootstrap techniques, as described in section 3.2. of \cite{scaillet}. More specifically, if $\widehat{C}_n^\star$ is the empirical copula built from a bootstrap sample, define
$$S_n^\star=\displaystyle{\sqrt{n}\sup_{(u,v)}\lbrace \widehat{C}_n^\star(u,v)-\widehat{C}_n(u,v) \rbrace}$$
and then $p_n^\star=\mathbb{P}[S_n^\star>S_n]$. In the application, a $500\times500$ uniform grid is used to approximate the supremium on the unit-square.} Using 5,000 bootstrap samples, we obtain a $p$-value lower than 1\textperthousand, with 90\% chance. So we can claim, with strong confidence, that in our data, joint lifes are PQD.

\subsubsection{From Joint Distributions to Insurance Premiums}

Consider two positive random variables $X$ and $Y$, with marginal cumulative distributions $F_X$ and $F_Y$ respectively, and survival functions $S_X$ and $S_Y$. From \cite{frechet}, without any further assumption,
$$
\max\lbrace0,F_X(x)+F_y(y)-1\rbrace \leq \mathbb{P}[X\leq x,Y\leq y] \leq \min\lbrace F_X(x),F_y(y)\rbrace, 
$$
or equivalently
$$
\max\lbrace0,S_X(x)+S_y(y)-1\rbrace \leq \mathbb{P}[X>x,Y>y] \leq \min\lbrace S_X(x),S_y(y)\rbrace, 
$$
for all $x,y\in\mathbb{R}_+$. In the context of joint lifes, the later can be written
\begin{equation}\label{eq:bounds:tpx:1}
    \max\lbrace0,{}_t p_{x_m}+{}_t p_{x_f}-1\rbrace \leq {}_t p_{x_m,x_f} \leq \min\lbrace {}_t p_{x_m},{}_t p_{x_f}\rbrace,
\end{equation}
for all time $t\in\mathbb{R}_+$. The upper bound is obtained when the associated copula is $C^+$.

Sharper bounds can be derived, at least for the lower bound, assuming some  positive association between variables, namely the PQD property. In that case, the lower bounds corresponds to the independent case, and the associated copula is $C^\perp$. Thus, as a consequence, if lifetimes of a husband age $(x_m)$ and spouse age $(x_f)$ are positively quadrant dependent, then Equation (\ref{eq:bounds:tpx:1}) becomes
\begin{equation}\label{eq:bounds:tpx:2}
    {}_t p_{x_m}\cdot {}_t p_{x_f} \leq {}_t p_{x_m,x_f} \leq \min\lbrace {}_t p_{x_m},{}_t p_{x_f}\rbrace,
\end{equation}

Consider some quantity of interest $\mathcal{I}$ (that could be the life expectancy, or some annuity), so that $\mathcal{I}(X,Y)$ can be written $ \mathbb{E}[\varphi(X,Y)]$, where $\varphi$ is a supermodular function, in the sense that 
\begin{equation}\label{eq:supermodular}
    \varphi(x_1,y_1)+\varphi(x_2,y_2) \geq \varphi(x_1,y_2)+\varphi(x_2,y_1)\text{ for all }x_2\geq x_1,y_2\geq y_1. 
\end{equation}
As proved in \cite{Lorentz} and \cite{Cambanis}, comonotonic vectors maximize $\mathcal{I}(X,Y)$, in the sense that
$$
\mathcal{I}(F_X^{-1}(U),F_Y^{-1}(1-U))\leq \mathcal{I}(X,Y) \leq \mathcal{I}(F_X^{-1}(U),F_Y^{-1}(U)),
$$
where $U$ is uniformly distributed. And in the case where $(X,Y)$ are positively quandrant dependent,
$$
\mathcal{I}(X^\perp,Y^\perp)\leq \mathcal{I}(X,Y) \leq \mathcal{I}(X^+,Y^+),
$$
where $(X^\perp,Y^\perp)$ denotes an independent version of vector $(X,Y)$, in the sense that $X^\perp$ has the same distribution as $X$, $Y^\perp$ has the same distribution as $Y$, and the copula of $(X^\perp,Y^\perp)$ is $C^\perp$ -- and similarly for some $(X^+,Y^+)$. In the case where $\varphi$ is a supermodular function (with a {\em less or equal} instead of a {\em greater or equal} in Equation \ref{eq:supermodular}), bounds are inverted. As mentioned in \cite{carrierechen} and \cite{Denuit_2004_JFE}, those bounds appear when calculating various annuities.

\subsection{Bounds for Insurance Premiums}

As discussed previously assuming positive association between life times $T(x_f)$ and $T(x_m)$, the independence case and the comonotonic cases will provide lower and upper bounds for various quantities. Hence, in the independent case,
$$
a^\perp_{\overline{x}_f,\overline{x}_m} =\sum_{k=1}^\infty \nu^k \left({}_kp_{x_f}+{}_kp_{x_m}-{}_kp_{x_f}\cdot {}_kp_{x_m}\right),
$$
$$
a^\perp_{{x}_f,{x}_m}=\sum_{k=1}^\infty \nu^k {}_kp_{x_f}\cdot {}_kp_{x_m},
$$
$$
a^\perp_{{x}_f|{x}_f}=\sum_{k=1}^\infty \nu^k {}_kp_{x_m}-\sum_{k=1}^\infty \nu^k {}_kp_{x_f}\cdot {}_kp_{x_m},
$$
while in the perfectly correlated case,
$$
a^+_{\overline{x}_f,\overline{x}_m} =\sum_{k=1}^\infty \nu^k \left(1-\min\lbrace{}_kp_{x_f},{}_kp_{x_m}\rbrace\right),
$$
$$
a^+_{{x}_f,{x}_m}=\sum_{k=1}^\infty \nu^k \min\lbrace{}_kp_{x_f},{}_kp_{x_m}\rbrace,
$$
$$
a^+_{{x}_m|{x}_f}=\sum_{k=1}^\infty \nu^k {}_kp_{x_m}-\sum_{k=1}^\infty \nu^k \min\lbrace{}_kp_{x_f},{}_kp_{x_m}\rbrace.
$$
And as shown in \cite{Denuit_2004_JFE}, if lifetimes are positively quadrant dependent,
$$
a^\perp_{{x}_f,{x}_m} \leq a_{{x}_f,{x}_m} \leq a^+_{{x}_f,{x}_m},
$$
while
$$
a^+_{\overline{x}_f,\overline{x}_m} \leq a_{\overline{x}_f,\overline{x}_m} \leq a^\perp_{\overline{x}_f,\overline{x}_m},
$$
and
$$
a^+_{{x}_m|{x}_f} \leq a_{{x}_m|{x}_f} \leq a^\perp_{{x}_m|{x}_f}.
$$
Thus, for the last-survivor and the widow's pension, the independence assumption is conservative (as soon as lifetimes are positively associated). In that case, using the independence assumption for pricing those annuities will incorporate some safety loading (as in Figure \ref{fig:widows_pension} where we plot the present value of a widow's pension, $a_{\text{m}|\text{f}}$ (relative to independent case $a^{\perp}_{\text{m}|\text{f}}$), as a function of $x_{\text{m}}$.).

\section{Parents and Children Dependencies}\label{apprendix:parents}

\begin{figure}[H]
  \centering
    \includegraphics[width=.8\textwidth]{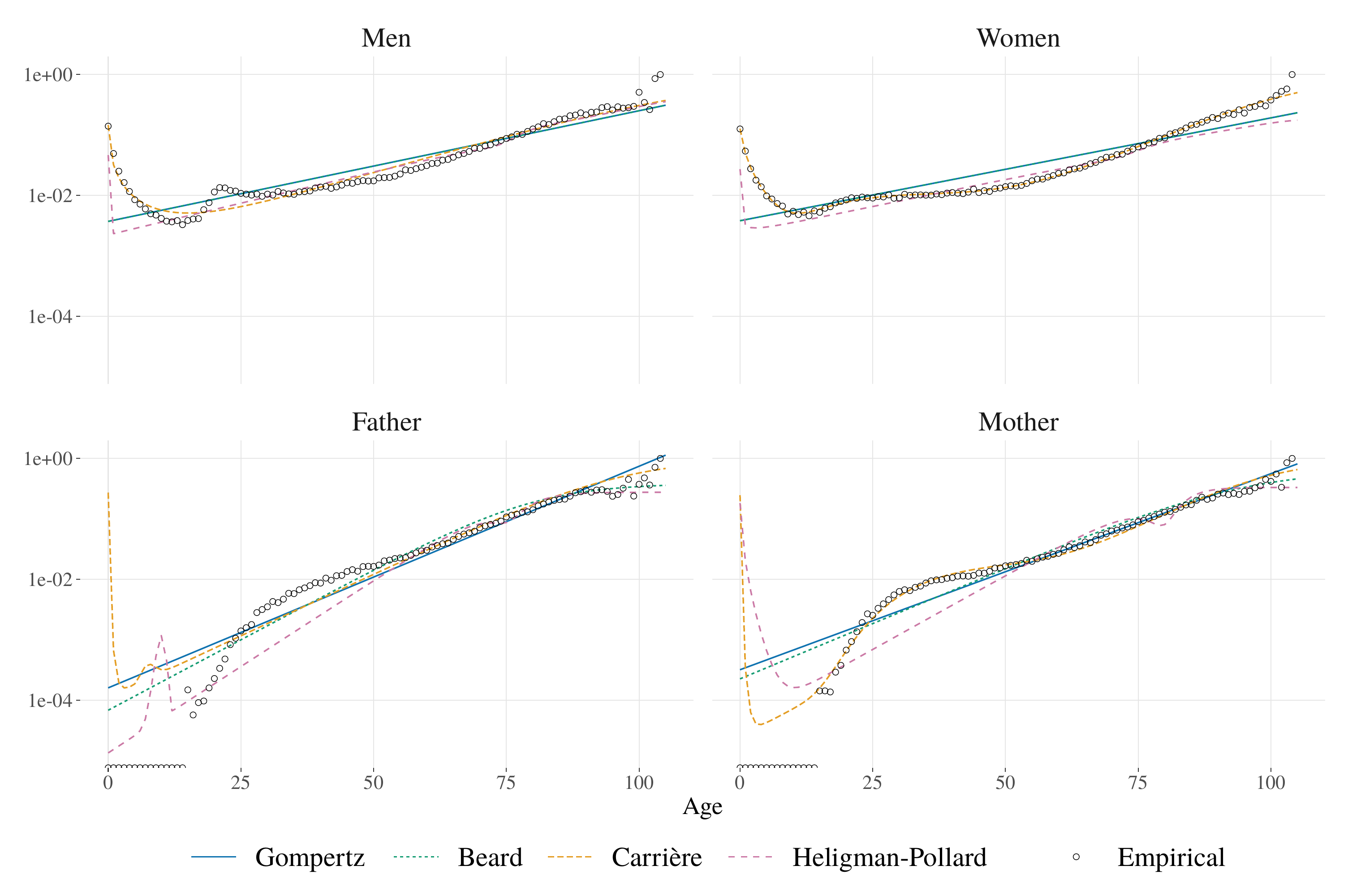}
\caption{Force of mortality (log scale) $\mu_x$ for individuals depending on their gender (top), and for their parents (bottom).}\label{fig:force_mortality_parents}
\end{figure}

\begin{figure}[H]
  \centering
    \includegraphics[width=.8\textwidth]{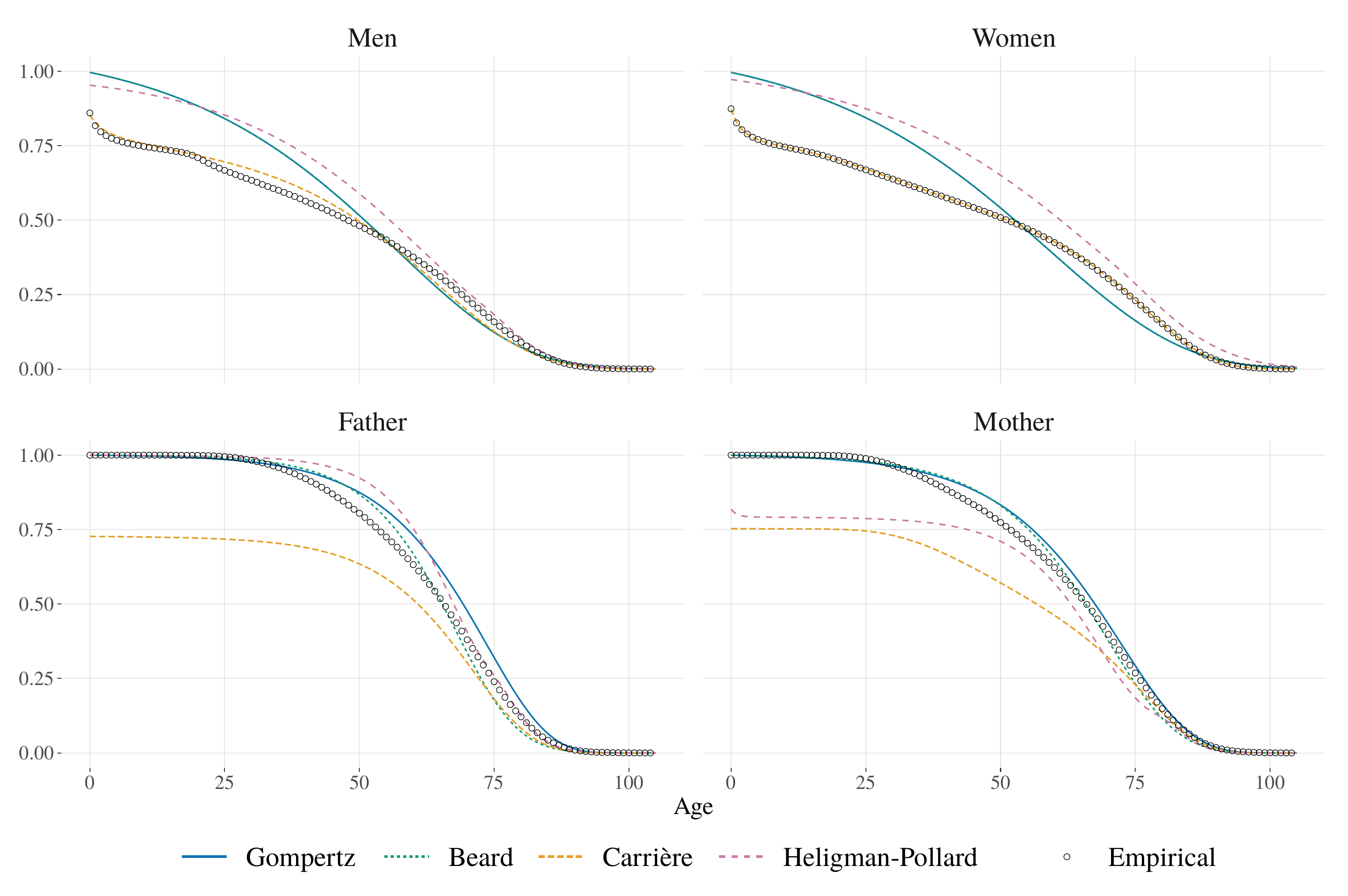}
\caption{Survival function for individuals depending on their gender (top), and for their parents (bottom).}\label{fig:survival_parents}
\end{figure}


\begin{figure}[H]
  \centering
    \includegraphics[width=\textwidth]{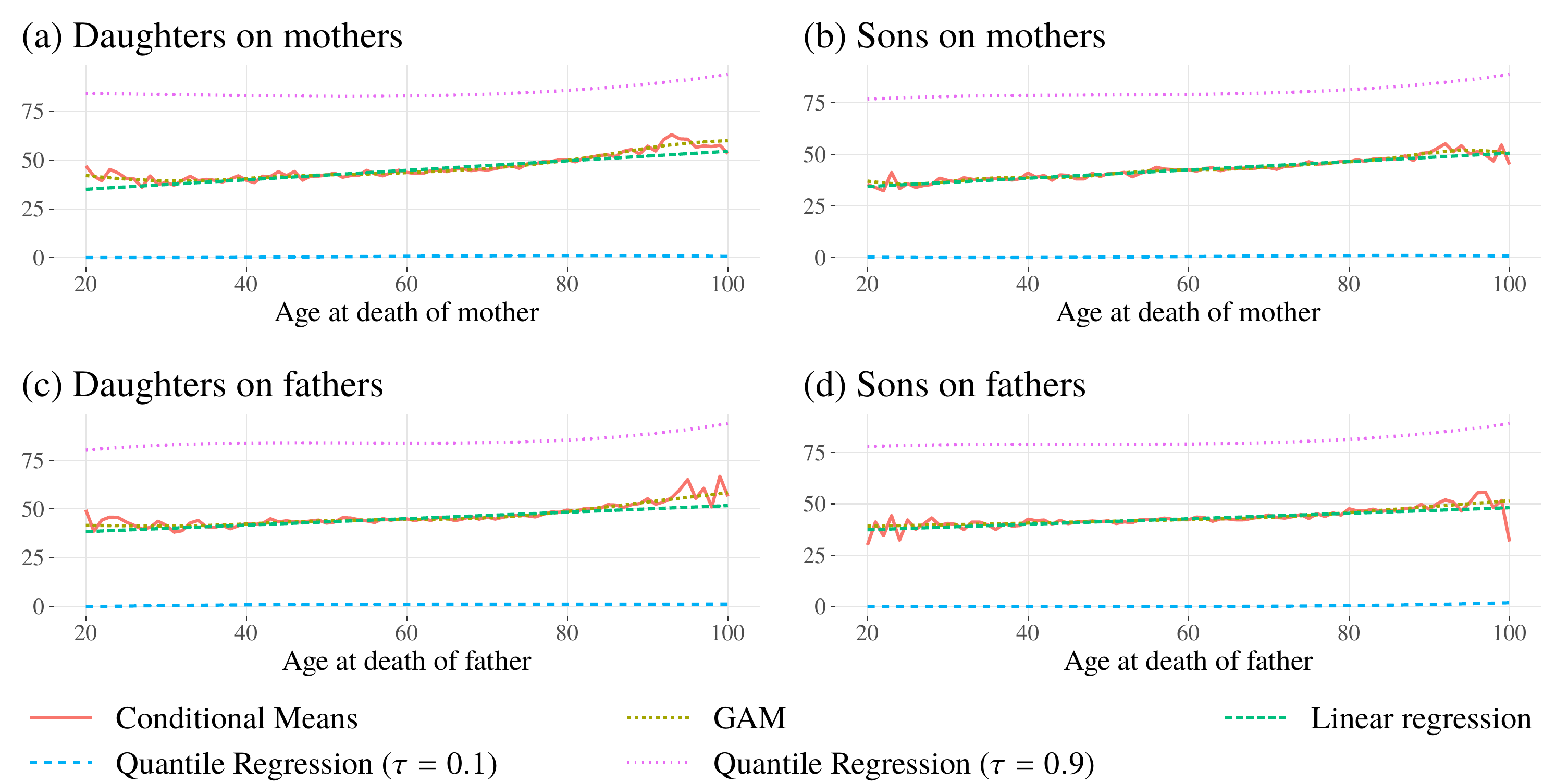}
  \begin{minipage}{\textwidth}
		\vspace{1ex}
		\scriptsize\underline{Note:} The blue and purple dashed lines correspond to the first and ninth deciles of quantile regression of age at death of an individual as a function of the age at death of their spouse.
\end{minipage}
\caption{Age of individuals (daughters and sons) as a function of age of parents.}\label{fig:p_parents_children_age}
\end{figure}

\begin{table}[H]
    \centering\scriptsize
\begin{tabular}{lrrr}
\hline\hline\\[-.75em]
 & \multicolumn{2}{c}{Variables}\\
\cmidrule(lr){2-3}\\[-.75em]
Relationship & Intercept & Age of the parents & R$^2$\\[.25em]
\cmidrule(lr){2-3}\cmidrule(lr){4-4}\\[-.75em]
Daughter vs. Mother & 30.17 [29.31,31.03] & 0.243 [0.230,0.256] & 0.0158\\
Son vs. Mother & 30.39 [29.60,31.18] & 0.202 [0.190,0.214] & 0.0120\\
Daughter vs. Father & 35.03 [34.06,36.00] & 0.167 [0.152,0.181] & 0.0059\\
Son vs. Father & 34.71 [33.82,35.59] & 0.134 [0.120,0.147] & 0.0042\\[.25em]
\hline\hline
\end{tabular}
\begin{minipage}{\textwidth}
		\vspace{1ex}
		\scriptsize\underline{Note:} The 95\% confidence intervals for each coefficient are provided between brackets next to the estimates.
\end{minipage}
    \caption{Linear regression coefficients of children's age at death as a function of parent's age at death.}
    \label{tab:lm_parents}
\end{table}

\begin{table}[H]
    \centering\tiny
    \begin{tabular}{rllrrrr}
\hline\hline\\[-.75em]
&&\multicolumn{2}{c}{Men} & \multicolumn{2}{c}{Women}\\[.25em]
\cmidrule(lr){3-4}\cmidrule(lr){5-6}\\[-.75em]
$x$ & Information used & $e_{x, \text{male}}$ & Dev. from Ref. & $e_{x, \text{female}}$ & Dev. from Ref.\\[.25em]
\cmidrule(lr){1-6}\\[-.75em]
\multicolumn{6}{c}{Information on parents when the child is 20 years old}\\[.25em]
\cmidrule(lr){1-6}\\[-.75em]
20 & Reference & 39.13 & 0.00 & 43.17 & 0.00\\
20 & Both parents still alive & 39.70 & 0.57 & 43.69 & 0.51\\
20 & Only mother still alive & 38.48 & -0.65 & 43.12 & -0.05\\
20 & Only father still alive & 38.02 & -1.11 & 41.37 & -1.81\\
20 & Only one parent still alive & 38.27 & -0.86 & 42.30 & -0.87\\
20 & Both parents deceased & 37.00 & -2.13 & 42.03 & -1.15\\
\addlinespace
30 & Reference & 33.31 & 0.00 & 36.87 & 0.00\\
30 & Both parents still alive & 33.71 & 0.40 & 37.27 & 0.40\\
30 & Only mother still alive & 32.97 & -0.34 & 36.73 & -0.14\\
30 & Only father still alive & 32.34 & -0.97 & 35.20 & -1.66\\
30 & Only one parent still alive & 32.67 & -0.64 & 36.02 & -0.84\\
30 & Both parents deceased & 31.85 & -1.45 & 36.71 & -0.16\\
\addlinespace
40 & Reference & 26.71 & 0.00 & 30.32 & 0.00\\
40 & Both parents still alive & 26.98 & 0.27 & 30.60 & 0.28\\
40 & Only mother still alive & 26.64 & -0.07 & 30.21 & -0.11\\
40 & Only father still alive & 25.81 & -0.90 & 29.13 & -1.19\\
40 & Only one parent still alive & 26.25 & -0.46 & 29.71 & -0.60\\
40 & Both parents deceased & 25.84 & -0.87 & 30.22 & -0.10\\[.25em]
\cmidrule(lr){1-6}\\[-.75em]
\multicolumn{6}{c}{Information on parents when the child is 30 years old}\\[.25em]
\cmidrule(lr){1-6}\\[-.75em]
30 & Reference & 33.31 & 0.00 & 36.87 & 0.00\\
30 & Both parents still alive & 34.13 & 0.82 & 37.99 & 1.12\\
30 & Only mother still alive & 33.27 & -0.03 & 37.07 & 0.20\\
30 & Only father still alive & 32.92 & -0.39 & 35.01 & -1.86\\
30 & Only one parent still alive & 33.12 & -0.19 & 36.18 & -0.69\\
30 & Both parents deceased & 31.68 & -1.63 & 35.78 & -1.09\\
\addlinespace
40 & Reference & 26.71 & 0.00 & 30.32 & 0.00\\
40 & Both parents still alive & 27.25 & 0.54 & 31.12 & 0.80\\
40 & Only mother still alive & 26.84 & 0.13 & 30.53 & 0.21\\
40 & Only father still alive & 26.22 & -0.49 & 28.70 & -1.62\\
40 & Only one parent still alive & 26.57 & -0.14 & 29.74 & -0.58\\
40 & Both parents deceased & 25.65 & -1.06 & 29.68 & -0.64\\[.25em]
\cmidrule(lr){1-6}\\[-.75em]
\multicolumn{6}{c}{Information on parents when the child is 40 years old}\\[.25em]
\cmidrule(lr){1-6}\\[-.75em]
40 & Reference & 26.71 & 0.00 & 30.32 & 0.00\\
40 & Both parents still alive & 28.04 & 1.33 & 31.90 & 1.58\\
40 & Only mother still alive & 27.26 & 0.55 & 31.54 & 1.22\\
40 & Only father still alive & 26.41 & -0.30 & 29.30 & -1.02\\
40 & Only one parent still alive & 26.91 & 0.20 & 30.66 & 0.34\\
40 & Both parents deceased & 25.77 & -0.94 & 29.10 & -1.22\\[.25em]
\hline\hline
\end{tabular}
\begin{minipage}{\textwidth}
		\vspace{1ex}
		\scriptsize\underline{Note:} The reference situation is one in which no information regarding the death's status of the parents is accounted for. $x$ is the age of children and $e_x$ the corresponding residual life expectancy expressed in years. The deviation from the reference is the difference between life expectancy when some information on parents death is used and life expectancy when such information is not accounted for.
\end{minipage}
    \caption{Residual life expectancy of children depending on information on parents.}
    \label{tab:ex_parents}
\end{table}

\clearpage

\section{Grandparents and Children Dependencies}\label{apprendix:grandparents}

\begin{table}[H]
    \centering\scriptsize
    \begin{tabular}{lrrrrrrrr}
\hline\hline\\[-.75em]
& Mean & SD & Min & Max & $Q_1$ & $Q_2$ & $Q_3$ & No Missing\\
\cmidrule(lr){2-9}\\[-.75em]
& \multicolumn{8}{c}{Men ($n=31,096$)}\\[.25em]
\cmidrule(lr){2-9}\\[-.75em]
Individual ($t_{\text{c}}$) & 39.1 & 29.5 & 0.0 & 104.0 & 3.8 & 43.3 & 65.7 & 0\\
Maternal grandfather & 64.6 & 14.3 & 15.0 & 104.1 & 55.2 & 66.7 & 75.4 & 0\\
Maternal grandmother & 63.3 & 15.3 & 15.0 & 104.0 & 53.0 & 65.8 & 75.0 & 0\\
Paternal grandfather & 64.4 & 14.2 & 15.0 & 104.1 & 55.1 & 66.5 & 75.2 & 0\\
Paternal grandmother & 62.9 & 15.2 & 15.0 & 102.5 & 53.0 & 65.1 & 74.5 & 0\\
Last survivor & 77.5 & 8.2 & 15.0 & 104.1 & 72.7 & 78.1 & 82.9 & 0\\
First to die & 48.4 & 12.2 & 15.0 & 100.0 & 39.2 & 48.1 & 57.6 & 0\\
Average grandparents & 63.8 & 8.0 & 15.0 & 100.0 & 58.6 & 64.2 & 69.5 & 0\\[.25em]
\cmidrule(lr){2-9}\\[-.75em]
& \multicolumn{8}{c}{Women ($n=28,367$)}\\[.25em]
\cmidrule(lr){2-9}\\[-.75em]
Individual ($t_{\text{c}}$) & 38.9 & 30.1 & 0.0 & 104.4 & 4.2 & 40.4 & 66.6 & 0\\
Maternal grandfather & 64.7 & 14.3 & 15.7 & 104.2 & 55.3 & 66.9 & 75.5 & 0\\
Maternal grandmother & 63.2 & 15.3 & 15.2 & 104.0 & 53.0 & 65.7 & 74.9 & 0\\
Paternal grandfather & 64.4 & 14.3 & 15.0 & 104.1 & 54.9 & 66.3 & 75.3 & 0\\
Paternal grandmother & 62.9 & 15.2 & 16.0 & 102.6 & 52.7 & 65.2 & 74.6 & 0\\
Last survivor & 77.5 & 8.2 & 25.0 & 104.2 & 72.6 & 78.1 & 83.0 & 0\\
First to die & 48.4 & 12.2 & 15.0 & 102.0 & 39.2 & 48.1 & 57.6 & 0\\
Average grandparents & 63.8 & 8.1 & 25.0 & 102.5 & 58.5 & 64.1 & 69.6 & 0\\[.25em]
\cmidrule(lr){2-9}\\[-.75em]
& \multicolumn{8}{c}{Men \& Women ($n=59,463$)}\\[.25em]
\cmidrule(lr){2-9}\\[-.75em]
Individual ($t_{\text{c}}$) & 39.0 & 29.8 & 0.0 & 104.4 & 4.0 & 42.0 & 66.1 & 0\\
Maternal grandfather & 64.6 & 14.3 & 15.0 & 104.2 & 55.2 & 66.7 & 75.4 & 0\\
Maternal grandmother & 63.3 & 15.3 & 15.0 & 104.0 & 53.0 & 65.7 & 74.9 & 0\\
Paternal grandfather & 64.4 & 14.3 & 15.0 & 104.1 & 55.0 & 66.4 & 75.3 & 0\\
Paternal grandmother & 62.9 & 15.2 & 15.0 & 102.6 & 52.9 & 65.1 & 74.6 & 0\\
Last survivor & 77.5 & 8.2 & 15.0 & 104.2 & 72.7 & 78.1 & 83.0 & 0\\
First to die & 48.4 & 12.2 & 15.0 & 102.0 & 39.2 & 48.1 & 57.6 & 0\\
Average grandparents & 63.8 & 8.0 & 15.0 & 102.5 & 58.6 & 64.1 & 69.5 & 0\\[.25em]
\hline\hline
\end{tabular}
\begin{minipage}{\textwidth}
		\vspace{1ex}
		\scriptsize\underline{Note:} this table provides descriptive statistics of the ages contained in the dataset of grandchildren and their grandparents. $n$ stands for the number of observations, $SD$ is the standard deviation, $Q_1$, $Q_2$, and $Q_3$ are the first, second, and third empirical quartiles, respectively, and No. Missing refers to the number of missing values.
\end{minipage}
    \caption{Age at death of the individuals and age of their grandparents, according to the gender of the grandchildren, keeping only grandchildren whose four grandparents are known.}
    \label{tab:desc_stat_gparents_4gp}
\end{table}

\begin{figure}[H]
  \centering
    \includegraphics[width=\textwidth]{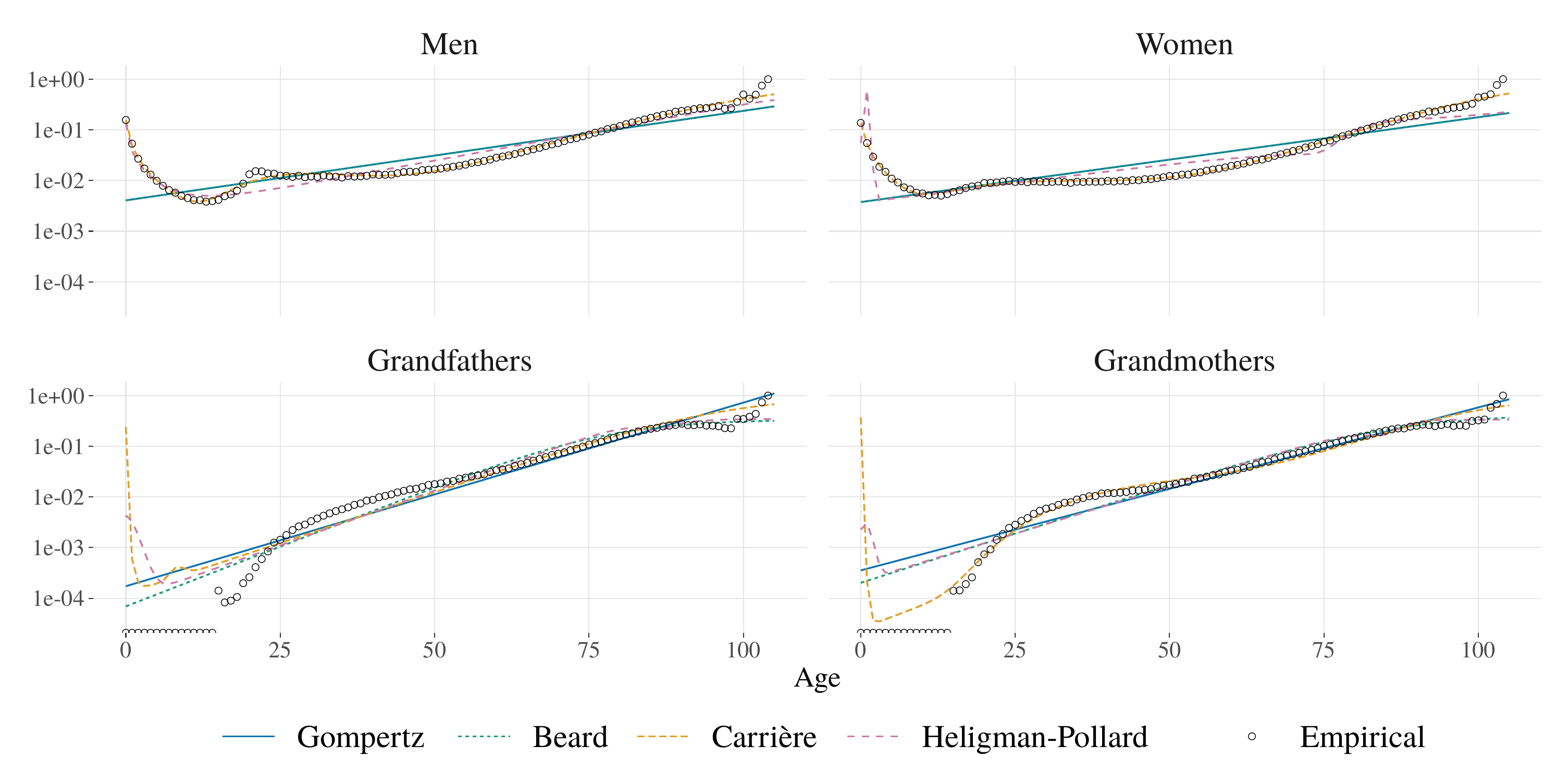}
  \begin{minipage}{\textwidth}
		\vspace{1ex}
		\scriptsize\underline{Note:} 
\end{minipage}
\caption{Force of mortality (log scale) $\mu_x$ for individuals depending on their gender (top), and for their grandparents (bottom).}\label{fig:force_mortality_gparents}
\end{figure}

\begin{figure}[H]
  \centering
    \includegraphics[width=\textwidth]{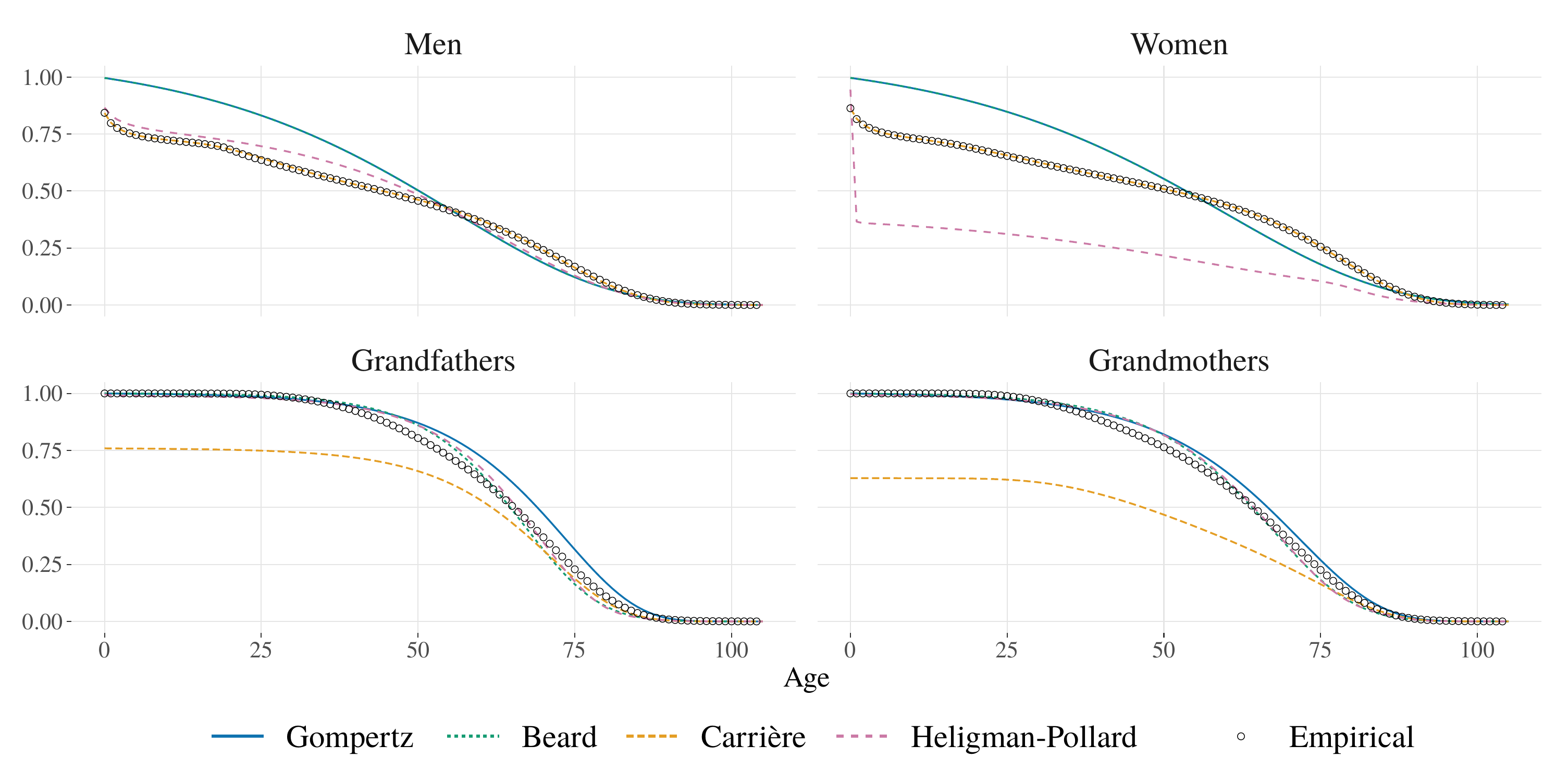}
  \begin{minipage}{\textwidth}
		\vspace{1ex}
		\scriptsize\underline{Note:} 
\end{minipage}
\caption{Survival function for individuals depending on their gender (top), and for their grandparents (bottom).}\label{fig:survival_gparents}
\end{figure}

\begin{figure}[H]
  \centering
    \includegraphics[width=\textwidth]{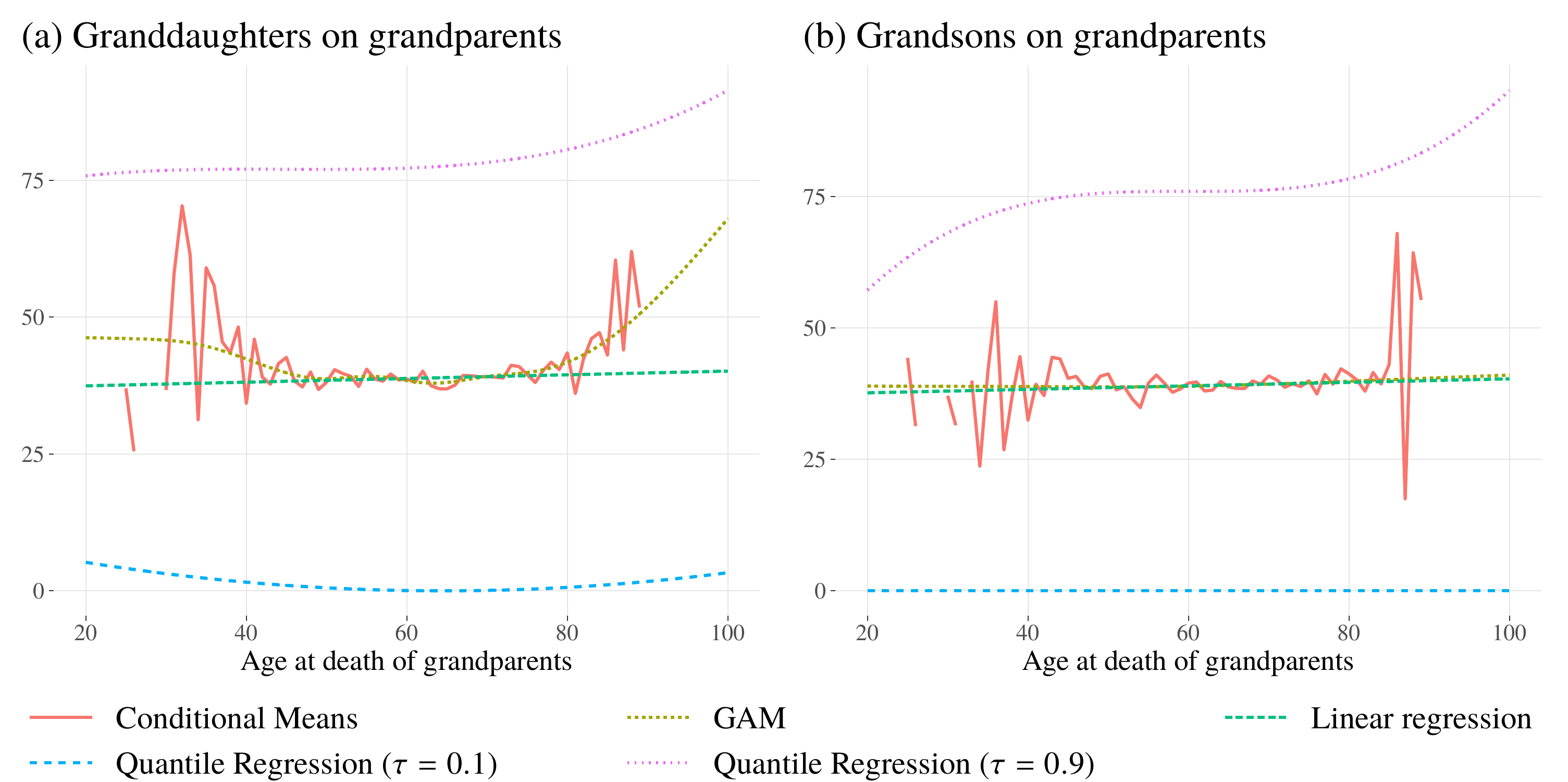}
  \begin{minipage}{\textwidth}
		\vspace{1ex}
		\scriptsize\underline{Note:} 
\end{minipage}
\caption{Age of individuals (daughters and sons) as a function of age of parents.}\label{fig:p_gparents_children_age}
\end{figure}

\begin{table}[H]
    \centering\scriptsize
\begin{tabular}{lrrr}
\hline\hline\\[-.75em]
 & \multicolumn{2}{c}{Variables}\\
\cmidrule(lr){2-3}\\[-.75em]
Grandchildren & Intercept & Age of the parents & R$^2$\\[.25em]
\cmidrule(lr){2-3}\cmidrule(lr){4-4}\\[-.75em]
Granddaughter & 37.113 [34.32,39.91] & 0.0280 [-0.0154,0.0715] & 0.000056\\
Grandson & 37.056 [34.43,39.69] & 0.0325 [-0.0084,0.0734] & 0.000078\\[.25em]
\hline\hline
\end{tabular}
\begin{minipage}{\textwidth}
		\vspace{1ex}
		\scriptsize\underline{Note:} The 95\% confidence intervals for each coefficient are provided between brackets next to the estimates.
\end{minipage}
    \caption{Linear regression coefficients of grandchildren's age at death as a function of grandparent's age at death.}
    \label{tab:lm_gparents}
\end{table}

\begin{table}[H]
    \centering\tiny
    \begin{tabular}{rllrrrr}
\hline\hline\\[-.75em]
&&\multicolumn{2}{c}{Men} & \multicolumn{2}{c}{Women}\\[.25em]
\cmidrule(lr){3-4}\cmidrule(lr){5-6}\\[-.75em]
$x$ & Information used & $e_{x, \text{male}}$ & Dev. from Ref. & $e_{x, \text{female}}$ & Dev. from Ref.\\[.25em]
\cmidrule(lr){1-6}\\[-.75em]
\multicolumn{6}{c}{Information on grandparents when the grandson is 10 years old}\\[.25em]
\cmidrule(lr){1-6}\\[-.75em]
10 & Reference & 44.84 & 0.00 & 44.60 & 0.00\\
10 & All grandparents deceases & 43.92 & -0.92 & 43.29 & -1.30\\
10 & Only 1 grandparent still alive & 45.06 & 0.22 & 45.04 & 0.44\\
10 & Only 2 grandparents still alive & 45.55 & 0.71 & 45.28 & 0.68\\
10 & Only 3 grandparents still alive & 45.25 & 0.41 & 45.82 & 1.23\\
10 & All grandparents still alive & 47.26 & 2.42 & 47.38 & 2.78\\
\addlinespace
15 & Reference & 40.88 & 0.00 & 41.06 & 0.00\\
15 & All grandparents deceases & 40.03 & -0.85 & 39.73 & -1.33\\
15 & Only 1 grandparent still alive & 41.07 & 0.19 & 41.58 & 0.52\\
15 & Only 2 grandparents still alive & 41.54 & 0.66 & 41.66 & 0.60\\
15 & Only 3 grandparents still alive & 41.18 & 0.30 & 42.46 & 1.40\\
15 & All grandparents still alive & 43.44 & 2.56 & 43.20 & 2.13\\
\addlinespace
20 & Reference & 37.08 & 0.00 & 37.71 & 0.00\\
20 & All grandparents deceases & 36.23 & -0.86 & 36.35 & -1.35\\
20 & Only 1 grandparent still alive & 37.26 & 0.18 & 38.27 & 0.57\\
20 & Only 2 grandparents still alive & 37.67 & 0.59 & 38.24 & 0.53\\
20 & Only 3 grandparents still alive & 37.46 & 0.38 & 39.14 & 1.44\\
20 & All grandparents still alive & 40.24 & 3.16 & 40.03 & 2.32\\[.25em]
\cmidrule(lr){1-6}\\[-.75em]
\multicolumn{6}{c}{Information on grandparents when the grandson is 15 years old}\\[.25em]
\cmidrule(lr){1-6}\\[-.75em]
15 & Reference & 40.88 & 0.00 & 41.06 & 0.00\\
15 & All grandparents deceases & 40.48 & -0.40 & 40.20 & -0.86\\
15 & Only 1 grandparent still alive & 41.12 & 0.24 & 41.63 & 0.57\\
15 & Only 2 grandparents still alive & 41.70 & 0.82 & 42.34 & 1.28\\
15 & Only 3 grandparents still alive & 40.94 & 0.06 & 42.67 & 1.61\\
15 & All grandparents still alive & 42.69 & 1.82 & 43.33 & 2.27\\
\addlinespace
20 & Reference & 37.08 & 0.00 & 37.71 & 0.00\\
20 & All grandparents deceases & 36.68 & -0.41 & 36.85 & -0.86\\
20 & Only 1 grandparent still alive & 37.28 & 0.19 & 38.21 & 0.51\\
20 & Only 2 grandparents still alive & 37.94 & 0.85 & 38.99 & 1.29\\
20 & Only 3 grandparents still alive & 37.42 & 0.33 & 39.44 & 1.74\\
20 & All grandparents still alive & 38.67 & 1.59 & 40.92 & 3.21\\[.25em]
\cmidrule(lr){1-6}\\[-.75em]
\multicolumn{6}{c}{Information on grandparents when the grandson is 20 years old}\\[.25em]
\cmidrule(lr){1-6}\\[-.75em]
20 & Reference & 37.08 & 0.00 & 37.71 & 0.00\\
20 & All grandparents deceases & 36.88 & -0.20 & 37.17 & -0.54\\
20 & Only 1 grandparent still alive & 37.59 & 0.51 & 38.49 & 0.78\\
20 & Only 2 grandparents still alive & 37.14 & 0.06 & 39.26 & 1.55\\
20 & Only 3 grandparents still alive & 37.98 & 0.90 & 41.02 & 3.32\\
20 & All grandparents still alive & 38.51 & 1.43 & 42.10 & 4.40\\[.25em]
\hline\hline
\end{tabular}
\begin{minipage}{\textwidth}
		\vspace{1ex}
		\scriptsize\underline{Note:} The reference situation is one in which no information regarding the death's status of the grandparents is accounted for. $x$ is the age of grandchildren and $e_x$ the corresponding residual life expectancy expressed in years. The deviation from the reference is the difference between life expectancy when some information on grandparents death is used and life expectancy when such information is not accounted for.
\end{minipage}
    \caption{Residual life expectancy of grandchildren depending on information on grandparents.}
    \label{tab:ex_gdparents}
\end{table}

\clearpage

\bibliographystyle{apa}
\bibliography{bibliography}

\end{document}